\documentclass[iop,twocolumn,tighetn,10pt,apj]{emulateapj}
\pdfoutput=1
\usepackage{graphicx}
\usepackage{amssymb}
\usepackage{amsmath}
\usepackage{color}
\usepackage{multirow}

\slugcomment{Accepted for publication in ApJS}
\shorttitle{AMR Computations with the PLUTO Code}
\shortauthors{Mignone et al.}
\begin{document}
\newcommand{\HALF}{\frac{1}{2}}
\newcommand{\pd}[2]{\frac{\partial #1}{\partial #2}}
\newcommand{\DS}{\displaystyle}
\renewcommand{\vec}[1]{\mathbf{#1}}
\newcommand{\Pivec}{\boldsymbol{\Pi}}
\newcommand{\Lvec}{\boldsymbol{\cal L}}
\newcommand{\hvec}[1]{\hat{\mathbf{#1}}}
\newcommand{\mm}[1]{\rm mm}
\newcommand{\tens}[1]{\mathsf{#1}}
\newcommand{\PLUTO}{PLUTO$\;$}
\newcommand{\PlCh}{PLUTO-CHOMBO$\;$}
\newcommand{\red}[1]{\color{red} #1 \color{black}}
\newcommand{\blue}[1]{\color{blue} #1 \color{black}}
\newcommand{\yes}{1}
\newcommand{\no}{0}
\newcommand{\vol}{{\cal V}}

\let\IncludeEps\yes    
\title{The PLUTO Code for Adaptive Mesh Computations in Astrophysical Fluid Dynamics}

\author{A. Mignone\altaffilmark{1}, C. Zanni\altaffilmark{2},
        P. Tzeferacos\altaffilmark{1}, B. van Straalen\altaffilmark{3}, P. Colella\altaffilmark{3}
        and G. Bodo\altaffilmark{2}}

\altaffiltext{1}{Dipartimento di Fisica Generale, Universit\'a di Torino,
                 via Pietro Giuria 1, 10125 Torino, Italy}
\altaffiltext{2}{INAF, Osservatorio Astronomico di Torino, Strada Osservatorio 20,
                 Pino Torinese, Italy}
\altaffiltext{3}{Lawrence Berkeley National Laboratory, 
                 1 Cyclotron Road, MS 50A-1148, Berkeley, CA 94720}

\begin{abstract}
We present a description of the adaptive mesh refinement (AMR) implementation 
of the PLUTO code for solving the equations of classical and special
relativistic magnetohydrodynamics (MHD and RMHD).
The current release exploits, in addition to the static grid version of the 
code, the distributed infrastructure of the CHOMBO library for multidimensional 
parallel computations over  block-structured, adaptively refined grids.
We employ a conservative finite-volume approach where primary flow quantities 
are discretized at the cell-center in a dimensionally unsplit fashion using the 
Corner Transport Upwind (CTU) method.
Time stepping relies on a characteristic tracing step where piecewise parabolic
method (PPM), weighted essentially non-oscillatory (WENO) or slope-limited 
linear interpolation schemes can be handily adopted. 
A characteristic decomposition-free version of the scheme is also illustrated.
The solenoidal condition of the magnetic field is enforced by augmenting the 
equations with a generalized Lagrange multiplier (GLM) providing propagation 
and damping of divergence errors through a mixed hyperbolic/parabolic explicit 
cleaning step.
Among the novel features, we describe an extension of the scheme to include 
non-ideal dissipative processes such as viscosity, resistivity and
anisotropic thermal conduction without operator splitting. 
Finally, we illustrate an efficient treatment of point-local, potentially 
stiff source terms over hierarchical nested grids by taking advantage of the 
adaptivity in time.
Several multidimensional benchmarks and applications to problems of 
astrophysical relevance assess the potentiality of the AMR version of 
PLUTO in resolving flow features separated by large spatial and temporal 
disparities.
\end{abstract}

\keywords{hydrodynamics - magnetohydrodynamics (MHD) - methods: numerical - relativity}

\section{Introduction}
%
%
%

Theoretical advances in modern astrophysics have largely benefited from 
computational models and techniques that have been improved over the past 
decades. 
In the field of gasdynamics, shock-capturing schemes represent the current 
establishment for reliable numerical simulations of high Mach-number, possibly 
magnetized flows in Newtonian or relativistic regimes.
As increasingly more sophisticated methods developed, a number of computer codes 
targeting complex physical aspects to various degrees have now become available 
to the community. 
In the field of magnetohydrodynamics (MHD), examples worth of notice are 
AstroBEAR \citep{AstroBEAR09}, Athena \citep{Athena08, SO10}, BATS-R-US \citep{Tothetal11},
ECHO \citep{ECHO07}, FLASH \citep{FLASH00}, NIRVANA \citep{NIRVANA08}, 
PLUTO \citep{PLUTO}, RAMSES \citep{RAMSES02, FHT06} and VAC \citep{VAC96,HKM08}. 
Some of these implementions provide additional capabilities that can approach the 
solution of the equations in the relativistic regimes: AMRVAC and PLUTO for 
special relativistic hydro, while the ECHO code allows to handle general 
relativistic MHD with a fixed metric. 
Other frameworks were specifically designed for special or general relativistic 
purposes, e.g. the RAM code \citep{RAM06}, HARM  \citep{HARM03} and RAISHIN 
\citep{RAISHIN06}.

In some circumstances, adequate theoretical modeling of astrophysical scenarios 
may become extremely challenging since great disparities in the spatial and 
temporal scales may simultaneously arise in the problem of interest. 
In these situations a static grid approach may become quite inefficient and, 
in the most extreme cases, the amount of computational time can make the 
problem prohibitive.
Typically, such conditions occur when the flow dynamics exhibit very localized 
features that evolve on a much shorter scale when compared to the rest of the 
computational domain.
To overcome these limitations, one possibility is to change or adapt the 
computational grid dynamically in space and time so that the features of 
interest can be adequately captured and resolved. 
Adaptive mesh refinement (AMR) is one such technique and can lead, for a certain 
class of problems, to a considerable speed up.
Some of the aforementioned numerical codes provide AMR implementations 
through a variety of different approaches. 
Examples worth of notice are the patch-based block-structured approach of 
\cite{BergerOli84,BergerCol89} (e.g. ASTROBEAR), 
the fully-octree approach described in \cite{DT93, Khokhlov98} (e.g.
RAMSES) or the block-based octree of \cite{MacNeice00} (e.g. FLASH) and
\cite{KNTG03, HK07} (e.g. BATS-R-US, AMRVAC). 

The present work focuses on the block-structured AMR implementation in 
the PLUTO code and its application to computational astrophysical gasdynamics.
\PLUTO is a Godunov-type code providing a flexible and versatile modular 
computational framework for the solution of the equations of gasdynamics under 
different regimes (e.g., classical/relativistic fluid dynamics, Euler/MHD).
A comprehensive description of the code design and implementation
may be found, for the static grid version, in \cite{PLUTO} (paper I henceforth).
Recent additions to the code include a relativistic version of the HLLD 
Riemann solver \citep{MUB09}, high-order finite difference schemes 
\citep{MTB10} and optically thin radiative losses with a non-equilibrium 
chemical network \citep{Tes08}.
Here we further extend the code description and show its performance on problems 
requiring significant usage of adaptively refined nested grids.
\PLUTO takes advantage of the CHOMBO library\footnote{
\texttt{https://seesar.lbl.gov/anag/chombo/}} that provides a distributed 
infrastructure for parallel computations over block-structured adaptively 
refined grids.  
The choice of block-structured AMR (as opposed to octree) is justified by the 
need of exploiting the already implemented modular skeleton introducing the 
minimal amount of modification and, at the same time, maximizing code 
re-usability.

The current AMR implementation leans on the Corner-Transport-Upwind 
\citep[CTU, ][]{Col90} method of \cite{MT10} (MT henceforth) in which a 
conservative finite volume discretization is adopted to evolve zone averages
in time.
The scheme is dimensionally unsplit, second-order accurate in space and time 
and can be directly applied to relativistic MHD as well.
Spatial reconstruction can be carried out in primitive or characteristic 
variables using high-order interpolation schemes such as the piecewise parabolic 
method \citep[PPM, ][]{CW84}, weighted essentially non-oscillatory (WENO) or 
linear Total Variation Diminishing (TVD) limiting.
The divergence-free constraint of magnetic field is enforced via a mixed 
hyperbolic/parabolic correction of \cite{Dedner02} that avoids the computational 
cost associated with an elliptic cleaning step, and the scrupulous treatment of 
staggered fields demanded by constrained transport algorithms \citep{Balsara04}.
As such, this choice provides a convenient first step in porting a 
considerable fraction of the static grid implementation to the AMR framework.
Among the novel features, we also show how to extend the time-stepping scheme to 
include dissipative terms describing viscous, resistive and thermally 
conducting flows.
Besides, we propose a novel treatment for efficiently computing the 
time-step in presence of cooling and/or reacting flows over hierarchical 
block-structured grids.

The paper is structured as follows. In Section \ref{sec:equation} we overview 
the relevant equations while in Section \ref{sec:numint} we describe the 
integration scheme used on the single patch.
In Section \ref{sec:CHOMBO} an overview of the block-structured AMR strategy as 
implemented in CHOMBO is given. 
Sections \ref{sec:MHD_Test} and \ref{sec:RMHD_Test} show the code performance on 
selected multidimensional test problems and astrophysical applications 
in classical and relativistic MHD, respectively.
Finally, in Section \ref{sec:summary} we summarize the main results of our work.

\section{Relevant Equations}\label{sec:equation}
%
%
%
%
%
%
%

The \PLUTO code has been designed for the solution of nonlinear systems of 
conservative partial differential equations of the mixed hyperbolic/parabolic 
type. In the present context we will focus our attention on the equations of
single-fluid magnetohydrodynamics, both in the Newtonian (MHD) and 
special relativistic (RMHD) regimes.

\subsection{MHD equations}
%
%

We consider a Newtonian fluid with density $\rho$, velocity 
$\vec{v}=(v_x,v_y,v_z)$ and magnetic induction $\vec{B} = (B_x, B_y, B_z)$ and 
write the single fluid MHD equations as 
\begin{equation}\label{eq:mhd}
\begin{array}{lcl}
\DS \pd{\rho}{t} + \nabla\cdot\left(\rho\vec{v}\right) &=& 0 \,,
 \\ \noalign{\medskip}
\DS \pd{\vec{(\rho\vec{v})}}{t} + \nabla\cdot\left[
   \rho\vec{v}\vec{v}^T 
 -     \vec{B}\vec{B}^T\right] 
 + \nabla p_t&=& \nabla\cdot\tens{\tau} + \rho\vec{g}\,,
\\ \noalign{\medskip}
\DS \pd{\cal E}{t} + \nabla\cdot\left[
  \left({\cal E} + p_t\right)\vec{v} - 
  \left(\vec{v}\cdot\vec{B}\right)\vec{B}\right]& =& 
    \nabla\cdot\tens{\Pi}_{\cal E} - \Lambda + \rho\vec{v}\cdot\vec{g}\,,
\\ \noalign{\medskip}
\DS \pd{\vec{B}}{t} - \nabla\times\left(\vec{v}\times\vec{B}\right) 
   &=& -\nabla\times\left(\tens{\eta}\vec{J}\right)\,,
\end{array}
\end{equation}
where $p_t = p + \vec{B}^2/2$ is the total (thermal+magnetic) pressure, 
${\cal E}$ is the total energy density, $\vec{g}$ is the gravitational 
acceleration term and $\Lambda$ accounts for optically thin radiative losses or heating.
Divergence terms on the right-hand side account for dissipative physical 
processes and are described in detail in Section \ref{sec:nonideal}.
Proper closure is provided by choosing an equation of state (EoS) which,
for an ideal gas, allows to write the total energy density as
\begin{equation}
 {\cal E} = \frac{p}{\Gamma - 1} + \frac{1}{2}\rho\vec{v}^2 
                                 + \frac{1}{2}\vec{B}^2\,,
\end{equation}
with $\Gamma$ being the specific heat ratio.
Alternatively, by adopting a barotropic or an isothermal EoS, the energy equation can be
discarded and one simply has, respectivelty, $p=p(\rho)$ or $p = c_s^2\rho$ 
(where $c_s$ is the constant speed of sound).

Chemical species and passive scalars are advected with the 
fluid and are described in terms of their number fraction $X_\alpha$
where $\alpha=1,\cdots,N_{\rm ions}$ label the particular ion.
They obey non-homogeneous transport equations of the form
\begin{equation}\label{eq:species}
 \pd{(\rho X_\alpha)}{t} + \nabla\cdot\left(\rho X_\alpha\vec{v}\right)  =
   \rho S_\alpha \,,
\end{equation}
where the source term $S_\alpha$ describes the coupling between different 
chemical elements inside the reaction network \cite[see for instance][]{Tes08}.

\subsubsection{Non-Ideal Effects}
\label{sec:nonideal}
%
%
%

Non-ideal effects due to dissipative processes are described by the
differential operators included on the right-hand side of Eq. (\ref{eq:mhd}).
Viscous stresses may be included through the viscosity tensor  $\tens{\tau}$ 
defined by 
\begin{equation}\label{eq:visc_tens}
 \tens{\tau} = \rho\nu\left[
      \nabla\vec{v} + (\nabla\vec{v})^T - \frac{2}{3} 
      \tens{I}\nabla\cdot\vec{v}\right]\,,
\end{equation}
where $\nu$ is the kinematic viscosity and $\tens{I}$ is the identity matrix. 
Similarly, magnetic resistivity is accounted for by prescribing the 
resistive $\eta$ tensor (diagonal).
Dissipative terms contribute to the net energy balance through the additional 
flux $\tens{\Pi}_{\cal E}$ appearing on the right hand side of Eq. (\ref{eq:mhd}):
\begin{equation}
\tens{\Pi}_{\cal E} = \vec{F}_c + \vec{v}\cdot\tens{\tau} -
    \tens{\eta}\cdot\vec{J}\times\vec{B}\,,
\end{equation}
where the different terms give the energy flux contributions due to, 
respectively, thermal conductivity, viscous stresses and magnetic resistivity.

The thermal conduction flux $\vec{F}_c$ smoothly varies between classical 
and saturated regimes and reads:
\begin{equation}\label{eq:tc_flux}
  \vec{F}_c = \frac{q}{|\vec{F}_{\rm class}| + q} \vec{F}_{\rm class}
\end{equation}
where $q = 5\phi\rho c_{\rm iso}^3$ is the magnitude of the saturated flux 
\citep{CMK77}, $\phi$ is a parameter of order unity accounting for uncertainties
in the estimate of $q$, $c_{\rm iso}$ is the isothermal speed of sound and 
\begin{equation}
\vec{F}_{\rm class} = \kappa_\parallel\hvec{b}\Big(\hvec{b}\cdot\nabla T\Big) 
+ \kappa_\perp\left[\nabla T - \hvec{b}\Big(\hvec{b}\cdot\nabla T\Big)\right]
\end{equation}
is the classical heat flux with conductivity coefficients $\kappa_\parallel$ 
and $\kappa_\perp$ along and across the magnetic field lines, respectively 
\citep{Orlando08}.
Indeed, the presence of a partially ordered magnetic field introduces a large 
anisotropic behavior by channeling the heat flux along the field lines
while suppressing it in the transverse direction (here 
$\hvec{b} = \vec{B}/|\vec{B}|$ is a unit vector along the field line).
We point out that, in the classical limit $q\to\infty$, thermal conduction is
described by a purely parabolic operator and flux discretization follows standard 
finite difference.
In the saturated limit ($|\nabla T|\to\infty$), on the other hand, the equation 
becomes hyperbolic and thus an upwind discretization of the flux is more 
appropriate \citep{BTH08}. 
This is discussed in more detail in Appendix \ref{app:satTC}.

%

\subsection{Relativistic MHD equations}
%
%

A (special) relativistic extension of the previous equations requires the 
solution of energy-momentum and number density conservation.
Written in divergence form we have
\begin{equation}
\label{eq:rmhd}
\begin{array}{lcl}
\DS \pd{(\rho\gamma)}{t}
   + \nabla\cdot\left(\rho\gamma\vec{v}\right) &=& 0 \,,
 \\ \noalign{\medskip}
\DS \pd{\vec{m}}{t}        
  + \nabla\cdot\left[
     w\gamma^2\vec{v}\vec{v} - \vec{B}\vec{B} - \vec{E}\vec{E}\right]
  + \nabla p_t &=&0  \,,
 \\ \noalign{\medskip}
\DS \pd{\vec{B}}{t} - \nabla\times\left(\vec{v}\times\vec{B}\right) &=& 0  \,,
\\ \noalign{\medskip}
\DS \pd{\cal E}{t}  
  + \nabla\cdot\left(\vec{m}-\rho\gamma\vec{v}\right) &=& 0  \,,
\end{array}
\end{equation}
where $\rho$ is the rest-mass density, $\gamma$ the Lorentz factor,
velocities are given in units of the speed of light ($c=1$)
and the fluid momentum $\vec{m}$ accounts for matter and electromagnetic 
terms: $\vec{m} = w\gamma^2\vec{v} + \vec{E}\times\vec{B}$,
where $\vec{E}=-\vec{v}\times\vec{B}$ is the electric field and $w$ is
the gas enthalpy.
Total pressure and energy include thermal and magnetic contributions and 
can be written as
\begin{equation}
 p_t = p + \frac{\vec{B}^2 + \vec{E}^2}{2} \,,\quad
 {\cal E} = w\gamma^2 - p +\frac{\vec{B}^2 + \vec{E}^2}{2} - \rho\gamma\,.
\end{equation}
Finally, the gas enthalpy $w$ is related to $\rho$ and $p$ via an equation 
of state which can be either the ideal gas law,
\begin{equation}\label{eq:gammaEos}
  w = \rho + \frac{\Gamma p}{\Gamma-1} \,,
\end{equation}
or the TM \citep[Taub-Mathews,][]{Mathews71} equation of state
\begin{equation}\label{eq:TMEos}
 w = \frac{5}{2}p + \sqrt{\frac{9}{4}p^2 + \rho^2} \,,
\end{equation}
which provides an analytic approximation of the Synge relativistic perfect
gas \citep{MMcK07}.

A relativistic formulation of the dissipative terms will not be presented
here and will be discussed elsewhere.

\subsection{General Quasi-Conservative Form}
%
%
%
%

In the following we shall adopt an orthonormal system of coordinates specified 
by the unit vectors $\hvec{e}_d$ ($d$ is used to label the direction, e.g.
$d=\{x,y,z\}$ in Cartesian coordinates) and conveniently assume that conserved
variables $\vec{U}=(\rho,\rho \vec{v},{\cal E}, \vec{B}, \rho X_\alpha)$ 
- for the MHD equations - and  
$\vec{U}=(\rho\gamma,\vec{m},{\cal E}, \vec{B})$ - for RMHD -
satisfy the following hyperbolic/parabolic partial differential equations
\begin{equation}\label{eq:ConsLaw}
  \pd{\vec{U}}{t} + \nabla\cdot\tens{F} = \nabla\cdot\tens{\Pi} + \vec{S}_p\,,
\end{equation}
where $\tens{F}$ and $\tens{\Pi}$ are, respectively, the hyperbolic 
and parabolic flux tensors. 
The source term $\vec{S}_p$ is a point-local source term which accounts for
body forces (such as gravity), cooling, chemical reactions and the source
term for the scalar multiplier (see Eq. \ref{eq:glm2x2} below).
We note that equations containing curl or gradient operators can always 
be cast in this form by suitable vector identities. 
For instance, the projection of $\nabla\times\vec{E}$ in the coordinate 
direction given by the unit vector $\hvec{e}_d$ can be re-written as
\begin{equation}\label{eq:curl2div}
  (\nabla\times\vec{E})\cdot\hvec{e}_d \equiv 
  \nabla\cdot\left(\vec{E}\times\hvec{e}_d\right)
      + \vec{E}\cdot\left(\nabla\times\hvec{e}_d\right)\,,
\end{equation}
where the second term on the right hand side should be included as an additional
source term in Eq. (\ref{eq:ConsLaw}) whenever different from zero (e.g. in 
cylindrical geometry).
Similarly one can re-write the gradient operator as 
$\nabla p = \nabla\cdot\left(\tens{I}p\right)$.
 

Several algorithms employed in \PLUTO are best implemented in terms
of primitive variables, $\vec{V} = (\rho, \vec{v}, \vec{B}, p)$.
In the following we shall assume a one-to-one mapping
between the two sets of variables, provided by appropriate 
conversion functions, that is  $\vec{V} = \vec{V}(\vec{U})$
and $\vec{U} = \vec{U}(\vec{V})$.

\section{Single Patch Numerical Integration}
\label{sec:numint}
%
%
%
%
%


\PLUTO approaches the solution of the previous sets of equations using 
either finite-volume (FV) or finite-difference (FD) methods both sharing a
flux-conservative discretization where volume averages (for the former) or 
point values (for the latter) of the conserved quantities are advanced in time.
The implementation is based on the well-established framework of Godunov type, 
shock-capturing schemes where an upwind strategy (usually a Riemann solver) is 
employed to compute fluxes at zone faces.
For the present purposes, we shall focus on the FV approach where 
volume-averaged primary flow quantities (e.g. density, momentum and energy) 
retain a zone-centered discretization.
However, depending on the strategy chosen to control the solenoidal constraint, 
the magnetic field can evolve either as a cell-average or as a face-average 
quantity (using Stoke's theorem).
As described in paper I, both approaches are possible in \PLUTO by choosing 
between Powell's eight wave formulation or the constrained transport (CT) 
method, respectively.

A third, cell-centered approach based on the generalized Lagrange multiplier 
(GLM) formulation of \cite{Dedner02} has recently been introduced in \PLUTO and a 
thorough discussion as well as a direct comparison with CT schemes can be 
found in the recent work by MT.
The GLM formulation easily builds in the context of MHD and RMHD equations by 
introducing an additional scalar field $\psi$ which couples the
divergence constraint to Faraday's law according to 
\begin{equation}\label{eq:glm2x2}
\left\{\begin{array}{lcl}
\DS \pd{\vec{B}}{t} - \nabla\times\left(\vec{v}\times\vec{B}\right) 
   + \nabla \psi  &=& 0  \,,
\\ \noalign{\medskip}
\DS \pd{\psi}{t} + c_h^2\nabla\cdot\vec{B} &=& \DS -\frac{c_h^2}{c_p^2}\psi\,,
\end{array}\right.
\end{equation}
where $c_h$ is the (constant) speed at which divergence errors are propagated 
while $c_p$ is a constant controlling the rate at which monopoles are damped.
The remaining equations are not changed and the conservative character is not 
lost. 
Owing to its ease of implementation, we adopt the GLM formulation as a 
convenient choice in the development of a robust AMR framework for Newtonian 
and relativistic MHD flows.
%

\subsection{Fully Unsplit Time Stepping}
%
%
%
%

The system of conservation laws is advanced in time using the 
Corner-Transport-Upwind \citep[CTU,][]{Col90} method recently described
by MT. Here we outline the algorithm in a more concise manner and
extend its applicability in presence of parabolic (diffusion) terms and higher
order reconstruction. Although the algorithms illustrated here are explicit in
time, we will also consider the description of more sophisticated and effective
approaches for the treatment of parabolic operators in a forthcoming paper.

We shall assume hereafter an equally-spaced grid with computational cells 
centered in $(x_i,y_j,z_k)$ having size $\Delta x\times\Delta y\times\Delta z$.
For the sake of exposition, we omit the integer subscripts $i,j$ 
or $k$ when referring to the cell center and only keep the half-increment 
notation in denoting face values, e.g., 
$\vec{F}_{d,\pm} \equiv \vec{F}_{y,i,j+\HALF,k}$ when $d=y$.
Following \cite{PLUTO}, we use $\Delta{\cal V}$ and $A_{d,\pm}$ to denote, 
respectively, the cell volume and areas of the lower and upper interfaces 
orthogonal to $\hvec{e}_d$.

An explicit second-order accurate discretization of Eqns. (\ref{eq:ConsLaw}), 
based on a time-centered flux computation, reads
\begin{equation}\label{eq:update}
 \bar{\vec{U}}^{n+1} = \bar{\vec{U}}^n 
  + \Delta t^n\sum_d\left(
  \Lvec_{H,d}^{n+\HALF} + \Lvec_{P,d}^{n+\HALF} \right)\,,
\end{equation}
where $\bar{\vec{U}}$ is the volume-averaged array of conserved values inside the
cell $(i,j,k)$, $\Delta t^n$ is the explicit time step whereas $\Lvec_{H,d}$ 
and $\Lvec_{P,d}$ are the increment operators corresponding to the hyperbolic and 
parabolic flux terms, respectively:
\begin{eqnarray}\label{eq:incrementop_hyp}
 \Lvec_{H,d}^{n+\HALF} &=& -\frac{
   A_{d,+}\vec{F}^{n+\HALF}_{d,+} - A_{d,-}\vec{F}^{n+\HALF}_{d,-}}
   {\Delta\vol_d} + \hat{\vec{S}}_d^{n+\HALF}
\,,\quad \\ \noalign{\medskip}
\label{eq:incrementop_par}
 \Lvec_{P,d}^{n+\HALF} &=& \frac{
   A_{d,+}\Pivec^{n+\HALF}_{d,+} - A_{d,-}\Pivec^{n+\HALF}_{d,-}}
   {\Delta\vol_d}   \,,
\end{eqnarray}
In the previous expression $\vec{F}_{d,\pm}$ and $\Pivec_{d,\pm}$ 
are, respectively, right ($+$) and left ($-$) face- and time-centered 
approximations to the hyperbolic and parabolic flux components in the 
direction of $\hvec{e}_d$.
The source term $\hat{\vec{S}}_d$ represents the directional contribution to the
total source vector 
$\sum\hat{\vec{S}}_d\equiv\hat{\vec{S}}_{\rm body}+\hat{\vec{S}}_{\rm geo}$ 
including body forces and geometrical terms implicitly arising when 
differentiating the tensor flux on a curvilinear grid. 
Cooling, chemical reaction terms and the source term in Eq. \ref{eq:glm2x2}
(namely $(c_h^2/c_p^2)\psi$) are treated separately in an operator-split fashion.

The computation of $\vec{F}_{d,\pm}$ requires solving, at cell interfaces, 
a Riemann problem between time-centered adjacent discontinuous states, i.e. 
\begin{equation}\label{eq:ctu_flux}
\begin{array}{lcl}
\vec{F}^{n+\HALF}_{x,i+\HALF} &=& {\cal R}\left(\vec{U}^{n+\HALF}_{i,+}, 
                                                \vec{U}^{n+\HALF}_{i+1,-}\right)
 \,, \\ \noalign{\medskip}
\vec{F}^{n+\HALF}_{y,j+\HALF} &=& {\cal R}\left(\vec{U}^{n+\HALF}_{j,+}, 
                                                \vec{U}^{n+\HALF}_{j+1,-}\right)
 \,, \\ \noalign{\medskip}
\vec{F}^{n+\HALF}_{z,k+\HALF} &=& {\cal R}\left(\vec{U}^{n+\HALF}_{k,+}, 
                                              \vec{U}^{n+\HALF}_{k+1,-}\right)
\,,
\end{array}
\end{equation}
where ${\cal R}(\cdot,\cdot)$ is the numerical flux resulting from the 
solution of the Riemann problem.
With \PLUTO, different Riemann solvers may be chosen at runtime depending on 
the selected physical module (see Paper I):
Rusanov (Lax-Friedrichs), HLL, HLLC and HLLD are common to both classical and 
relativistic MHD modules while the Roe solver is available for 
hydro and MHD.
The input states in the Eq. (\ref{eq:ctu_flux}) are obtained by first 
carrying an evolution step in the normal direction, followed by 
correcting the resulting values with a transverse flux gradient.
This yields the corner-coupled states which, in absence of parabolic terms 
(i.e. when $\tens{\Pi}=\tens{0}$), are constructed exactly as illustrated 
by MT. For this reason, we will not repeat it here.

When $\tens{\Pi} \neq 0$, on the other hand, we adopt a slightly different
formulation that does not require any change in the computational stencil.
At constant $x$-faces, for instance, we modify the corner-coupled states
to
\begin{equation}\label{eq:transv}
\vec{U}^{n+\HALF}_{i,\pm} = \vec{U}^{*}_{i,\pm} 
 + \frac{\Delta t^n}{2}\left[\sum_{d\neq x}\Lvec_{H,d}^n 
 +                           \sum_d \Lvec_{P,d}^n\right]\,,
\end{equation}
where, using Godunov's first-order method, we compute for example
\begin{equation}
\vec{F}^n_{y,j+\HALF} = {\cal R}(\bar{\vec{U}}^n_j, \bar{\vec{U}}^n_{j+1})\,,
\end{equation}
i.e., by solving a Riemann problem between cell-centered states.
States at constant $y$ and $z$-faces are constructed in a similar manner.
The normal predictors $\vec{U}^{*}_{\pm}$ can be obtained 
either in characteristic or primitive variables as outlined in 
Section \ref{sec:np_char} and Section \ref{sec:np_prim}, respectively.
Parabolic (dissipative) terms are discretized in a flux-conservative 
form using standard central finite difference approximations to the
derivatives. 
For any term in the form $\Pi = g(\vec{U})\partial_x f(\vec{U)}$, 
for instance, we evaluate the right interface flux appearing in Eq. 
(\ref{eq:incrementop_par}) with the second-order accurate expression
\begin{equation}
 \Pi_{x,+} \approx g\left(\frac{\vec{U}_i + \vec{U}_{i+1}}{2}\right)
             \frac{\DS f\left(\vec{U}_{i+1}\right)-f\left(\vec{U}_{i}\right)}
                  {\Delta x}\,,
\end{equation}
and similarly for the other directions
\citep[a similar approach is used by][]{TMG08}.
We take the solution available at the cell center at $t=t^n$ in Eq. 
(\ref{eq:transv}) and at the half time step $n+\HALF$ in Eq. (\ref{eq:update}). 
For this latter update, time- and cell-centered conserved quantities may be 
readily obtained as
\begin{equation} 
\vec{U}^{n+\HALF} = \bar{\vec{U}}^n
 + \frac{\Delta t^n}{2}\sum_d\left(\Lvec_{H,d}^n + \Lvec_{P,d}^n\right)\,.
\end{equation}
%

The algorithm requires a total of $6$ solutions to the Riemann problems 
per zone per step and it is stable under the Courant-Friedrichs-Levy
(CFL) condition $C_a \le 1$ (in 2D) or $C_a \le 1/2$ (in 3D), where
\begin{equation}\label{eq:Ca}
 C_a = \Delta t^{n+1}\max_{ijk}\left[\max_d\left(
 \frac{\lambda^{\max}_d}{\Delta x_d}\right)
 + \sum_d\frac{2{\cal D}^{\max}_d}{\Delta x_d^2}\right] 
\end{equation}
is the CFL number, $\lambda^{\max}_d$ and ${\cal D}^{\max}_d$ are the (local) 
largest signal speed and diffusion coefficient in the direction given by 
$d=\{x,y,z\}$, 
respectively.
Equation (\ref{eq:Ca}) is used to retrieve the time step $\Delta t^{n+1}$ 
for the next time level if no cooling or reaction terms are present. 
Otherwise we further limit the time step so that the relative change in 
pressure and chemical species remains below a certain threshold $\epsilon_c$:
\begin{equation}\label{eq:dt_cool}
 \Delta t^{n+1} \to \min\left[\Delta t^{n+1}, \frac{\epsilon_c\Delta t^n}
  {\max(|\delta p/p|,|\delta X_\kappa|)}\right]
\end{equation}
where $\delta p/p$ and $\delta X_\kappa$ are the maximum fractional 
variations during the source step \citep{Tes08}.
We note that the time step limitation given by equation (\ref{eq:dt_cool}) does
not depend on the mesh size and can be estimated on unrefined cells only.
This allows to take full advantage of the adaptivity in time as 
explained in Section \ref{sec:cool_source}.

\subsection{Normal predictors in characteristic variables}
\label{sec:np_char}
%
%
%
%

The computation of the normal predictor states can be carried out in 
characteristic variables by projecting the vector $\vec{V}$ of primitive 
variables onto the left eigenvectors 
$\vec{l}^{\kappa}_i\equiv\vec{l}^\kappa(\vec{V}_i)$ of the primitive system.
Specializing to the x direction:
\begin{equation}\label{eq:char_var}
  w^\kappa_{i,l} = \vec{l}_i^\kappa\cdot \vec{V}_{i+l}\,,\quad
  l = -S,\cdots,S\,,
\end{equation}
where $\kappa = 1,\cdots,N_{\rm wave}$ labels the characteristic wave with 
speed $\lambda^\kappa$ and the projection extends to all neighboring zones 
required by the interpolation stencil of width $2S+1$. 
The employment of characteristic fields rather than primitive variables 
requires the additional computational cost associated with the full spectral 
decomposition of the primitive form of the equations.
Nevertheless, it has shown to produce better-behaved solutions for highly 
nonlinear problems, notably for higher order methods.

For each zone $i$ and characteristic field $\kappa$, we first interpolate 
$w^\kappa_{i,l}$ to obtain $w^\kappa_{i,\pm}$, that is, the 
rightmost ($+$) and leftmost ($-$) interface values from within the cell. 
The interpolation can be carried out using either fourth-, third- or 
second-order piecewise reconstruction as outlined later in this section.

Extrapolation in time to $t^n+\Delta t^n/2$ is then carried out by 
using an upwind selection rule that discards waves not reaching a given 
interface in $\Delta t/2$. The result of this construction, omitting 
the $\kappa$ index for the sake of exposition, reads
\begin{equation}\label{eq:wnorm_predp}
 w^{*}_{i,+} = w^{\rm ref}_{i,+} + \beta_{+}\left\{w_{i,+} - \frac{\nu}{2}
 \left[\delta w_i + \delta^2w_i\left(3 - 2\nu\right)\right] 
  - w^{\rm ref}_{i,+}\right\}\,,
\end{equation}
\begin{equation}\label{eq:wnorm_predm}
 w^{*}_{i,-} = w^{\rm ref}_{i,-} + \beta_{-}\left\{w_{i,-} - \frac{\nu}{2}
 \left[\delta w_i - \delta^2w_i\left(3 + 2\nu\right)\right] 
  - w^{\rm ref}_{i,-}\right\}\,,
\end{equation}
where $\nu \equiv \lambda\Delta t/\Delta x$ is the Courant 
number of the $\kappa$-th wave, $\beta_\pm = (1\pm{\rm sign}(\nu))/2$,
whereas $\delta w_i$ and $\delta^2 w_i$ are defined by
\begin{equation}
 \delta w_i   = w_{i,+} - w_{i,-}\,,\qquad
 \delta^2 w_i = w_{i,+} - 2w_i  + w_{i,-}\,.
\end{equation}
The choice of the reference state $w^{\rm ref}_{i,\pm}$ is somewhat 
arbitrary and one can simply set $w^{\rm ref}_{i,\pm} = w_{i,0}$ 
\citep{RGK07} which has been found to work well for flows containing 
strong discontinuities. Alternatively, one can use the original prescription
\citep{CW84}
\begin{equation}
w^{\rm ref}_{i,+} = w_{i,+} - \frac{\nu_{\max}}{2}
  \left[\delta w_i + \delta^2w_i\left(3 - 2\nu_{\max}\right)\right]\,,
\end{equation}
\begin{equation}
w^{\rm ref}_{i,-} = w_{i,-} - \frac{\nu_{\min}}{2}
  \left[\delta w_i - \delta^2w_i\left(3 + 2\nu_{\min}\right)\right]\,,
\end{equation}
where $\nu_{\max}=\max(0,\max_\kappa(\nu_\kappa))$ and
$\nu_{\min}=\min(0,\min_\kappa(\nu_\kappa))$ are chosen so as to minimize
the size of the term susceptible to characteristic limiting 
\citep{CW84, MC02, MPB05}.
However we note that, in presence of smooth flows, both choices may reduce 
the formal second order accuracy of the scheme since, in the limit of 
small $\Delta t$, contributions carried by waves not reaching a zone edge 
are not included when reconstructing the corresponding interface value
\citep{Athena08}.
%
%
In these situations, a better choice is to construct the normal predictors 
without introducing a specific reference state or, equivalently, by 
assigning $w^{\rm ref}_{i,\pm} = w_{i,\pm}$ in Eq. (\ref{eq:wnorm_predp}) 
and (\ref{eq:wnorm_predm}).
 



The time-centered interface values obtained in characteristic variables
through Eq. (\ref{eq:wnorm_predp}) and (\ref{eq:wnorm_predm}) are 
finally used as coefficients in the right-eigenvector expansion 
to recover the primitive variables:
%
\begin{equation}
 \vec{V}^{*}_{i,\pm} = \sum_\kappa w^{\kappa,*}_{i,\pm} \vec{r}^\kappa_{i} 
  + \frac{\Delta t^n}{2}\left(\vec{S}_{g,i}^n 
  + \vec{S}_{B_x,i}^n\frac{\Delta B_x}{\Delta x}
  + \vec{S}_{\psi,i}^n\frac{\Delta \psi}{\Delta x}\right)\,,
\end{equation}
where $\vec{r}^{\kappa}$ is the right eigenvector associated to the 
$\kappa$-th wave, $\vec{S}_g$ is a source term accounting for body forces 
and geometrical factors while $\vec{S}_{\rm B_x}$ and $\vec{S}_{\rm\psi}$
arise from calculating the interface states in primitive variables rather 
than conservative ones and are essential for the accuracy of the scheme
in multiple dimensions.
See MT for a detailed discussion on the implementation of
these terms.

As mentioned, the construction of the left and right interface values
can be carried out using different interpolation techniques.
Although some of the available options have already been presented in the 
original paper \citep{PLUTO}, here we briefly outline the implementation 
details for three selected schemes providing (respectively) fourth-, 
third- and second-order spatially accurate interface values in the limit
of vanishing time step.
Throughout this section we will make frequent usage of the undivided 
differences of characteristic variables (\ref{eq:char_var}) such as
\begin{equation}
 \Delta w_{i,+\HALF} = w_{i,+1} - w_{i,0}\,,\quad
 \Delta w_{i,-\HALF} = w_{i,0} - w_{i,-1}\,.
\end{equation}
for the $i$-th zone.

\paragraph{Piecewise Parabolic Method}
%
%

The original PPM reconstruction by \cite{CW84} \citep[see also][]{MC02}
can be directly applied to characteristic variables giving 
the following fourth-order limited interface values:
\begin{equation}\label{eq:ppm1}
 w_{i,\pm} = \frac{w_{i,0} + w_{i,\pm 1}}{2} \mp 
             \frac{\overline{\Delta w}_{i,\pm 1} 
                 - \overline{\Delta w}_{i,0}}{6}\,,
\end{equation}
where slope limiting is used to ensure that $w_{i,\pm}$ are bounded between 
$w_i$ and $w_{i\pm 1}$:
\begin{equation}\label{eq:MClim}
 \overline{\Delta w}_{i,0} = {\rm mm}\left[
  \frac{\Delta w_{i,+\HALF}+\Delta w_{i,-\HALF}}{2},
   2{\rm mm}\left(\Delta w_{i,-\HALF}, \Delta w_{i,+\HALF}\right)\right]
\end{equation}
and
\begin{equation}\label{eq:minmod}
 {\rm mm}(a,b) = \frac{\rm{sign}(a) + \rm{sign}(b)}{2}
                  \min\left(|a|, |b|\right) \,,
\end{equation}
is the MinMod function.
The original interface values defined by Eq. (\ref{eq:ppm1}) must then be 
corrected to avoid the appearance of local extrema. 
By defining $\delta_\pm = w_{i,\pm} - w_{i,0}$, we further apply the following
parabolic limiter
\begin{equation}\label{eq:ppm_parabolic_lim}
 \delta_\pm = \left\{\begin{array}{ll}
   0          & \quad \textrm{if}\quad  \delta_+\delta_- > 0 \\ \noalign{\medskip}
 -2\delta_\mp & \quad \textrm{if}\quad  |\delta_\pm| \ge 2|\delta_\mp|\,,
\end{array}\right.
\end{equation}
where the first condition flattens the distribution when $w_{i,0}$ is a local 
maximum or minimum whereas the second condition prevents the appearance of
an extremum in the parabolic profile.
Once limiting has been applied, the final interface values are obtained as
\begin{equation}
 w_{i,\pm} = w_{i,0} + \delta_\pm\,.
\end{equation}
In some circumstances, we have found that further application of the 
parabolic limiter (\ref{eq:ppm_parabolic_lim}) to primitive variables 
may reduce oscillations.

\paragraph{Third-order improved WENO}
%

As an alternative to the popular TVD limiters, the third-order improved 
weighted essentially non-oscillatory (WENO) reconstruction proposed by 
\cite{YC09} (see also \citealp{MTB10}) may be used. 
The interpolation still employs a three-point stencil but provides 
a piecewise parabolic profile that preserves the accuracy at 
smooth extrema, thus avoiding the well known clipping of classical 
second-order TVD limiters. 
Left and right states are recovered by a convex combination of $2^{\rm nd}$ 
order interpolants into a weighted average of order $3$.
The nonlinear weights are adjusted by the local smoothness of the solution 
so that essentially zero weights are given to non smooth stencils while 
optimal weights are prescribed in smooth regions. In compact notation:
\begin{equation}\label{eq:WENO3}
\begin{array}{ccc}
 w_{i,+} &=& \DS w_{i,0} + 
 \frac{a_+\Delta w_{i,+\HALF} + \HALF a_-\Delta w_{i,-\HALF}}{2a_++a_-}\,,
 \\ \noalign{\medskip}
 w_{i,-} &=& \DS w_{i,0} -
 \frac{a_-\Delta w_{i,-\HALF} + \HALF a_+\Delta w_{i,+\HALF}}{2a_-+a_+}\,,
\end{array}
\end{equation}
where
\begin{equation}\label{eq:WENO3a}
 a_\pm = 1 + \frac{\left(\Delta w_{i,+\HALF} - \Delta w_{i,-\HALF}\right)^2}
                  {\Delta x^2 + \Delta w_{i,\pm\HALF}^2}\,.
\end{equation}
As one can see, an attractive feature of WENO reconstruction consists in 
completely avoiding the usage of conditional statements.
The improved WENO scheme has enhanced accuracy with respect to the traditional
3rd order scheme of \cite{JS96} in regions where the solution is smooth and 
provides oscillation-free profiles near strong discontinuities .

\paragraph{Linear reconstruction}
Second-order traditional limiting is provided by 
\begin{equation}\label{eq:Linear}
  w_{i,\pm} = w_{i,0} \pm \frac{\overline{\Delta w}_{i,0}}{2}\,,
\end{equation}
where $\overline{\Delta w}_{i,0}$ is a standard limiter function such as the 
monotonized-central (MC) limiter (Eq. \ref{eq:MClim}). 
Other, less steep forms of limiting are the harmonic mean
\citep{VL74}:
\begin{equation}\label{eq:harmonic_lim}
  \overline{\Delta w}_{i,0} = \left\{\begin{array}{ll}
\DS  \frac{2\Delta w_{i,+\HALF}\Delta w_{i,-\HALF}}
       {\Delta w_{i,+\HALF}+\Delta w_{i,-\HALF}} \
  & \quad {\rm if}\quad \Delta w_{i,+\HALF}\Delta w_{i,-\HALF} > 0 \,,
  \\ \noalign{\medskip}
  0 & \quad{\rm otherwise}\,;
\end{array}\right.
\end{equation}
the Van Albada limiter \citep{VA82}:
\begin{equation}\label{eq:vanalbada_lim}
  \overline{\Delta w}_{i,0} = \left\{\begin{array}{ll}
\DS  \frac{\Delta w_{i,+\HALF}\Delta w_{i,-\HALF}(\Delta w_{i,+\HALF}+\Delta w_{i,-\HALF})}
       {\Delta w_{i,+\HALF}^2+\Delta w_{i,-\HALF}^2}
  & \quad {\rm if}\quad \Delta w_{i,+\HALF}\Delta w_{i,-\HALF} > 0 
  \\ \noalign{\medskip}
  0 & \quad{\rm otherwise}
\end{array}\right.
\end{equation}
or the MinMod limiter, Eq. (\ref{eq:minmod}).

Linear reconstruction may also be locally employed in place of a higher
order method whenever a strong shock is detected.
This is part of a built-in hybrid mechanism that selectively identifies 
zones within a strong shock in order to introduce additional dissipation
by simultaneously switching to the HLL Riemann solver.
Even if the occurrence of such situations is usually limited to very few 
grid zones, this fail-safe mechanism does not sacrifice the second-order
accuracy of the scheme and has been found to noticeably improve the 
robustness of the algorithm avoiding the occurrence of unphysical states.
This is described in more detail in Appendix \ref{app:MULTID_Flattening}.
However, in order to assess the robustness and limits of applicability 
of the present algorithm, it will not be employed for the
test problems presented here unless otherwise stated. 

\subsection{Normal predictors in primitive variables}
\label{sec:np_prim}
%
%
%
%
%

The normal predictor states can also be directly computed in primitive 
variables using a simpler formulation that avoids the characteristic 
projection step.
In this case, one-dimensional left and right states are obtained at 
$t^{n+\HALF}$ using 
\begin{equation}\label{eq:MUSCL-Hancock}
 \vec{V}^*_{i,\pm} = \vec{V}_i^n + \beta_\pm\left[
                      \vec{V}\left(\vec{U}_i^{*}\right)-\vec{V}_i^n \pm
 \frac{\Delta\vec{V}_i}{2}\right]\,,
\end{equation}
where $\beta_+ = (1+\mathrm{sgn}(\lambda_{\max}))/2$ and
$\beta_- = (1-\mathrm{sgn}(\lambda_{\min}))/2$ may be used to 
introduce a weak form of upwind limiting, in a similar fashion to 
Section \ref{sec:np_char}.
Time-centered conservative variables $\vec{U}^*_i$ follow from a simple 
conservative MUSCL-Hancock step:
\begin{equation}
  \vec{U}^*_i = \vec{U}_i^n - \frac{\Delta t^n}{2\Delta\vol_x} 
  \left[A_{x,+}\vec{F}\left(\vec{V}^n_{i,+}\right) -
        A_{x,-}\vec{F}\left(\vec{V}^n_{i,-}\right)\right] + 
      \frac{\Delta t^n}{2}\hat{\vec{S}}_d\,,
\end{equation}
where $A_{x,\pm}$ and $\Delta\vol$ are the area and volume elements
and $\vec{V}^n_{i,\pm}$ are obtained using linear reconstruction 
(Eq. \ref{eq:Linear}) of the primitive variables.
This approach offers ease of implementation over the characteristic 
tracing step as it does not require the eigenvector decomposition of
the equations nor their primitive form.
Notice that, since this step is performed in conservative variables, 
the multidimensional terms proportional to $\partial B_x/\partial x$ 
do not need to be included.

\section{AMR Strategy - CHOMBO}
\label{sec:CHOMBO}
%
%
%
%
%

\ifx\IncludeEps\yes
\begin{figure}[!h]
 \includegraphics*[width=\columnwidth]{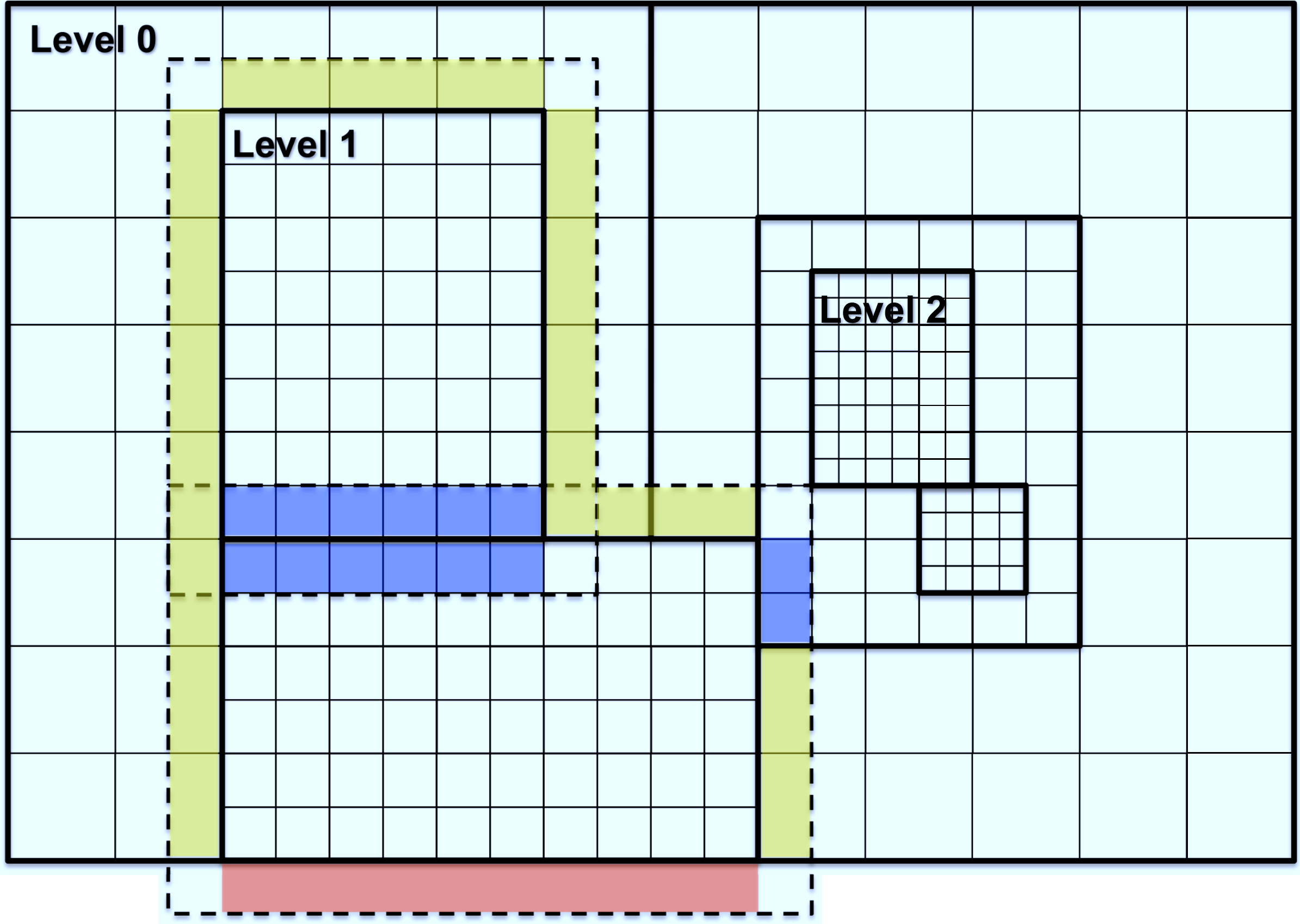}
  \caption{\footnotesize Two-dimensional example of a three-level AMR hierarchy, with the base level ($\ell=0$) covering the entire computational
           domain. Solid lines are representative of the level resolution. Dashed lines contour the ghost zones of two patches of 
           level $\ell=1$. Colors indicate different filling methods: physical outer boundaries (red), boundaries filled by exchanging values
           with adjacent patches on the same level (blue), boundaries filled by interpolating from the next coarser level (yellow).}
 \label{fig:AMRGrid}
\end{figure}
\fi

The support for Adaptive Mesh Refinement (AMR) calculations in \PLUTO is provided by the CHOMBO library.
CHOMBO is a software package aimed at providing a distributed infrastructure for serial and parallel calculations 
over block-structured, adaptively refined grids in multiple dimensions. It is written in a combination of C++ and Fortran77
with MPI and is developed and distributed by the Applied Numerical Algorithms Group of Lawrence Berkeley National 
Laboratory (https://seesar.lbl.gov/anag/chombo/).

In the block-structured AMR approach, once the solution has been computed over a rectangular grid which discretizes
the entire computational domain, it is possible to identify the cells which require additional resolution and cover them 
with a set of rectangular grids (also referred to as blocks or patches), characterized by a finer mesh spacing.
CHOMBO follows the \cite{BergerRig95} strategy to determine the most efficient patch layout to cover the cells
that have been tagged for refinement. 
This process can be repeated recursively to define the solution over a hierarchy of
$\ell=0, \dots, \ell_\mathrm{max}$ levels of refinement whose spatial resolutions satisfy
the relation $\Delta x_d^\ell = r^\ell \; \Delta x_d^{\ell+1}$, where the integer 
$r^\ell$  is the refinement ratio between level $\ell$ and level $\ell+1$. A level of refinement is composed by a 
union of rectangular grids which has to be: {\it disjointed}, i.e. two blocks of the same level can be adjacent without 
overlapping; {\it properly nested}, i.e. a cell of level $\ell$ cannot be only partially covered  by cells
of level $\ell+1$ and cells of  level $\ell+1$ must be separated from cells of level $\ell-1$ at least
by a row of cells of level $\ell$.
A simple example of a bidimensional adaptive grid distributed over a hierarchy of three levels of
refinement is depicted in Fig. \ref{fig:AMRGrid}.

Following the notation of \cite{Pember96}, a global mapping is employed on all levels: in three 
dimensions, cells on level $L$ are identified with global indexes $i,j,k$
($0\leq i < N_1^\ell$, $0\leq j < N_2^\ell$, $0\leq k < N_3^\ell$, with $N_{1,2,3}^\ell$ being the equivalent global 
resolution of level $\ell$ in the three directions). 
Correspondingly, the cell $i,j,k$ of level $\ell$ is covered by $(r^\ell)^3$ cells of level $\ell+1$ identified by global
indexes $l,m,n$ satisfying the conditions $r^\ell i \leq l \leq r^\ell(i+1)-1$, $r^\ell j \leq m \leq r^\ell(j+1)-1$, 
$r^\ell k \leq n \leq r^\ell(k+1)-1$.  
Taking direction 1 as an example, the expressions $x_{1,-}^\ell=i \Delta x_1^\ell$,  $x_1^\ell = (i+1/2) \Delta x_1^\ell$, 
$x_{1,+}^\ell=(i+1) \Delta x_1^\ell$, define the physical coordinates of the left edge, center and right edge of a cell respectively.

\ifx\IncludeEps\yes
\begin{figure} [!h]
\includegraphics*[width=\columnwidth]{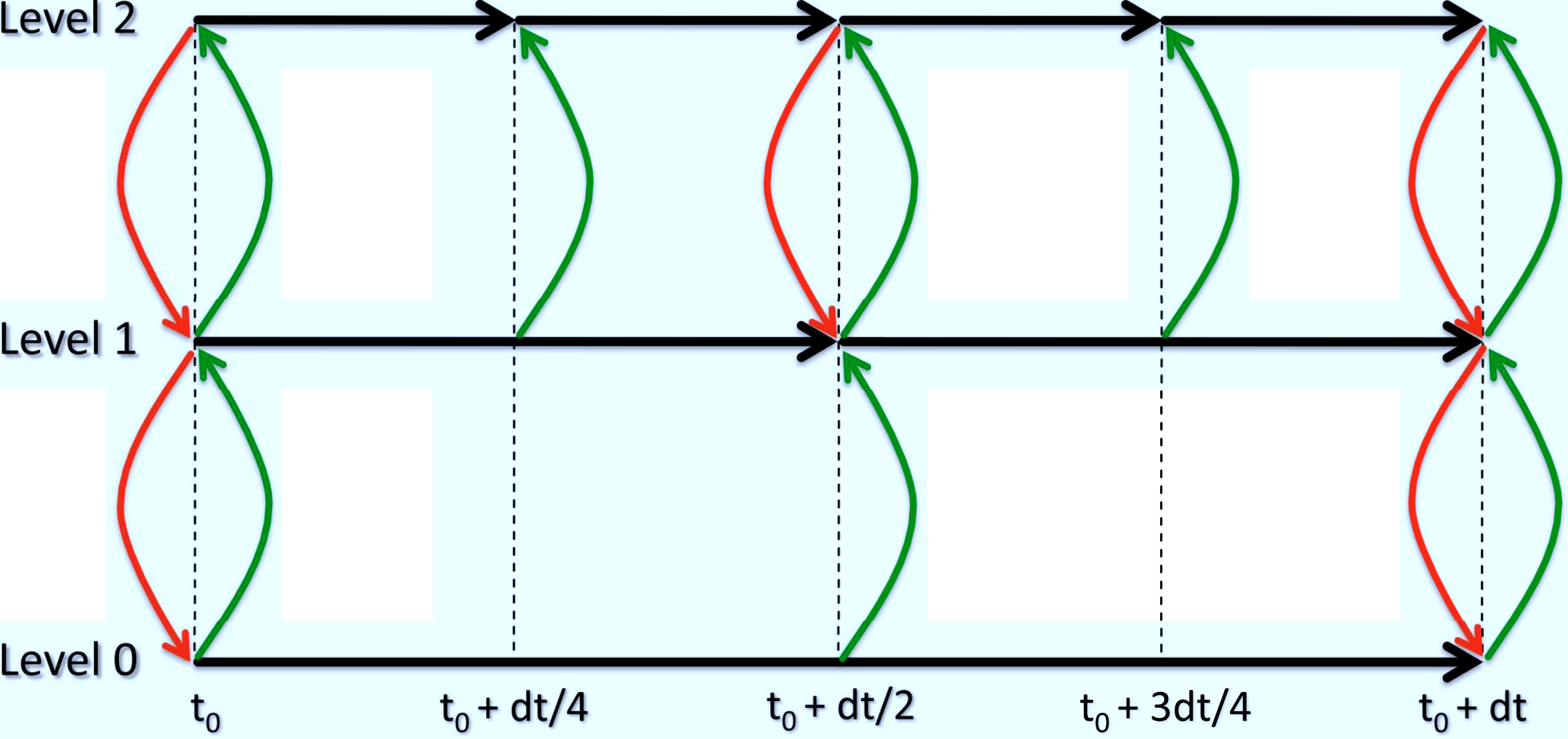} 
\caption{\footnotesize Schematic representation of the time evolution of an AMR hierarchy 
         composed by three levels with a refinement ratio $r=2$. 
         The length of the horizontal black arrows is proportional to the
         timestep size $\Delta t^\ell$. Curved vertical arrows indicate interlevel 
         communications. Red arrows represent fine-to-coarse communications 
         between synchronized adjacent levels, including conservative averaging 
         (Eq. \ref{eq:avgdown}) and refluxing (Eq. \ref{eq:fluxcorr}, 
         \ref{eq:addreflux}). Green arrows represent coarse-to-fine 
         communications, including the conservative interpolation (Eq.
         \ref{eq:prolong}) needed to fill ghost zones and to define the 
         solution on newly generated cells of the finer level.}
\label{fig:Time}
\end{figure}
\fi
If the adaptive grid is employed to evolve time-dependent hyperbolic partial differential equations, 
the CFL stability condition allows to apply refinement in time as well as in space,
as first proposed by \cite{BergerOli84} and further developed in \cite{BergerCol89}.
In fact, each level $\ell$ advances in time with a time-step $\Delta t^\ell=\Delta t^{\ell-1}/r^{\ell-1}$
which is $r^{\ell-1}$ times smaller than the time-step of the next coarser level $\ell-1$.
Starting the integration at the same instant, two adjacent levels synchronize every $r^\ell$
timesteps, as schematically illustrated in Fig. \ref{fig:Time} for a refinement ratio $r^\ell = 2$. 
Even though in the following discussion we will assume that level $\ell+1$ completes $r^\ell$
timesteps to synchronize with level $\ell$, CHOMBO allows the finer level $\ell+1$
to advance with smaller substeps if the stability condition requires it.
Anyway, the additional substeps must guarantee that levels $\ell$ and $\ell+1$ are
synchronized again at time $t^\ell + \Delta t^\ell$.

The time evolution of single patches is handled by \PLUTO, as illustrated in 
Section \ref{sec:numint}. Before starting the time evolution, the ghost cells 
surrounding the patches must be filled according to one of these three possibilities: (1) 
assigning ``physical'' boundary conditions to the ghost cells which lie outside the 
computational domain (e.g. the red area in Fig. \ref{fig:AMRGrid}); (2) exchanging boundary 
conditions with the adjacent patches of the same level (e.g. the blue area in Fig. 
\ref{fig:AMRGrid}); (3) interpolating values from the parent coarser level for 
ghost cells which cover cells of a coarser patch (e.g. the yellow area in Fig. 
\ref{fig:AMRGrid}). 

\begin{deluxetable}{c c c}
\tablecaption{Systems of coordinates adopted in PLUTO-CHOMBO\label{table:coord}}
\tablewidth{0pt}
\tablehead{\colhead{} & \colhead{Cartesian} & \colhead{Cylindrical} }
\startdata
$x_1$      &  $x$                          &    $r$                 \\
$x_2$      &  $y$                          &    $z$                 \\
$x_3$      &  $z$                          &    /                   \\
$V_1$      &  $x$                          &    $r^2/2$             \\
$V_2$      &  $y$                          &    $z$                 \\
$V_3$      &  $z$                          &    /                   \\
$A_{1,+}$  &  $\Delta y \Delta z$          &   $r_+ \Delta z$       \\
$A_{2,+}$  &  $\Delta z \Delta x$          &   $r \Delta r$         \\
$A_{3,+}$  &  $\Delta x \Delta y$          &   /                    \\
$\vol$     &  $\Delta x \Delta y \Delta z$ &  $r \Delta r \Delta z$ 
\enddata
\end{deluxetable}
 
\ifx\IncludeEps\yes
\begin{figure}[!h]
\includegraphics*[width=\columnwidth]{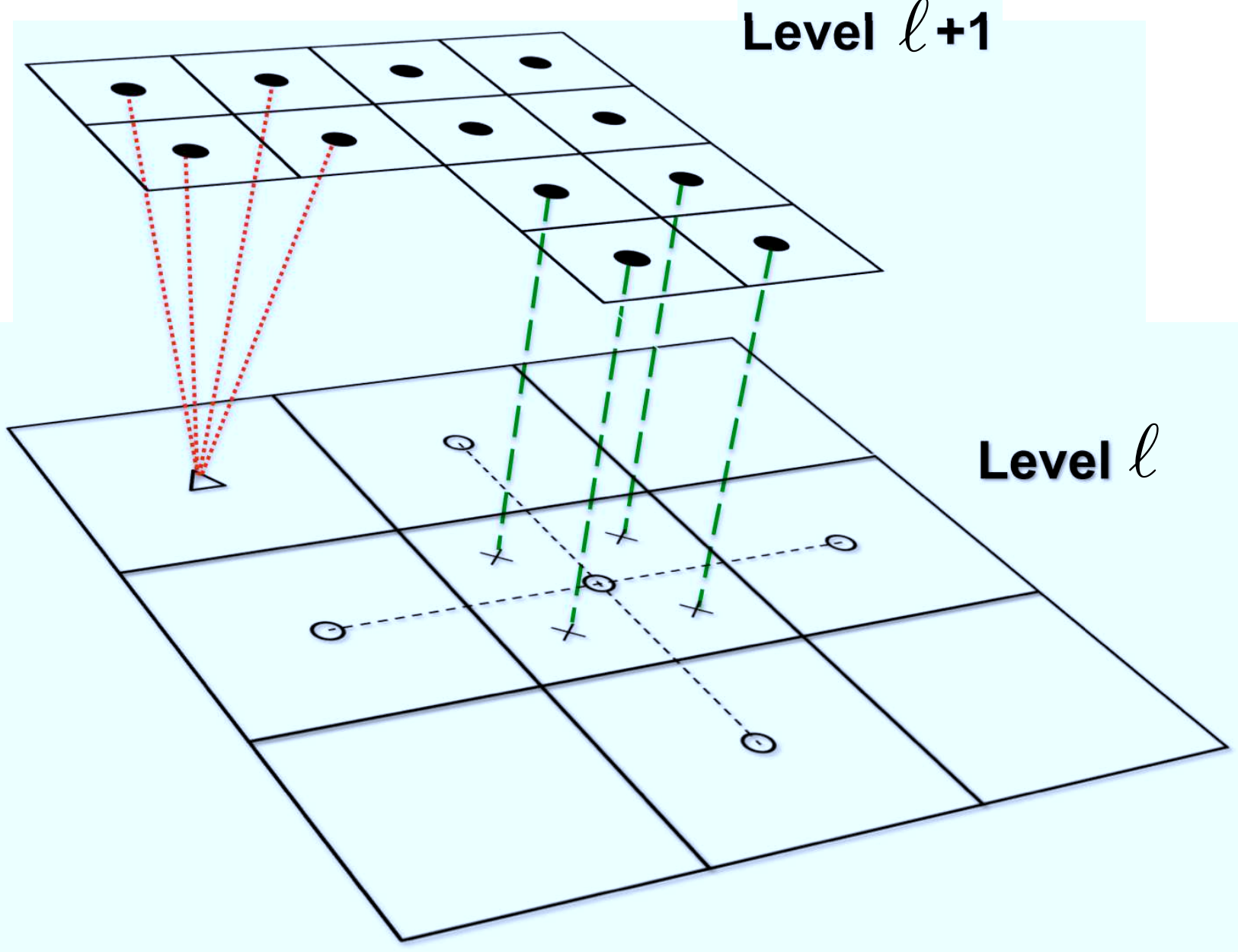}
\caption{\footnotesize Illustrative bidimensional example of prolongation and restriction operations between two levels with
         a refinement ratio $r^\ell$ = 2. Cells on level $\ell+1$ can be filled by linearly interpolating the coarse 
         values on level $\ell$ (empty circles) at cross-marked points (prolongation, green dashed lines, Eq. \ref{eq:prolong}).
         Cells on level $\ell$ can be filled by averaging down values from level $\ell+1$ in a conservative way
         (restriction, red dotted lines, Eq. \ref{eq:avgdown}). }
\label{fig:Projection}
\end{figure}
\fi
\ifx\IncludeEps\yes
\begin{figure}
\includegraphics*[width=\columnwidth]{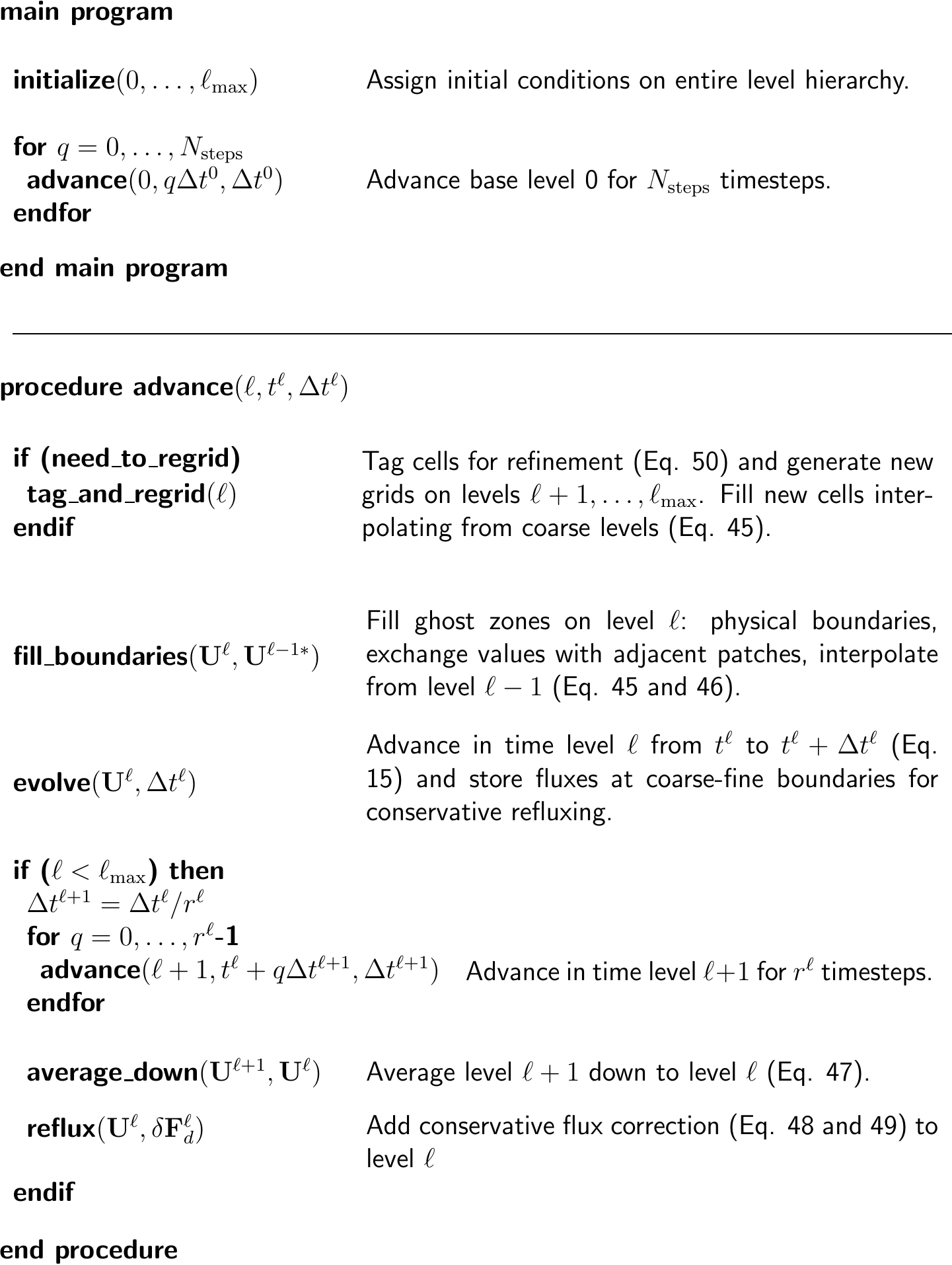} 
\caption{\footnotesize Pseudocode for the recursive level integration in the
          block structured AMR.}
\label{fig:pseudocode}
\end{figure}
\fi

As schematically illustrated in Fig. \ref{fig:Projection} in two dimensions, the coarse-to-fine prolongation needed in case (3) 
(green dashed lines) is based on a piecewise linear interpolation at points marked by crosses using the linear slopes computed from 
the surrounding coarse cells. In three dimensions the interpolant has the general form:
\begin{equation}
\vec{U}_{l,m,n}^{\ell+1} = \vec{U}_{i,j,k}^{\ell}  + \sum_{d=1}^3 
  \frac{\vol_{d}^{\ell+1}-\vol_{d}^{\ell}}{\vol_{d,+}^{\ell}-\vol_{d,-}^{\ell}} \Delta_d \vec{U}_{i,j,k}^{\ell} \; ,
\label{eq:prolong}
\end{equation}
where $\vol_{d}^{\ell}$ is the volume coordinate of cell centers of level $\ell$ in direction $d$
and $\vol_{d,\pm}^{\ell}$ is its value on the right and left
faces of the cell respectively (see Table \ref{table:coord} for definitions). 
The linear slopes $\Delta_d \vec{U}_{i,j,k}^{\ell}$ are calculated as central differences, except for cells touching
the domain boundary, where one-sided differences are employed. 
The monotonized-central limiter (Eq \ref{eq:MClim}) is applied to the linear slopes so that no 
new local extrema are introduced.
 
Notice that, since two contiguous levels are not always synchronized (see Fig. \ref{fig:Time}),
coarse values at an intermediate time are needed to prolong the solution from a coarse level to the
ghost zones of a finer level. 
Coarse values of level $\ell$ are therefore linearly interpolated between time $t^\ell$ and time
$t^\ell + \Delta t^\ell$ and the piecewise linear interpolant Eq. (\ref{eq:prolong})
is applied to the coarse solution \begin{equation}
\vec{U}_{i,j,k}^{\ell*} =  (1-\alpha)\vec{U}^{\ell}_{i,j,k}(t^{\ell}) +
                          \alpha \vec{U}^{\ell}_{i,j,k}(t^{\ell}+\Delta t^{\ell}) \; ,
\label{eq:intalpha}
\end{equation}
where $\alpha = (t^{\ell+1}-t^{\ell})/\Delta t^{\ell}$.
This requires that, everytime level $\ell$ and level $\ell+1$ are synchronized, a timestep on level
$\ell$ must be completed before starting the time integration of level $\ell+1$.
Therefore, the time evolution of the entire level hierarchy is performed recursively from level $\ell=0$
up to $\ell=\ell_{\max}$, as schematically illustrated by the pseudo-code
in Fig. \ref{fig:pseudocode}.

\ifx\IncludeEps\yes
\begin{figure}[!b]
 \includegraphics*[width=\columnwidth]{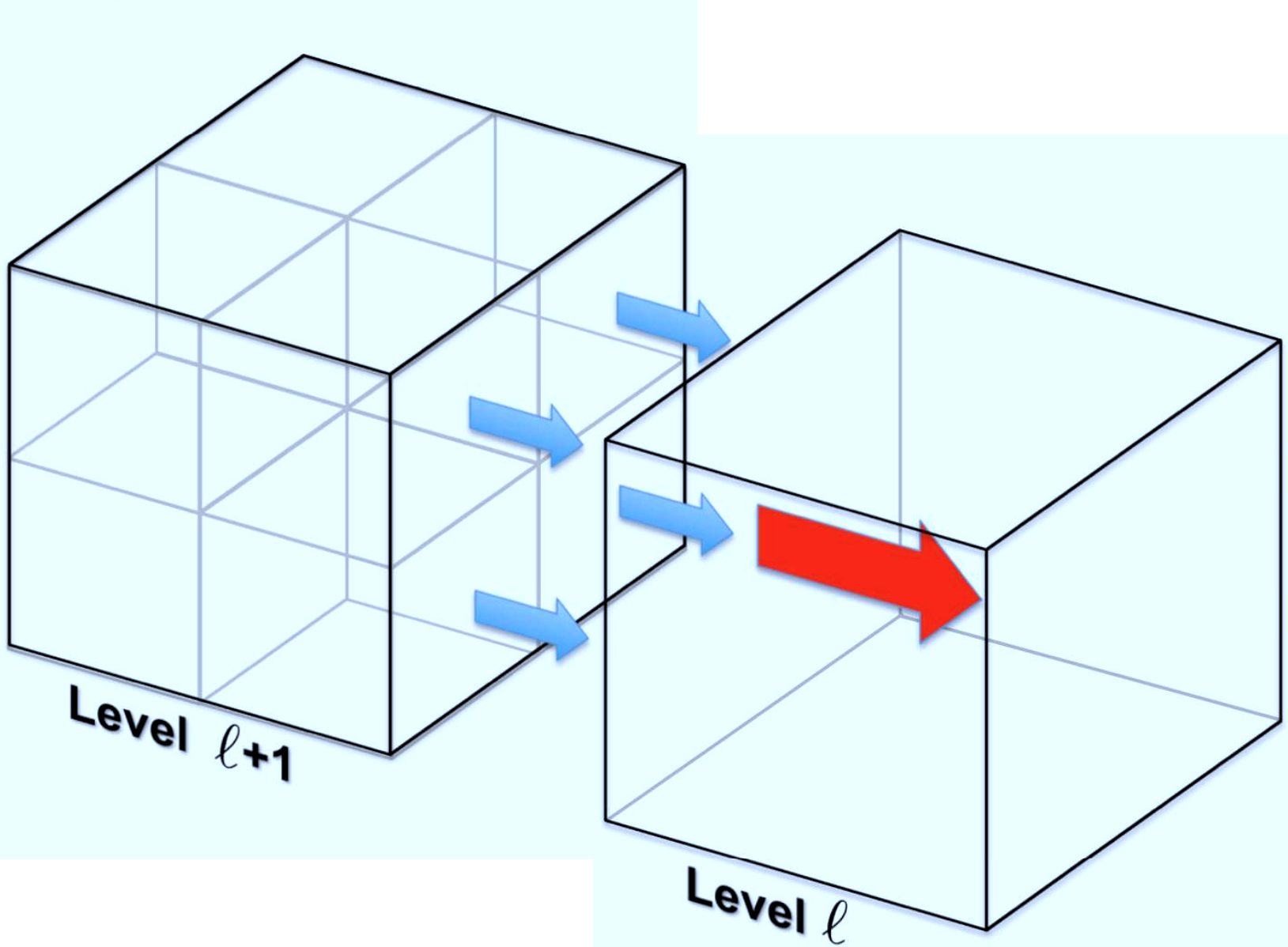}
 \caption{\footnotesize Schematic visualization of the refluxing operation needed at fine-coarse interfaces to preserve the conservative properties
          of the solution. One cell on the coarser level $\ell$ and the cells of level $\ell+1$ adjacent on one side are represented, assuming 
          a refinement ratio $r^\ell=2$. Whenever level $\ell$ and $\ell+1$ are synchronized, the coarse flux (red arrow) must be replaced by 
          the spatial and temporal average of the finer fluxes crossing the fine-coarse interface (blue arrows, Eq. \ref{eq:fluxcorr})
           and the solution must be corrected accordingly (Eq. \ref{eq:addreflux}). }
  \label{fig:Reflux}
\end{figure}  
\fi
When two adjacent levels are synchronized, some corrections to the solutions are performed
to enforce the conservation condition on the entire level hierarchy.
To maintain consistency between levels, the solution on the finer level $\ell+1$ is restricted to the lower
level $\ell$ by averaging down the finer solution in a conservative way (red dotted lines in
Fig. \ref{fig:Projection})
\begin{equation}
\vec{U}_{i,j,k}^{\ell} = \sum_{l=r^\ell i}^{r^\ell(i+1)-1} \sum_{m=r^\ell j}^{r^\ell(j+1)-1} 
                         \sum_{n=r^\ell k}^{r^\ell(k+1)-1} \frac{\vol_{l,m,n}^{\ell+1} \vec{U}_{l,m,n}^{\ell+1}}{\vol_{i,j,k}^{\ell}} \; ,
\label{eq:avgdown}
\end{equation}
where $\vol_{i,j,k}^{\ell}$ is the volume of cell $i,j,k$ of level $\ell$.  

Moreover, the flux through an edge which is shared between a cell of level $\ell$ and 
$(r^\ell)^2$ cells of level $\ell+1$ must be corrected to maintain the conservative form of the equations. 
For example if the cell $i,j,k$ on level $\ell$ shares its left boundary with $(r^\ell)^2$ cells of level $\ell+1$ 
(see Fig. \ref{fig:Reflux}), the flux calculated during the coarse integration must be
replaced with the average in time and space of the fluxes crossing the $(r^\ell)^2$ faces of the finer level cells. 
In this particular example, the flux correction is defined as:
\begin{equation}
\delta \vec{F}_d^\ell = - A_{d,-}^{\ell} \vec{F}_{d,-}^\ell + \frac{1}{r^\ell} \sum_{q=1}^{r^\ell} \sum_{m=r^\ell j}^{r^\ell(j+1)-1} 
\sum_{n=r^\ell k}^{r^\ell(k+1)-1} A_{d,+}^{\ell+1} \vec{F}_{d,m,n,+}^{\ell+1,q} \; ,
\label{eq:fluxcorr}
\end{equation}
where the index $q$ sums over the timsteps of level $\ell+1$, while $m$ and $n$ are the indexes transverse to direction $d$.
The flux correction is added to the solution on level $\ell$ after the time integration of level $\ell$ and $\ell+1$ has been completed:
\begin{equation}
\vec{U}_{i,j,k}^{\ell} = \vec{U}_{i,j,k}^{\ell} + \Delta t^\ell \frac{\delta \vec{F}_d^\ell}{\vol_{i,j,k}^{\ell}} \; .
\label{eq:addreflux}
\end{equation}
  
Finally, when levels from $\ell$ up to $\ell_\mathrm{max}$ are synchronized, it is possible to tag the cells which need refinement and 
generate a new hierarchy of grids on levels from $\ell+1$ up to $\ell_\mathrm{max}$ which covers the tags at each level. 
Whenever new cells are created on level $\ell+1$ it is possible 
to fill them interpolating from level $\ell$ according to Eq. (\ref{eq:prolong}). 
It is important to notice that this interpolant preserves the
conservative properties of the solution.

\subsection{Refinement Criteria}
%
%

In PLUTO-CHOMBO zones are tagged for refinement whenever a prescribed
function $\chi(\vec{U})$ of the conserved variables and of its derivatives 
exceeds a prescribed threshold, i.e., $\chi(\vec{U}) > \chi_r$.
Generally speaking, the refinement criterion may be problem-dependent thus 
requiring the user to provide an appropriate definition of $\chi(\vec{U})$. 
The default choice adopts a criterion based on the second derivative error 
norm \citep{Lohner87}, where 
\begin{equation}\label{eq:RefCrit}
  \chi(\vec{U}) = \sqrt{
 \frac{\sum_d |\Delta_{d,+\HALF}\sigma  - \Delta_{d,-\HALF}\sigma|^2}
      {\sum_d \left(|\Delta_{d,+\HALF}\sigma| + |\Delta_{d,-\HALF}\sigma| 
              +\epsilon \sigma_{d,{\rm ref}}\right)^2 }
 }
\end{equation}
where $\sigma \equiv \sigma(\vec{U})$ is a function of the conserved variables,
$\Delta_{d,\pm\HALF}\sigma$ are the undivided forward and backward differences in 
the direction $d$, e.g., $\Delta_{x,\pm\HALF}\sigma = \pm(\sigma_{i\pm1} - \sigma_i)$.
The last term appearing in the denominator, $\sigma_{d,{\rm ref}}$, prevents 
regions of small ripples from being refined \citep{FLASH00} and it is
defined by
\begin{equation}
  \sigma_{x,{\rm ref}} = |\sigma_{i+1}| + 2|\sigma_i| + |\sigma_{i-1}|
\end{equation}
Similar expressions hold when $d=y$ or $d=z$. 
In the computations reported in this paper we use $\epsilon = 0.01$
as the default value.

\subsection{Time Step Limitation of Point-Local Source Terms}
\label{sec:cool_source}
%
%
%

In the usual AMR strategy, grids belonging to level $\ell$ are advanced 
in time by a sequence of steps with typical size
\begin{equation}\label{eq:level_dt}
  \Delta t^{\ell,n+1} = \frac{\Delta t^{0}_{\min}}{(2)^\ell}\,,
\end{equation}
where we assume, for simplicity, a grid jump of 2.
Here $\Delta t^{0}_{\min}$ is chosen by collecting and re-scaling 
to the base grid the time steps from all levels available from the previous 
integration step:
\begin{equation}
 \Delta t^{0}_{\min} = \min_{0\le \ell\le \ell_{\max}}
 \left[(2)^\ell\Delta t^{\ell,n}\right]\,,
\end{equation}
where $\Delta t^{\ell,n}$ is computed using (\ref{eq:dt_cool}).
However, this procedure may become inefficient in presence of source terms 
whose time scale does not depend on the grid size. 
As an illustrative example, consider a strong radiative shock propagating 
through a static cold medium. In the optically thin limit, radiative losses 
are assumed to be local functions of the state vector, but they do not involve
spatial derivatives. 
If the fastest time scale in the problem is dictated by the cooling process, 
the time step should then become approximately the same on all levels,
$\Delta t\approx \Delta t^{\ell}_{\rm rad} \approx 
\Delta t^{0}_{\rm rad}$, regardless of the mesh size. 
However from the previous equations, one can see that finer levels with 
$\ell>0$ will advance with a time step $(2)^\ell$ smaller than required by the single 
grid estimate. Eq. (\ref{eq:level_dt}) is nevertheless essential for proper 
synchronization between nested levels.

Simple considerations show that this deficiency may be cured by 
treating split and leaf cells differently. Split zones in a given level $\ell$ 
are, in fact, overwritten during the projection step using the more accurate 
solution computed on children cells belonging to level $\ell+1$.
Thus, accurate integration of the source term is not important for
these cells and could even be skipped.
From these considerations, one may as well evaluate the source term-related 
time step on leaf cells only, where the accuracy and stability of the computed 
solution is essential.
This trick should speed up the computations by a factor of approximately $(2)^\ell$, 
thus allowing to take full advantage of the refinement offered by the AMR 
algorithm without the time step restriction. 
Besides, this should not alter nor degrade the solution computed during this 
single hierarchical integration step as long as the projection step precedes 
the regrid process.

The proposed modification is expected to be particularly efficient in those 
problems where radiative losses are stronger in proximity of steep gradients.

\subsection{Parallelization and load balancing}
%
%
%

Both \PLUTO and \PlCh support calculations in parallel computing 
environments through the Message Passing Interface (MPI).
Since the time evolution of the AMR hierarchy is performed recursively,
from lower to upper levels, each level of refinement is parallelized
independently by distributing its boxes to the set of processors.
The computation on a single box has no internal parallelization.
Boxes are assigned to processors by balancing the computational load
on each processor. Currently, the workload of a single box is estimated
by the number of grid points of the box, considering that the integration
requires approximately the same amount of flops per grid point.
This is not strictly true in some specific case, e.g in the presence of
optically thin radiative losses, and a strategy to improve the load
balance in such situations is currently under development. On the basis
of the box workloads, CHOMBO's load balancer uses the Kernighan-Lin
algorithm for solving knapsack problems.

CHOMBO {\it weak} scaling performance has been thoroughly benchmarked:
defining the initial setup by spatially replicating a 3D hydrodynamical problem
proportionally to the number of CPUs employed, the execution time stays
constant with excellent approximation \citep[see][ for more details]{VanStraalen09}.

To test the parallel performance of PLUTO-CHOMBO in real applications, we
performed a number of {\it strong} scaling tests by computing the execution
time as a function of the number of processors for a given setup. While this is
a usual benchmark for static grid calculations, in the case of AMR computations
this diagnostic is strongly problem-dependent and difficult to interpret.

In order to find some practical rule to improve the scaling of AMR
calculations, we investigated the dependency of the parallel performance on
some parameters characterizing the adaptive grid structure: the maximum
box size allowed and the number of levels of refinement employed using different
refinement ratios.
As a general rule, the parallel performance deteriorates when the number of blocks
per level becomes comparable to the number of processors or, alternatively,
when the ideal workload per processor (i.e. the number of grid cells of a level
divided by the number of CPUs) becomes comparable to the maximum box size.
As we will show, decreasing the maximum possible block size can sensibly increase
the number of boxes of a level and therefore improves the parallel performance.
On the other hand, using less refinement levels with larger refinement ratios to
achieve the same maximum resolution can lower the execution time, reducing the
parallel communication volume and avoiding the integration of intermediate
levels.

In Section \ref{sec:MHD_Test} and \ref{sec:RMHD_Test} we will present several parallel scaling
tests and their dependence on the aforementioned grid parameters.

\section{MHD Tests}
\label{sec:MHD_Test}
%
%
%
%

In this section we consider a suite of test problems specifically designed
to assess the performance of \PlCh for classical MHD flows.
The selection includes one, two and three-dimensional standard numerical
benchmarks already presented elsewhere in the literature as well as applications
of astrophysical relevance.
The single-patch integrator adopts the characteristic tracing step described in
section \ref{sec:np_char} with either PPM, WENO or linear interpolations carried
out in characteristic variables.

\subsection{Shock tube problems}
%
%

The shock tube test problem is a common benchmark for an 
accurate description of both continuous and discontinuous 
flow features. In the following we consider one and three dimensional 
configurations of standard shock tubes proposed 
by \cite{Torrilhon03} and \cite{RJ95}.

\subsubsection{One-Dimensional Shock tube}
\label{sec:Compound}
%
%
\ifx\IncludeEps\yes
\begin{figure}
\centering
\includegraphics*[width=\columnwidth]{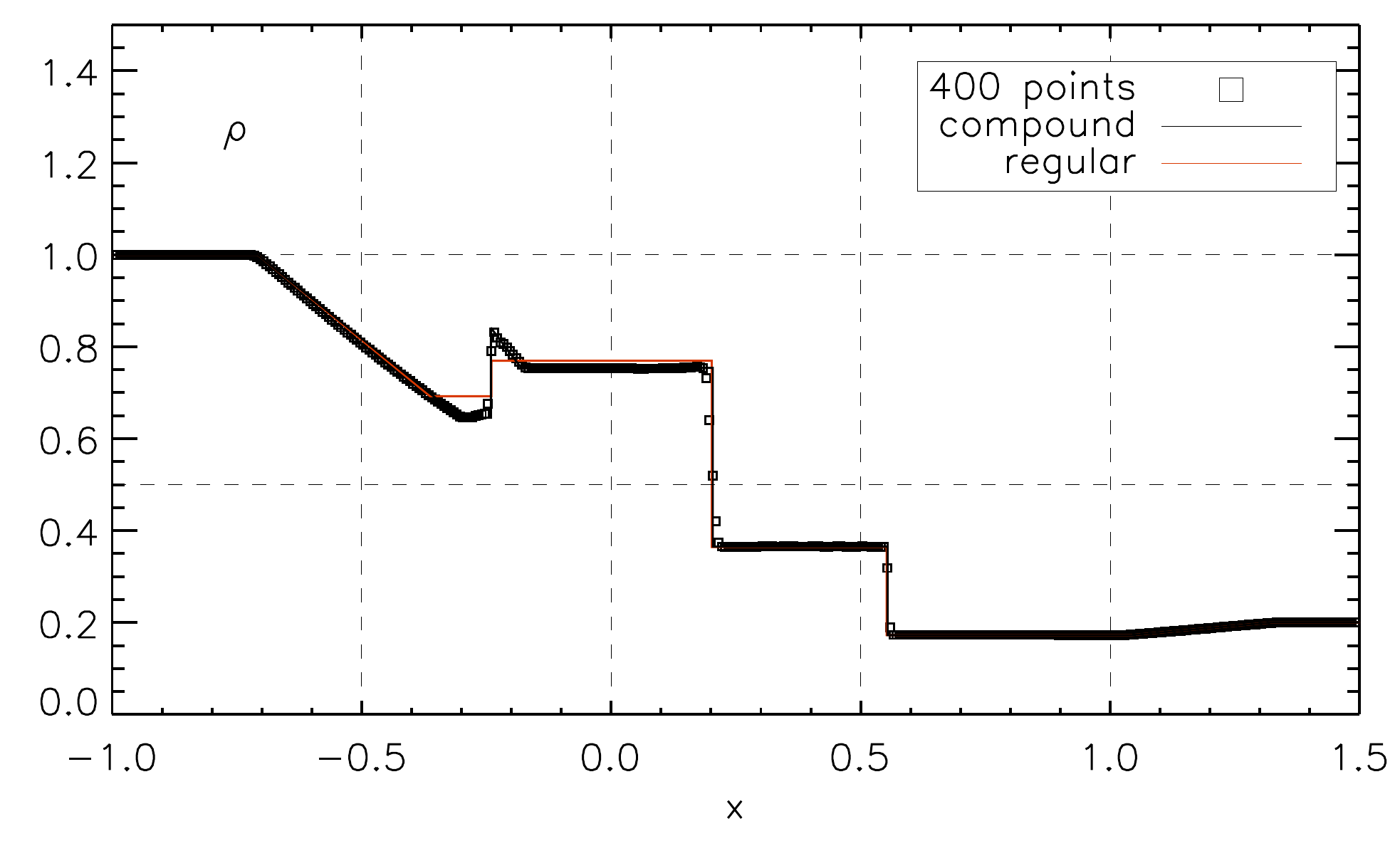}
 \caption{\footnotesize Density profiles for the 1D shock tube problem at $t=0.4$ with $\alpha=\pi$.
 Solid lines correspond to admissible analytic solutions with (black) and 
 without (red) a compound wave. The numerical solution (symbols) is obtained with 400 
 grid points.}
 \label{fig:sod1d}
\end{figure}
\begin{figure}[!h]
\centering
\includegraphics*[width=\columnwidth]{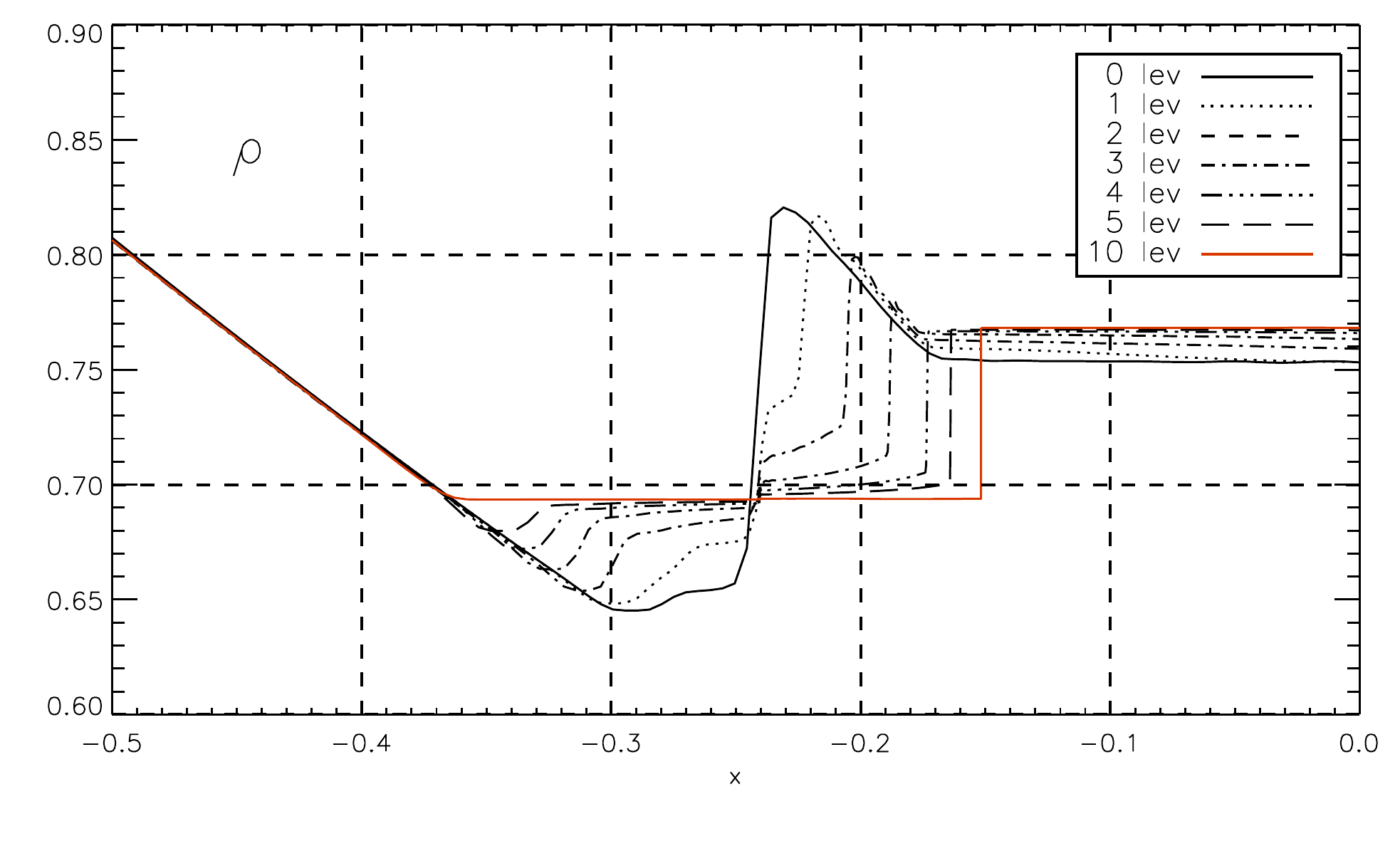}
 \caption{\footnotesize Density profiles for the 1D shock tube problem at $t=0.4$ at the 
 vicinity of the compound wave locus, with $\alpha=3$. The AMR levels vary 
 from $0$ to $10$ as reported in the legend, exploring cases of equivalent resolution
 of $512$ (solid), $1024$ (dot), $2048$ (dash), $4096$ (dot-dash), $8192$ ($3$ dot-dash), 
 $16,384$ (long dash) and $1,048,576$ (solid red) points. At high resolution the solution 
 converges to the regular one. This figure is analogous to fig. 3 of \cite{FHT06}.}
 \label{fig:sod1d_amr}
\end{figure}
\fi
Following \cite{Torrilhon03} we address the capability of the AMR scheme
to handle and refine discontinuous features as well as to correctly resolve 
the non uniqueness issue of MHD Riemann problems in finite volume 
schemes (\citealp{Torrilhon03, TB04} and references therein).
Left and right states are given by 
\begin{equation}
\left\{
\begin{array}{lclr}
\vec{V}_L & = & \DS
 \left(1,0,0,0,1,1,0,1\right)^T 
              & \mathrm{for}\quad  x_1 < 0 \,, \\ \noalign{\medskip}
\vec{V}_R & = & \DS \left(0.2,0,0,0,1,\cos(\alpha),\sin(\alpha),0.2\right)^T 
              & \mathrm{for}\quad  x_1 > 0  \,,
\end{array}\right.
\end{equation}
where $\vec{V} = \left(\rho,v_x,v_y,v_z, B_x, B_y, B_z, p\right)$ is
the vector of primitive variables and $\Gamma = 2$. 
As discussed by \cite{Torrilhon03}, for a wide range of initial conditions 
MHD Riemann problems have unique solutions, consisting of Lax shocks, 
contact discontinuities or usual rarefaction waves. 
Nevertheless, there exist certain sets of initial values that can result in 
non unique solutions. 
When this occurs, along with the regular solution arises one that allows for 
irregular MHD waves, for example a compound wave. 
A special case where the latter appears is when the initial transverse 
velocity components are zero and the magnetic field vectors are 
anti-parallel. 
Such a case was noted by \cite{BW88} and can be reproduced by simply choosing 
$\alpha = \pi$ in our initial condition.

A one dimensional non unique solution is calculated using a static grid 
with $400$ zones, $x\in[-1,1.5]$. 
Left and right boundaries are set to outflow and the evolution stops 
at time $t = 0.4$, before the fast waves reach the borders.     
The resulting density profile is shown in Fig. (\ref{fig:sod1d}). 
The solid lines denote the two admissible exact solutions: the regular 
(red) and the one containing a compound wave (black), the latter situated 
at $x\,\sim\,-0.24$. 
It is clear that the solution obtained with the Godunov-type code is the 
one with the compound wave (symbols).

The crucial problematic of this test occurs when $\alpha$ is close to but 
not exactly equal to $\pi$. 
\cite{Torrilhon03} has proven that regardless of scheme, the numerical 
solution will erroneously tend to converge to an ``irregular'' one 
similar to $\alpha=\pi$ (pseudo-convergence), even if the initial conditions 
should have a unique, regular solution.
This pathology can be cured either with high order schemes \citep{TB04} or 
with a dramatic increase in resolution on the region of interest, proving
AMR to be quite a useful tool. 
To demonstrate this we choose $\alpha=3$ for the field's twist. 

\begin{deluxetable}{cccccc}
\tablecaption{CPU running time for the one-dimensional MHD shock-tube using
              both static and AMR computations.}
\tablewidth{0pt}
\tablehead{\multicolumn{2}{c}{Static Run} &  \multicolumn{3}{c}{AMR Run} &
           \colhead{Gain} \\ 
           \cline{1-2} \cline{3-5} \\
           \colhead{$N_x$} & \colhead{Time (s)} & \colhead{Level} & 
           \colhead{Ref ratio} & \colhead{Time (s)} & \colhead{}}
\startdata
512     & 0.5                      &  0  & 2    & 0.5    &  1\\
1024    & 1.9                      &  1  & 2    & 1.1    &  1.7  \\
2048    & 7.5                      &  2  & 2    & 2.3    &  3.3 \\
4096    & 31.6                     &  3  & 2    & 4.5    &  7.0 \\ 
8192    & 138.0                    &  4  & 2    & 8.7    &  15.9 \\
16384   & 546.1                    &  5  & 2    & 16.9   &  32.3 \\
1048576 & 2.237 $\cdot 10^6$$^\dagger$ & 10  & 2 (4)& 1131.1 &  1977.8
\enddata
\tablecomments{The first and second columns give the number of points 
              $N_x$ and corresponding CPU for the static grid run (no AMR).
              The third, fourth and fifth columns give, respectively, the number  
              of levels, the refinement ratio and CPU time for the AMR run at 
              the equivalent resolution. The last row refers to the solid red
              line of Fig. \ref{fig:sod1d_amr}, where a jump ratio of four was
              introduced between levels $6$ and $7$ to reach an equivalent of
              $\sim 10^6$ grid points. 
              The last column shows the corresponding gain factor calculated
              as the ratio between static and AMR execution time.}
\tablenotetext{$\dagger$}{CPU time has been inferred from ideal scaling.}
\label{tab:mhd_shocktube_timing}              
\end{deluxetable}
Starting from a coarse grid of $512$ computational zones, we vary the 
number of refinement levels, with a consecutive jump ratio of two. The
$10$ level run (solid red line) incorporates also a single jump ratio of
four between the sixth and seventh refinement levels, reaching a maximum
equivalent resolution of $1,048,576$ zones (see Fig. \ref{fig:sod1d_amr}). 
The refinement criterion is set upon the variable 
$\sigma = (B_x^2 + B_y^2+B_z^2)/\rho$ using Eq. (\ref{eq:RefCrit}) 
with a threshold $\chi_r = 0.03$, whereas integration is performed using PPM 
with a Roe Riemann solver and a Courant number $C_a = 0.9$.
As resolution increases the compound wave disentangles and the solution 
converges to the expected regular form \citep{Torrilhon03,FHT06}. 
In Table \ref{tab:mhd_shocktube_timing} we compare the CPU running time
of the AMR runs versus static uniform grid computations at the same
effective resolution. With 5 and 10 levels of refinement (effective
resolutions 16,384 and 1,048,576 zones, respectively) the AMR approach is
$\sim 32$ and $\sim 1978$ times faster than the uniform mesh
computation, respectively.

\subsubsection{Three-Dimensional Shock tube}
%
%
\ifx\IncludeEps\yes
\begin{figure*}\centering
\includegraphics[width=0.95\textwidth]{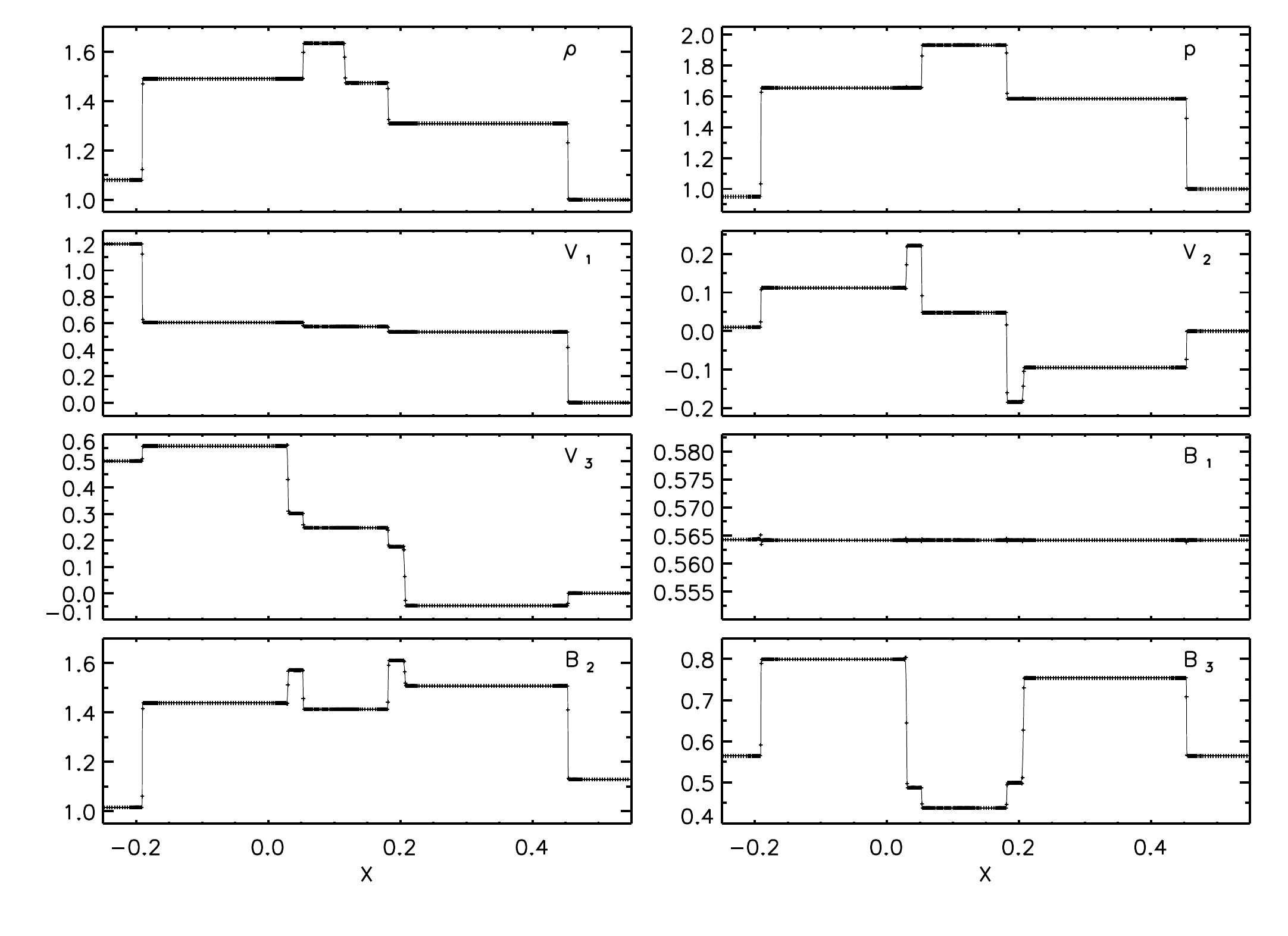}
 \caption{\footnotesize Primitive variable profiles for the 3D shock tube problem at $t=0.02\cos\alpha\cos\gamma$, 
          along the $x$ direction. Density and thermal pressure are plotted in 
          the top panels while vector field components normal (``1'') and transverse 
          (``2'' and ``3'') to the initial surface of discontinuity are shown in the middle 
          and bottom panels. 
          We show a smaller portion of the domain, $x\in[-0.25,0.55]$, in 
          order to emphasize the change of resolution by symbol density.}
 \label{fig:sod3db}
\end{figure*}
\begin{figure}\centering
\includegraphics*[width=\columnwidth]{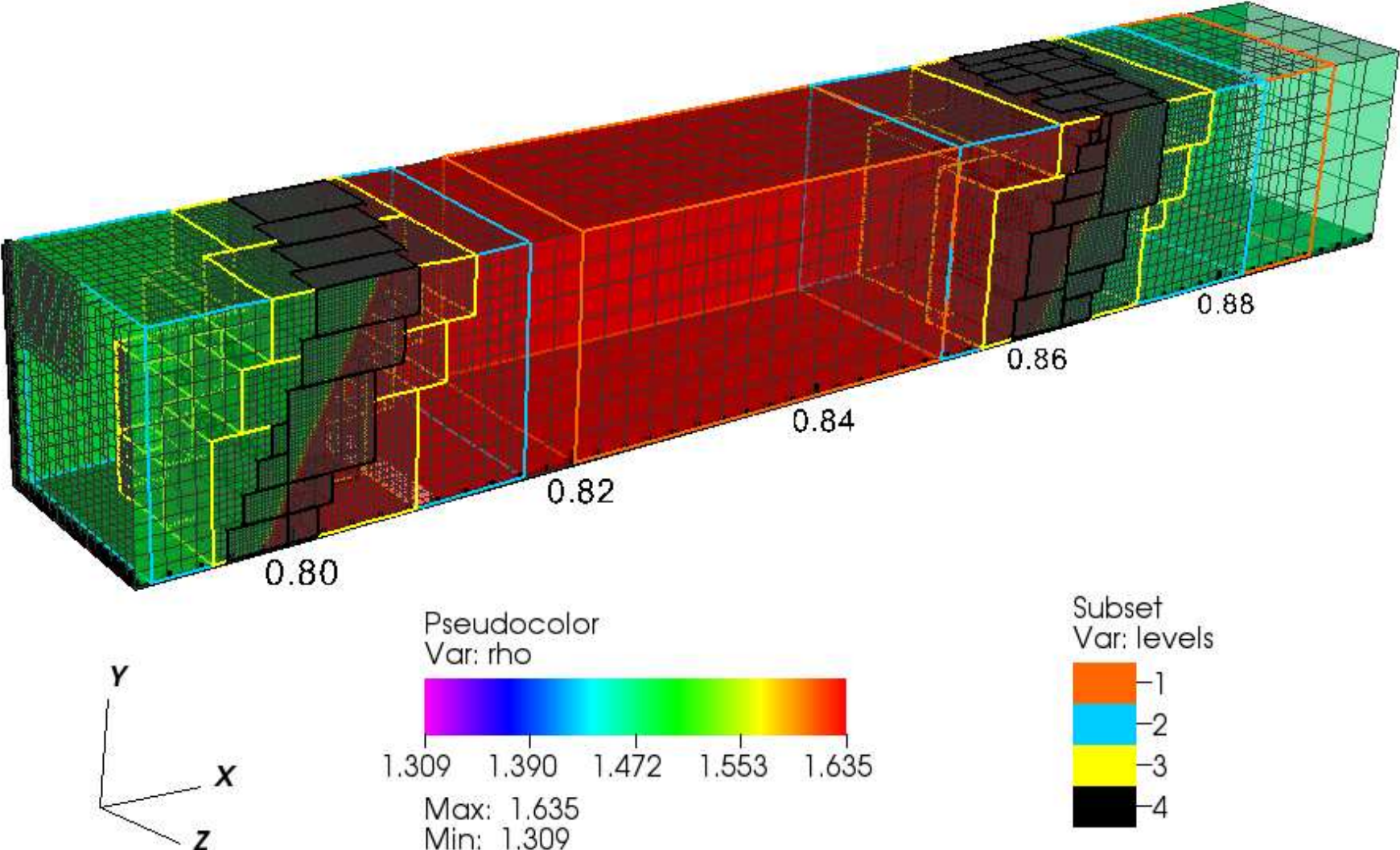}
 \caption{\footnotesize Closeup of the top-hat feature in the density profile for the 3D shock tube problem, 
          along with AMR level structure and the mesh.
          Different colors are used to distinguish grid levels.}
 \label{fig:sod3da}
\end{figure}
\fi
The second Riemann problem was proposed by \cite{RJ95} and later
considered also by \cite{Toth00}, \cite{BS00}, \cite{GS08}, \cite{MT10}, \cite{MTB10}. 
An initial discontinuity is described in terms of primitive variables as 
\begin{equation}\label{eq:ic_3dst}
\left\{\begin{array}{lclr}
\vec{V}_L & = & \DS
 \left(1.08,1.2,0.01,0.5,\frac{2}{\sqrt{4\pi}},\frac{3.6}{\sqrt{4\pi}} 
                         \frac{2}{\sqrt{4\pi}},0.95\right)^T \\ \noalign{\medskip}
          &   &\quad\quad\quad\quad\quad\quad\quad\quad\quad\quad\quad\quad \mathrm{for}\quad  x_1 < 0  \\ \noalign{\medskip}
\vec{V}_R & = & \DS \left(1,0,0,0,\frac{2}{\sqrt{4\pi}},\frac{4}{\sqrt{4\pi}},
                                  \frac{2}{\sqrt{4\pi}},1\right)^T \\ \noalign{\medskip}
          &   &\quad\quad\quad\quad\quad\quad\quad\quad\quad\quad\quad\quad \mathrm{for}\quad  x_1 > 0  
\end{array}\right.
\end{equation}
where $\vec{V} = \left(\rho,v_1,v_2,v_3, B_1, B_2, B_3, p\right)$ is the vector
of primitive variables.
The subscript ``1'' gives the direction perpendicular to the initial surface
of discontinuity whereas ``2'' and ``3'' correspond to the transverse directions.
We first obtain a one-dimensional solution on the domain $x\in[-0.75,0.75]$ 
using 6144 grid points, stopping the computations at $t=0.2$.

In order to test the ability of the AMR scheme to maintain the translational
invariance and properly refine the flow discontinuities, the shock tube is 
rotated in a three dimensional Cartesian domain. 
The coarse level of the computational domain consists of 
$[384\times4\times4]$ zones and spans $[-0.75,0.75]$ in the $x$ direction 
while $y,z\,\in [0,0.015625]$. 
The rotation angles, $\gamma$ around the $y$ axis and $\alpha$ around the 
$z$ axis, are chosen so that the planar symmetry is satisfied by an integer 
shift of cells $(n_x,n_y,n_z)$. 
The rotation matrix can be found in \cite{MTB10}. 
By choosing $\tan\alpha= -r_1/r_2$ and 
$\tan\beta = \tan\gamma/\cos\alpha = r_1/r_3$, one can show 
\citep{GS08} that the three integer shifts $n_x, n_y, n_z$ must obey
\begin{equation}
n_x - n_y \frac{r_1}{r_2} + n_z \frac{r_1}{r_3} = 0 \,,
\end{equation}
where cubed cells have been assumed and $(r_1,r_2,r_3) = (1,2,4)$ will be used.
Computations stop at $t=0.2\cos\alpha\cos\gamma$, once again before the 
fast waves reach the boundaries. 
We employ 4 refinement levels with consecutive jumps of two, 
corresponding to an equivalent resolution of $6144\times64\times64$ zones.
The refinement criterion is based on the normalized second derivative of 
$\chi = (|B_x| + |B_y|)\rho$ with a threshold value $\chi_r = 0.1$.
Integration is done with PPM reconstruction, a Roe Riemann 
solver and a Courant number of $C_a = 0.4$.

The primitive variable profiles (symbols) are displayed in Fig. \ref{fig:sod3db}
along the $x$ direction\footnote{
Note that similar plots were produced in MT and \cite{MTB10} but 
erroneously labeled along the ``rotated direction'' rather than the $x$ axis.}, 
together with the one 
dimensional reference solution in the $x\in[-0.25,0.55]$ region. 
In agreement with the solution of \cite{RJ95}, the wave pattern produced consists 
of a contact discontinuity that separates two fast shocks, two slow shocks and a 
pair of rotational discontinuities. 
A three dimensional closeup of the top-hat feature in the density profile is 
shown in Fig. \ref{fig:sod3da}, along with AMR levels and mesh.
The discontinuities are captured correctly, and the AMR grid structure respects
the plane symmetry. 
Our results favorably compare with those of \cite{GS08, MT10, MTB10} and 
previous similar 2D configurations.

The AMR computation took approximately $3\,{\rm hours}$ and $53$ minutes on two $2.26\,\textrm{GHz}$
Quad-core Intel Xeon processors ($8$ cores in total).
For the sake of comparison, we repeated the same computation on a uniform mesh 
of $768\times8\times8$ zones ($1/8$ of the effective resolution) with the 
static grid version of \PLUTO employing $\approx 79$ seconds.
Thus, extrapolating from ideal scaling, the computational cost of the fixed grid 
calculation is expected to increase by a factor $2^{12}$ giving an overall 
gain of the AMR over the uniform grid approach of  $\sim 23$.

\subsection{Advection of a magnetic field loop}
%
%

\ifx\IncludeEps\yes
\begin{figure*}\centering
\includegraphics[width=0.9\textwidth]{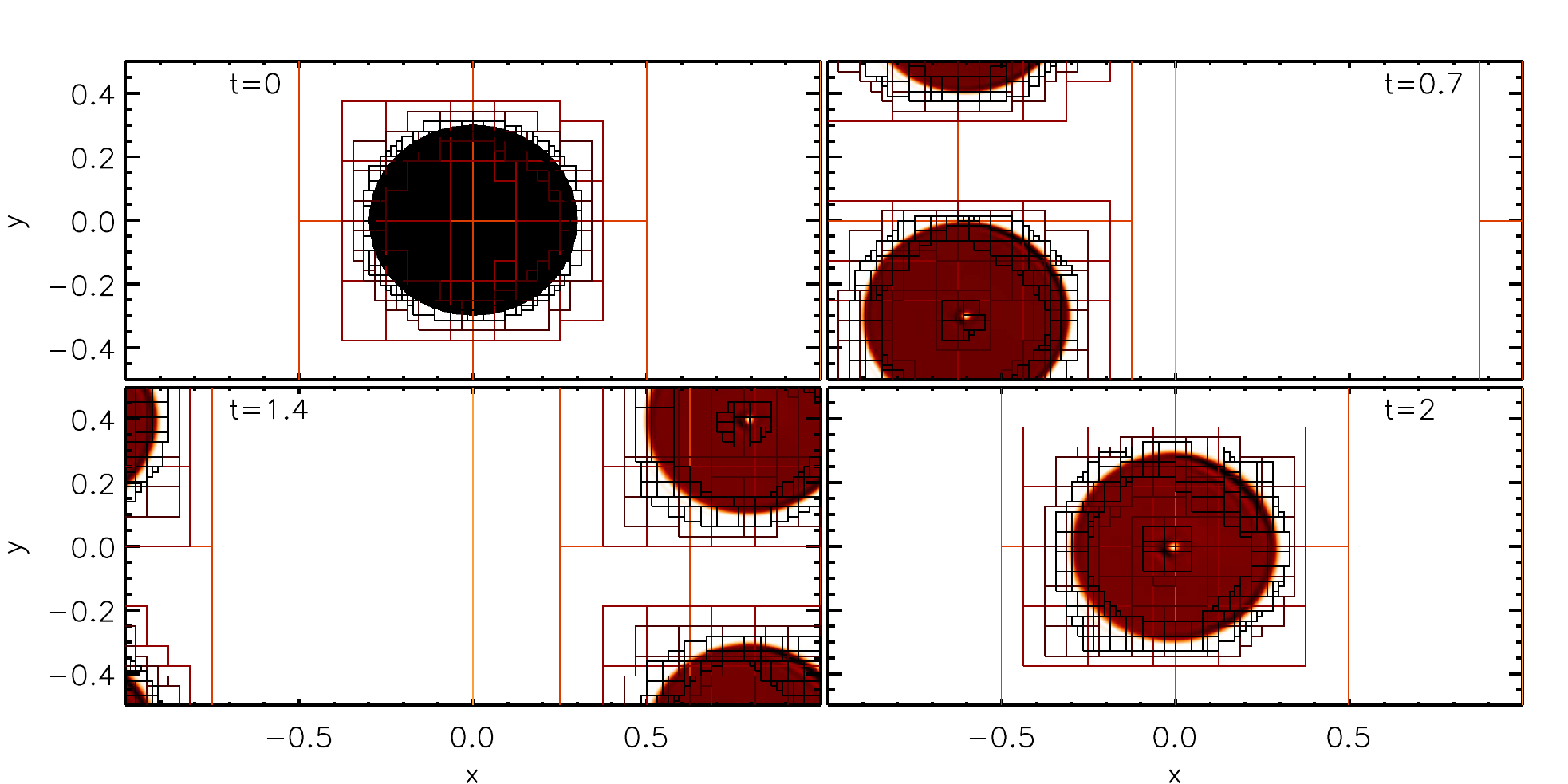}
 \caption{\footnotesize Magnetic energy density for the 2D field loop problem at $t=0,\,0.7,\,1.4,\,2$. 
  Overplotted are the refinement levels.}
 \label{fig:fl2d}
\end{figure*}
\begin{figure}[t]\centering
\includegraphics*[width=0.7\columnwidth]{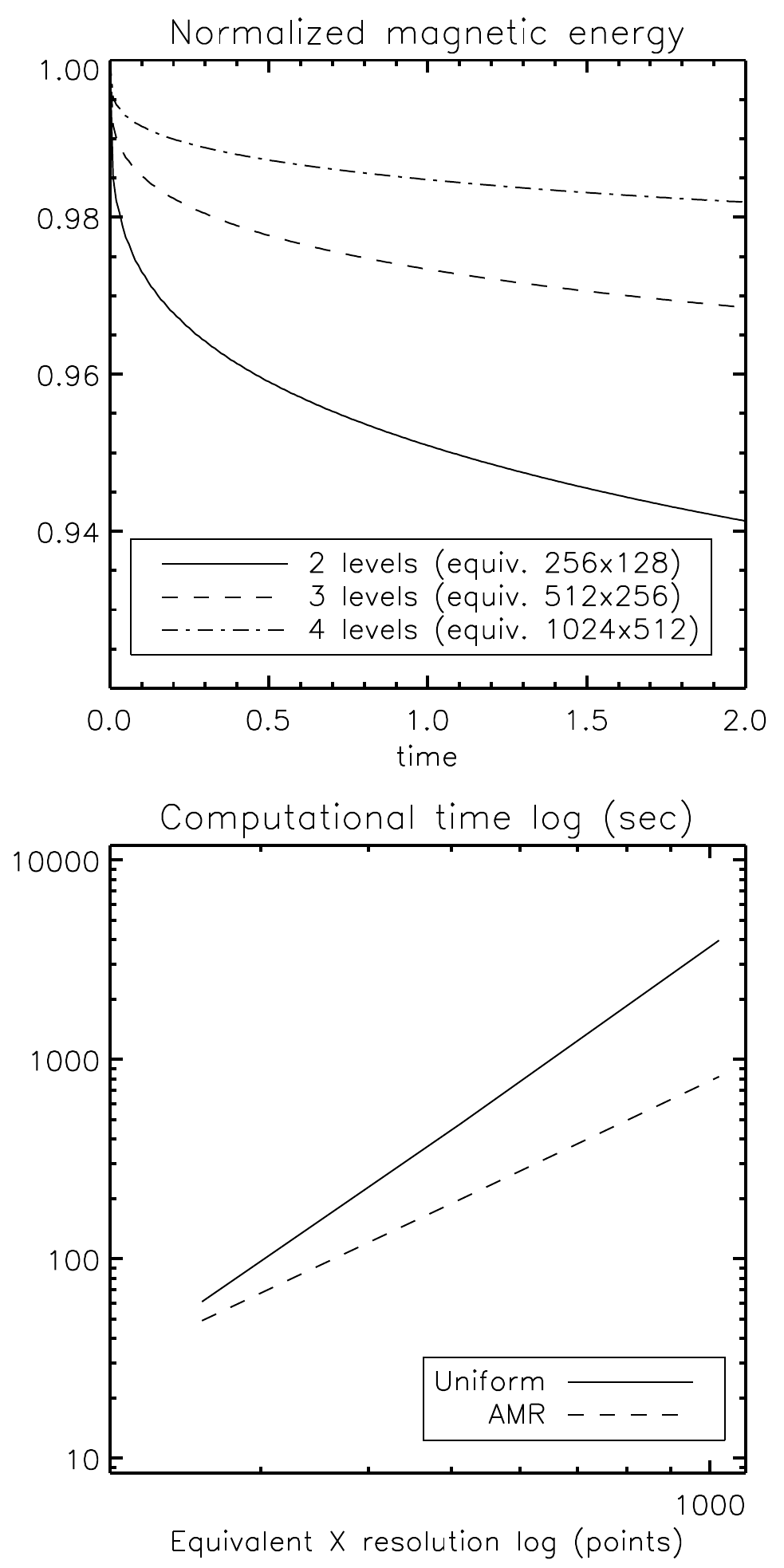}
 \caption{\footnotesize Upper panel: Normalized magnetic energy for the field loop
          advection problem as a function of time. 
          Lower panel: computational time as a function of equivalent resolution.}
 \label{fig:fl2d_energy}
\end{figure}
\fi
The next test problem considers the two dimensional advection of a magnetic 
field loop. This test, proposed by \cite{GS05}, aims to benchmark the scheme's 
dissipative properties and the correct discretization balance of 
multi-dimensional terms through monitoring the preservation of the initial 
circular shape of the loop. 

As in  \cite{GS05} and \cite{FHT06}, we define the computational domain by $x\in[-1,1]$ 
and $y\in[-0.5,0.5]$ discretized on a coarse grid of $64\times 32$ grid cells.  
In the initial condition, both density and pressure are uniform and equal to $1$, while 
the velocity of the flow is given by 
$\vec{v} = V_0\cos\alpha\hvec{e}_x + V_0\sin\alpha\hvec{e}_y$
with $V_0 = \sqrt{5}$, $\sin \alpha = 1/\sqrt{5}$ and $\cos \alpha = 2/\sqrt{5}$. 
The magnetic field is then defined through its magnetic vector potential as 
\begin{equation}  
A_z = \left\{ \begin{array}{ll}
    A_0(R-r) & \textrm{if} \quad r \leq R \,, \\ \noalign{\medskip}
    0   & \textrm{if} \quad r > R \,,
  \end{array} \right.
\end{equation}     
where $A_0 = 10^{-3}$, $R=0.3$, and $r=\sqrt{x^2 + y^2}$. 
The simulation evolves until $t=2$, when the loop has thus performed two crossing 
through the periodic boundaries. The test is repeated with $2$, $3$ and $4$
levels of refinement (jump ratio of two), resulting to equivalent resolutions of $[256\times128]$,
$[512\times256]$ and $[1024\times512]$ respectively. 
Refinement is triggered whenever the second derivative error norm of $(B_x^2 + B_y^2)\times10^6$, 
computed via Eq. (\ref{eq:RefCrit}), exceeds the threshold $\chi_r = 0.1$.
The integration is carried out utilizing WENO reconstruction and the Roe 
Riemann solver, with $C_a = 0.4$.

The temporal evolution of magnetic energy density is seen in Fig. (\ref{fig:fl2d}), 
with $4$ levels of refinement. 
As the field loop is transported inside the computational domain, the grid 
structure changes to follow the evolution and retain the initial circular form. 
An efficient way to quantitatively measure the diffusive properties of the 
scheme is to monitor the dissipation of the magnetic energy. 
In the top panel of Fig. (\ref{fig:fl2d_energy}) we plot the normalized mean 
magnetic energy as a function of time.
By increasing the levels of refinement the dissipation of magnetic energy 
decreases, with $<{\bf B}^2>$ ranging from $\sim 94\%$ to $\sim 98\%$ of the 
initial value.
In order to quantify the computational gain of the AMR scheme we repeat the simulations
with a uniform grid resolved onto as many points as the equivalent resolution, without 
changing the employed numerical method. The speed-up is reported in the
bottom  panel of Fig. \ref{fig:fl2d_energy}. 

\subsection{Resistive Reconnection}
%
%

\ifx\IncludeEps\yes
\begin{figure*}\centering
\includegraphics[width=0.9\textwidth]{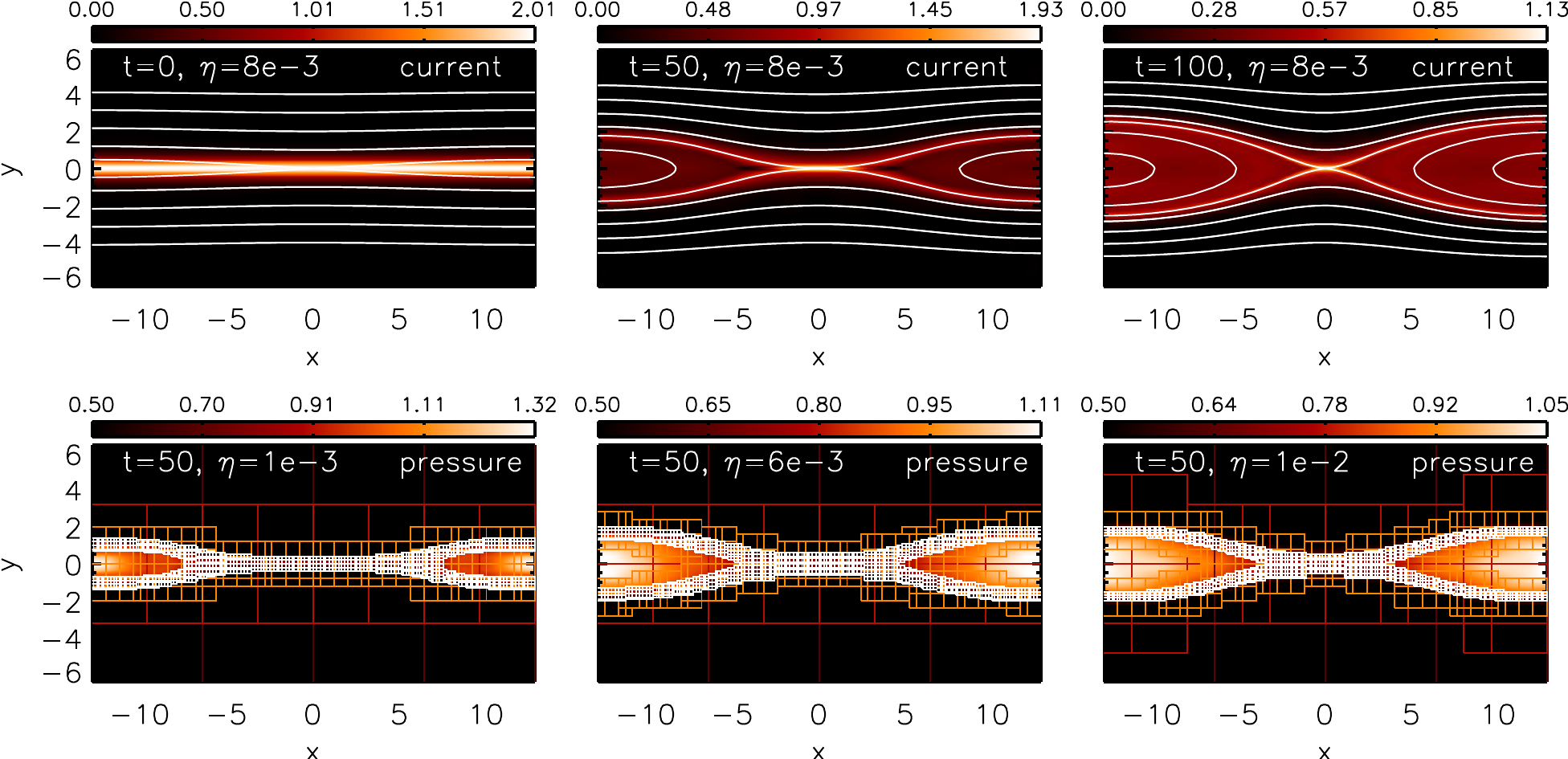}
 \caption{\footnotesize Upper row: temporal evolution of the current density and magnetic field lines 
  for the resistive reconnection problem, with $\eta=8\,10^{-3}$. Snapshots refer to $t=0,\,50,\,100$.
  Lower row: Pressure profiles for various values of resistivity $\eta$, along with the AMR 
  level structure at $t=50$. The refinement strategy consists of $3$ levels with a jump ratio of $2:4:4$
  (equivalent resolution of $2048\times 1024$ mesh points).}
 \label{fig:gem_evolution}
\end{figure*}
\begin{figure}\centering
\includegraphics*[width=\columnwidth]{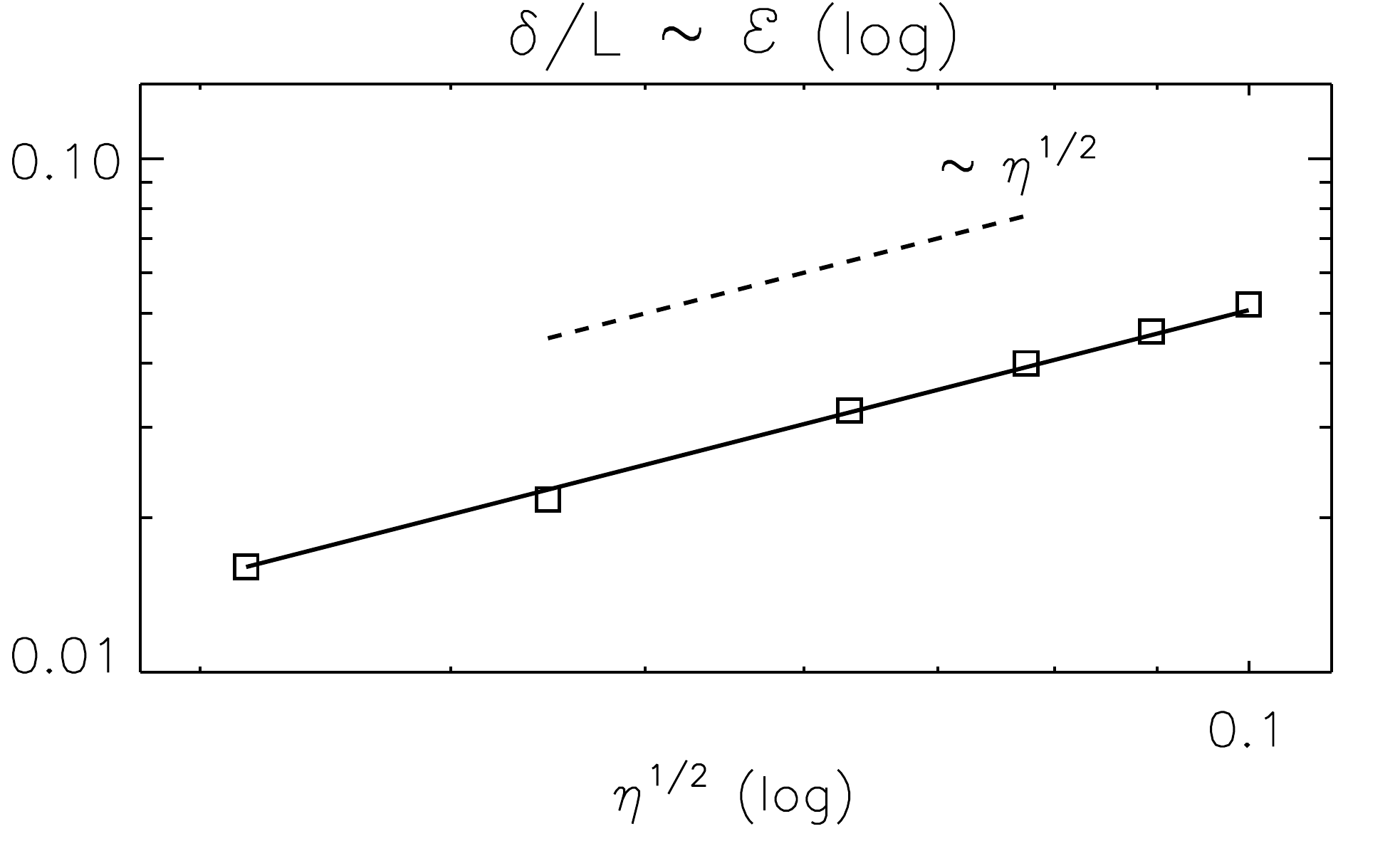}
 \caption{\footnotesize Time average of $\delta/L$, analogous to the magnetic reconnection rate ${\cal E}$, 
 as a function of resistivity. Symbols represent the actual data, whereas over-plotted
 (dashed line) is the Sweet-Parker scaling $\sim \sqrt{\eta}$, along with a best fit (solid line). 
 The numerical results are in agreement with the theoretical scaling.}
 \label{fig:gem_rate}
\end{figure}
\begin{figure}\centering
\includegraphics*[width=\columnwidth]{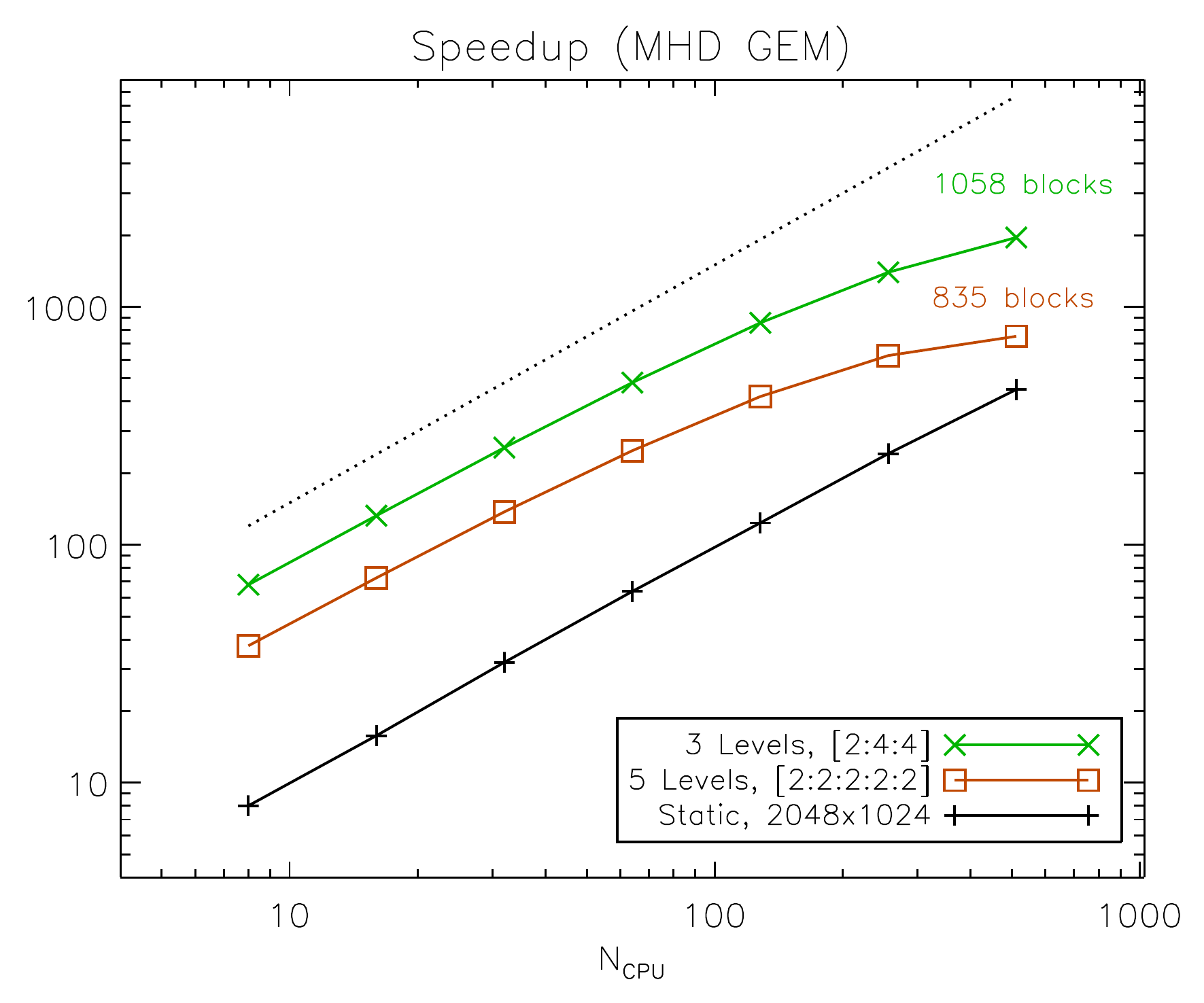}
 \caption{\footnotesize Parallel speedup at $t=50$ as a function of the number of 
          processors ($N_{\rm CPU}$) for the resistive reconnection problem.
          The different lines refer to the execution times obtained
          with 5 levels of refinement with consecutive jumps of 2 (red 
          squares), 3 levels with jump ratios $2:4:4$ (green crosses) and a 
          fixed uniform grid with $2048\times1024$ zones (black plus signs).
          The dotted line gives the ideal scaling whereas the number of blocks on
          the finest level at the end of integration is reported above each curve.}
 \label{fig:mhd_GEMScaling}
\end{figure}
\fi
Magnetic reconnection refers to the process of breaking and reconnection of 
magnetic field lines with opposite directions, accompanied with a
conversion of magnetic energy into kinetic and thermal energy of the plasma. 
This is believed to be the basic mechanism behind energy release during solar
flares.
The first solution to the problem was given independently by \cite{Sweet58} and
\cite{Parker57}, treating it as a two dimensional boundary layer problem in
the laminar limit. 

According to the Sweet-Parker model, the magnetic field's convective inflow is
balanced by Ohmic diffusion. Along with the assumption of continuity, this yields
a relation between reconnection and plasma parameters. If $L$ and $\delta$ are
the boundary layer's half length and width respectively, we can write the
reconnection rate ${\cal E}$ as
\begin{equation}
{\cal E}\equiv \frac{u_{\textrm{in}}}{u_{\textrm{out}}}\sim \frac{\delta}{L}\sim\frac{1}{\sqrt{S}}.
\label{eq:rec_rate}
\end{equation}
With $u_{\textrm{in}}$ and $u_{\textrm{out}}$ we denote the inflow and outflow
speeds, into and out of the boundary layer, respectively. 
The Lundquist number for the boundary layer is defined as $S = u_{A}L/\eta$, 
with $u_A$ being the Alfv\'en velocity directly upstream of the layer, 
$\eta$ the magnetic resistivity and $L$ the layer's half length. 
This dependency of the reconnection rate with the square root of magnetic 
resistivity is called the Sweet-Parker scaling and has been verified both 
numerically \citep{Biskamp86, UK00} and experimentally \citep{Ji98}. 


Following the guidelines of the GEM Magnetic Reconnection Challenge \citep{Birn01}, 
the computational domain is a two dimensional Cartesian box, with
$x\in[-L_x/2,L_x/2]$ and $y\in[-L_y/2,L_y/2]$ where we choose $L_x\,=\,25.6$ and
$L_y\,=\,12.8$.
The initial condition consists of a Harris current sheet: the magnetic field
configuration is described by $B_x(y)=B_0\tanh(y/\lambda)$ whereas the flow's
density is $\rho = \rho_0 \textrm{sech}^2(y/\lambda) + \rho_{\infty}$, where
$\lambda=0.5$, $\rho_0=1$ and $\rho_{\infty}=0.2$. 
The flow's thermal pressure is deduced assuming equilibrium with magnetic pressure, 
$P = B_0^2/2=0.5$. 
The initial magnetic field components are perturbed via
\begin{equation}
\begin{array}{ll}
dB_x  = - \Psi_0(\pi/L_y)\sin(\pi y/L_y)\cos(2\pi x/L_x)\\\noalign{\medskip}
dB_y  =   \Psi_0(2\pi/L_x)\sin(2\pi x/L_x)\cos(\pi y /L_y).
\end{array}
\end{equation}
where $\Psi_0 = 0.1$. The coarse grid consists of $[64\times32]$ points and
additional levels of refinement are triggered using the following criterion 
based on the current density:
\begin{equation}\label{eq:MHD_GEM_tag}
  \frac{\left|\Delta_xB_y - \Delta_yB_x\right|}
              {|\Delta_xB_y|+|\Delta_yB_x|+\sqrt{\rho}/\xi} > 
   \frac{\chi_{\min} - \chi_{\max}}{1 + (1-\xi)^2} + \chi_{\max}
\end{equation}
where $\Delta_xB_y$ and $\Delta_yB_x$ are the undivided central differences 
of $B_y$ and $B_x$ in the $x$ and $y$ direction, respectively,
and $\xi = \Delta x^0/\Delta x^\ell \ge 1$ is the ratio of grid spacings 
between the base ($0$) and current level ($l$).
The threshold values $\chi_{\min}=0.2$ and $\chi_{\max} = 0.37$ are chosen 
in such a way that refinement becomes increasingly harder for higher 
levels. 
We perform test cases using either $5$ levels of refinement with a consecutive
jump ratio of two or $3$ levels with jump ratio $2:4:4$ reaching, in both cases,
an equivalent resolution of $2048\times 1024$ mesh points.
Boundaries are periodic in the $x$ direction, whereas perfectly conducting 
boundary walls are set at $y=\pm L_y/2$. 
We follow the computations until $t=100$ using PPM reconstruction with a Roe Riemann
solver and a Courant number $C_a = 0.8$.
In Fig. \ref{fig:gem_evolution} we display the temporal evolution of the  
current density for a case where the uniform resistivity is set to 
$\eta=8\cdot10^{-3}$. A reconnection layer is created in the center of the 
domain, which predisposes resistive reconnection \citep{Biskamp86}. 
In agreement with \cite{RSW06} the maximum value of the current density 
decreases with time, (fig. 4 of that study). 
As seen in the first two panels ($t=0,\,50$), the refinement criterion is 
adequate to capture correctly both the boundary layer and the borders of 
the magnetic island structure. 
In the rightmost panel ($t=100$) we also draw sample magnetic field lines 
to better visualize the reconnection region. 

In order to compare our numerical results with theory, we repeated the 
computation varying the value of the magnetic resistivity $\eta$. 
For small values of resistivity (large Lundquist numbers $S$), the boundary 
layer is elongated and presents large aspect ratios $A\equiv L/\delta$. 
\cite{Biskamp86} reports that for $A$ beyond the critical value of 
$A_{\textrm{crit}}\simeq100$ the boundary layer is tearing unstable, 
limiting the resistivity range in which the Sweet-Parker reconnection 
model operates. 
Since in the Sun's corona $S$ can reach values of 
$\sim10^{14} >>S_{\max}$, secondary island formation must be taken 
into account \citep{CD09}.
In this context, ensuring that we respect Biskamp's stability criterion, we 
calculate the temporal average of $\delta / L$, analogous to the 
reconnection rate ${\cal E}$, for various $\eta$ and reproduce the 
Sweet-Parker scaling (Fig. \ref{fig:gem_rate}). 
The boundary layer's half -width ($\delta$) and -length ($L$) are estimated 
from the e-folding distance of the peak of the electric current, while the AMR scheme 
allows us to economically resolve the layer's thickness with enough grid 
points.


Parallel performance for this problem is shown in Fig. \ref{fig:mhd_GEMScaling},
where we plot the speedup $S=T_1/T_{N_{\rm CPU}}$ as a function of the number 
of processors $N_{\rm CPU}$ for the 3- and 5-level AMR computations 
as well as for fixed uniform grid runs carried out at the equivalent 
resolution of $2048\times1024$ zones.
Here $T_1$ is the same reference constant for all calculations and equal to 
the (inferred) running time of the single processor static mesh computation 
while $T_{N_{\rm CPU}}$ is the execution time measured with $N_{\rm CPU}$ 
processors.
The scaling reveals an efficiency (defined as $S/N_{\rm CPU}$) larger than
$0.8$ for less than $256$ processors with the 3- and 5-level computations being,
respectively, $8$ to $9$ and $4$ to $5$ times faster than the fixed grid approach.
The number of blocks on the finest level is maximum at the end of integration
and is slightly larger for the 3-level run (1058 vs. 835).
This result indicates that using fewer levels of refinement with larger 
grid ratios can be more efficient than introducing more levels with consecutive 
jumps of 2, most likely because of the reduced integration cost due to the 
missing intermediate levels and the decreased overhead associated with coarse-fine 
level communication and grid generation process.
Efficiency quickly drops when the number of CPU tends to become, within
a factor between $2$ and $3$, comparable to the number of blocks.

\subsection{Current Sheet}
%
%
\ifx\IncludeEps\yes
\begin{figure*}\centering
\includegraphics[width=0.9\textwidth]{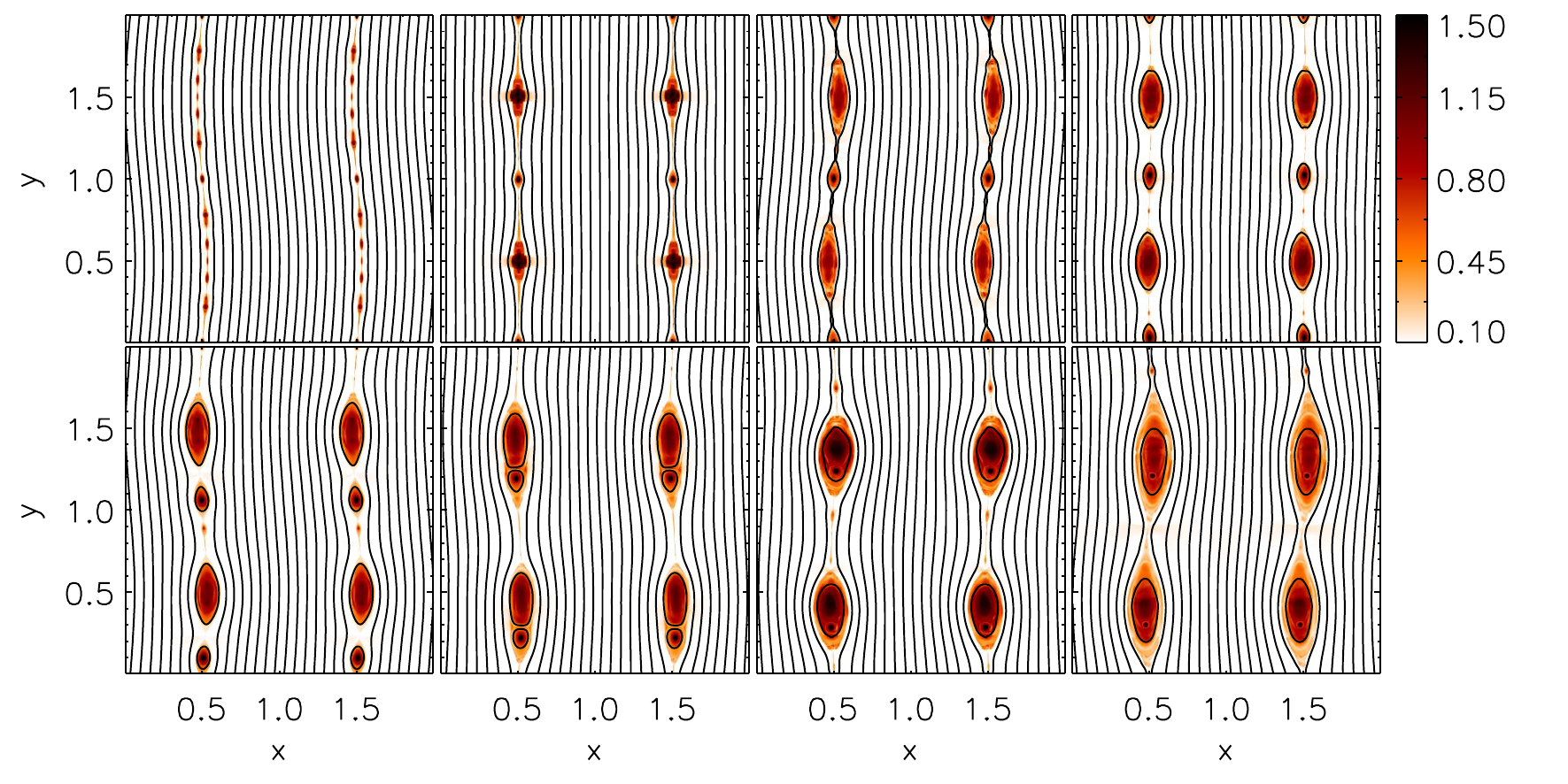}
 \caption{\footnotesize Time evolution of the pressure profiles along with sample magnetic field lines, 
  for the current sheet problem. Temporal snapshots refer to $t = 0.5,\,1,\,1.5,\,2$ (upper four) and
  $t = 2.5,\,3,\,3.5,\,4$ (lower four).}
 \label{fig:current_evolution}
\end{figure*}
\begin{figure}\centering
\includegraphics*[width=\columnwidth]{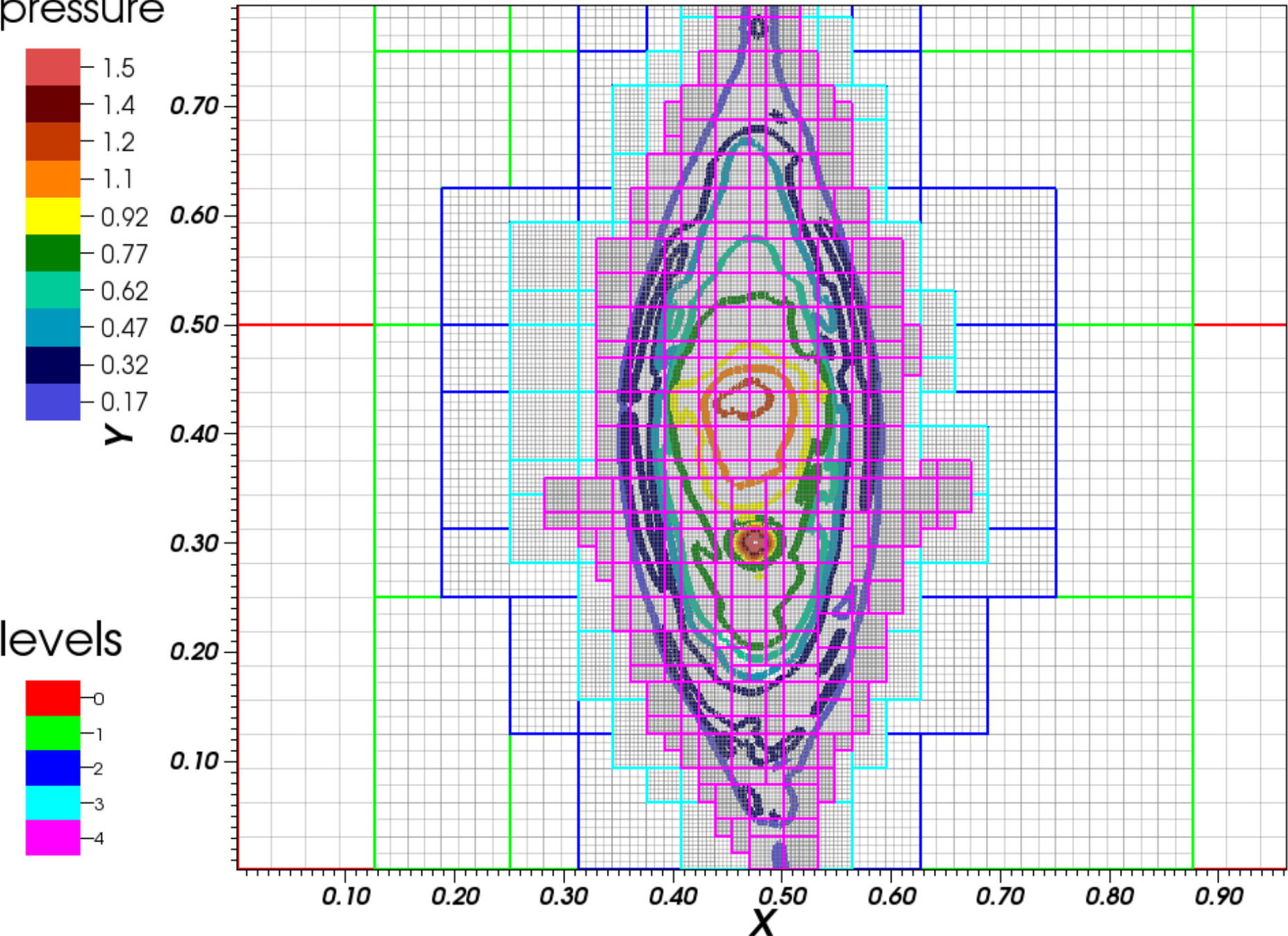}
 \caption{\footnotesize Closeup of the bottom left island at $t=4$ for the current sheet problem.
 Pressure contours, along with the refinement levels and grid are shown.}
 \label{fig:current_grid}
\end{figure}
\fi
The current sheet problem proposed by \cite{GS05} and later considered by 
\cite{FHT06} in the AMR context is particularly sensitive to numerical 
diffusion. 
The test problem follows the evolution of two current sheets, initialized 
through a discontinuous magnetic field configuration. 
Driven solely by numerical resistivity, reconnection processes take place, 
making the resulting solution highly susceptible to grid resolution.

The initial condition is discretized onto a Cartesian two dimensional 
grid $x,\,y\in[0,2]$, with $64\times64$ zones at the coarse level. 
The fluid has uniform density $\rho=1$ and thermal pressure $P=0.1$. 
Its bulk flow velocity ${\bf v}$ is set to zero, allowing only for a small 
perturbation in $v_x\,=\,v_0\,\sin(\pi\,y)$, where $v_0=0.1$. 
The initial magnetic field has only one non-vanishing component in the 
vertical direction, 
\begin{equation}  
B_y = \left\{ \begin{array}{ll}
        -B_0 & \textrm{if} \quad |x-1| \leq 0.5 \,, \\ \noalign{\medskip}
 \,\,\,\,B_0 & \textrm{otherwise}\,,
  \end{array} \right.
\end{equation}     
where $B_0=1$, resulting in a magnetically dominated configuration. 
Boundaries are periodic and the integration terminates at $t=4$.
We activate refinement whenever the maximum between the two error norms 
(given by Eq. \ref{eq:RefCrit}) computed with the specific internal energy 
and the $y$ component of the magnetic field exceeds the threshold value $\chi_r = 0.2$.
In order to filter out noise between magnetic island we set $\epsilon=0.05$.
We allow for $4$ levels of refinement and carry out integration with the Roe 
Riemann solver, the improved WENO reconstruction and a CFL number of $0.6$.

In Fig. \ref{fig:current_evolution} we show temporal snapshots of pressure 
profiles for $t=0.5,\,1.0,\,1.5,\,2,\,3.5$ and $4.0$, along with sample 
magnetic field lines. 
Since no resistivity has been specified, the elongated current sheets are 
prone to tearing instability as numerical resistivity is small with respect 
to Biskamp's criterion \citep{Biskamp86}. 
Secondary islands, ``plasmoids'', promptly form and propagate parallel to the 
field in the $y$ direction.
As reconnection occurs, the field is dissipated and its magnetic energy is 
transformed into thermal energy, driving Alfv\'en and compressional waves which 
further seed reconnection \citep{GS05, LD09}. 
Due to the dependency of numerical resistivity on the field topology, 
reconnection events are most probable at the nodal points of $v_x$. 
At the later stages of evolution, the plasmoids eventually merge in proximity 
of the anti-nodes of the transverse speed, forming four larger islands, in 
agreement with the results presented in \cite{GS05}. 
A closeup on the bottom left island is shown in Fig. \ref{fig:current_grid}, at 
the end of integration. The refinement criterion on the second derivative of 
thermal pressure, as suggested by \cite{FHT06}, efficiently 
captures the island features of the solution.

\subsection{Three-dimensional Rayleigh Taylor Instability}
%
%

\ifx\IncludeEps\yes
\begin{figure*}\centering
\includegraphics*[width=0.33\textwidth]{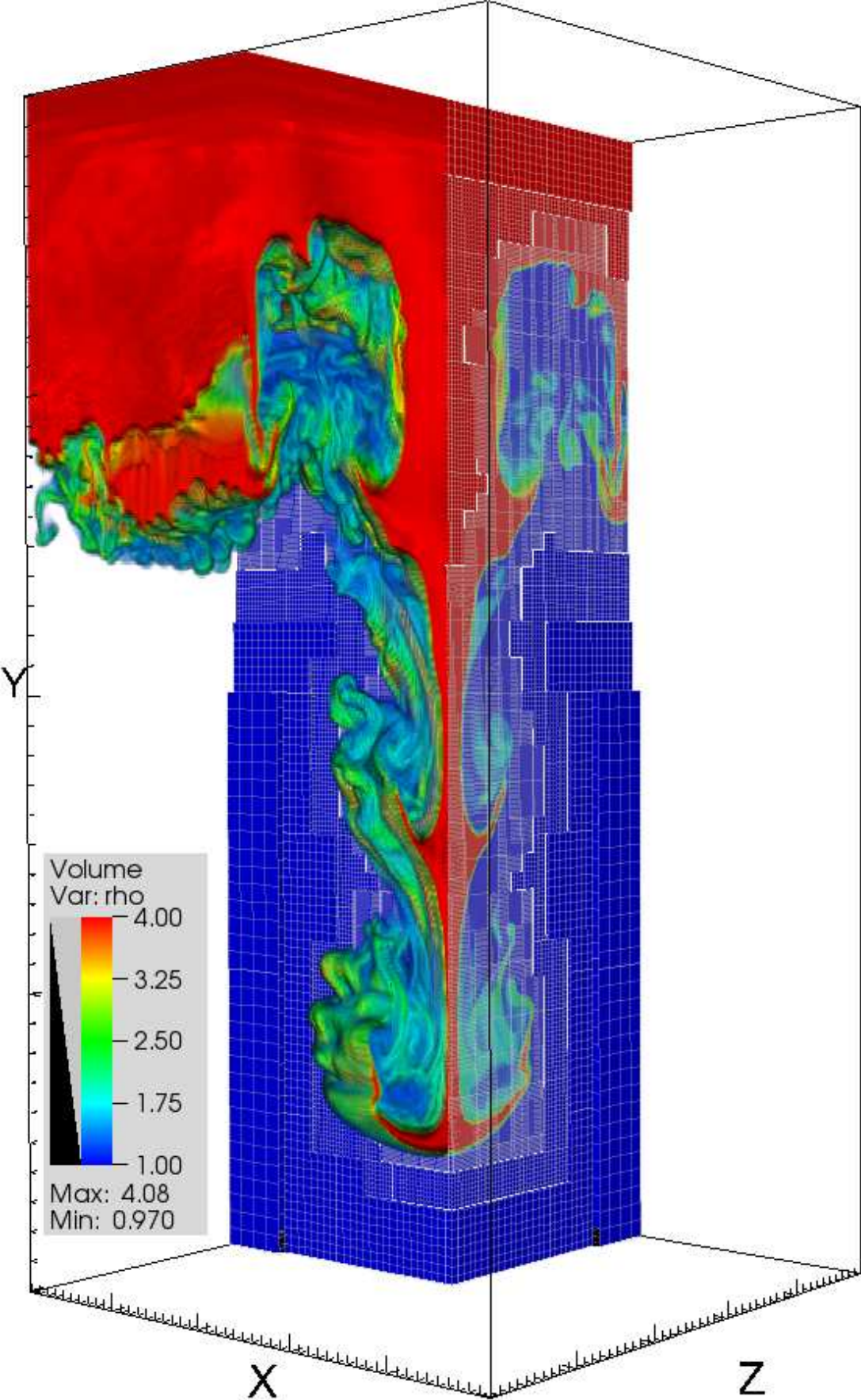}%
\includegraphics*[width=0.33\textwidth]{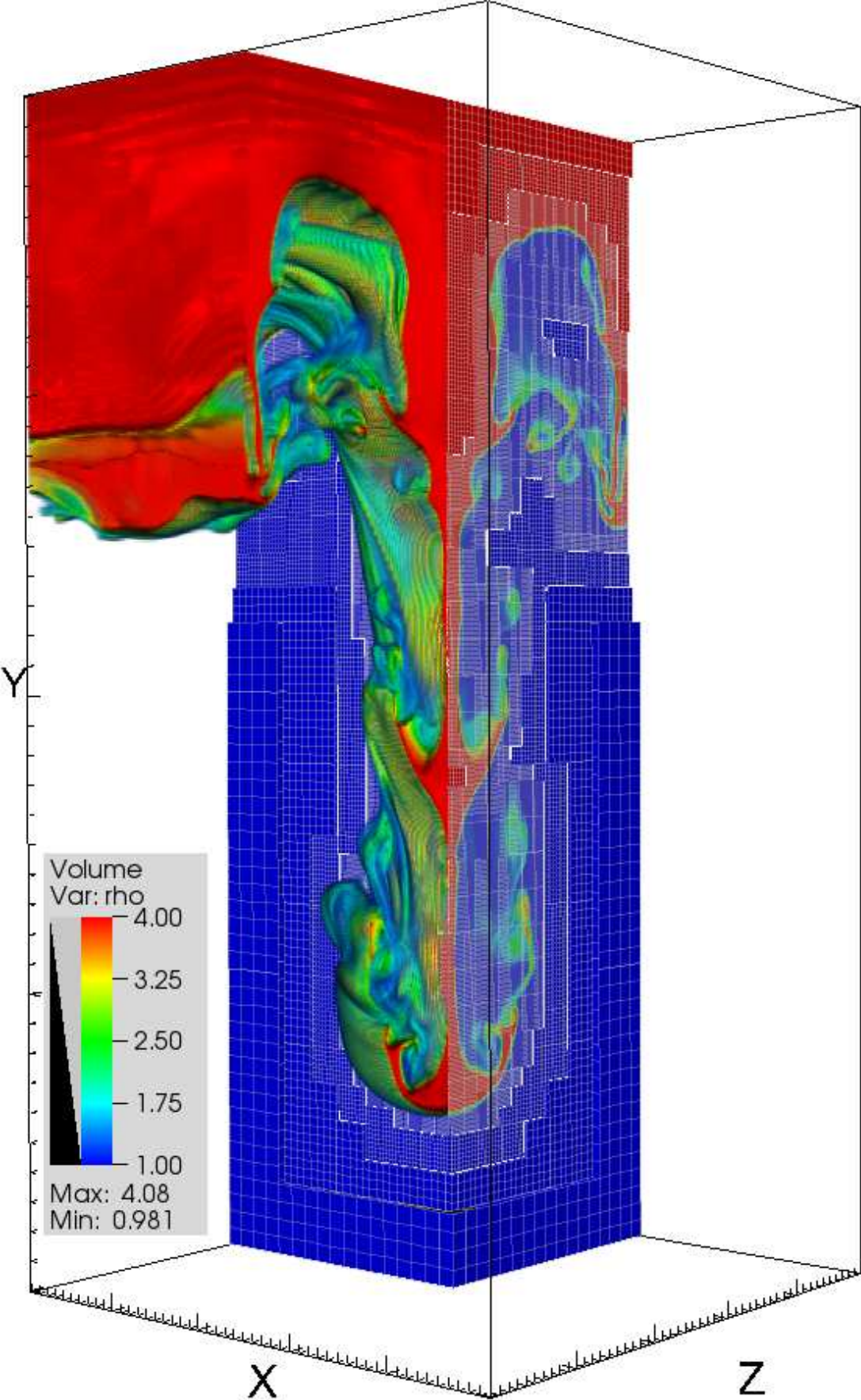}%
\includegraphics*[width=0.33\textwidth]{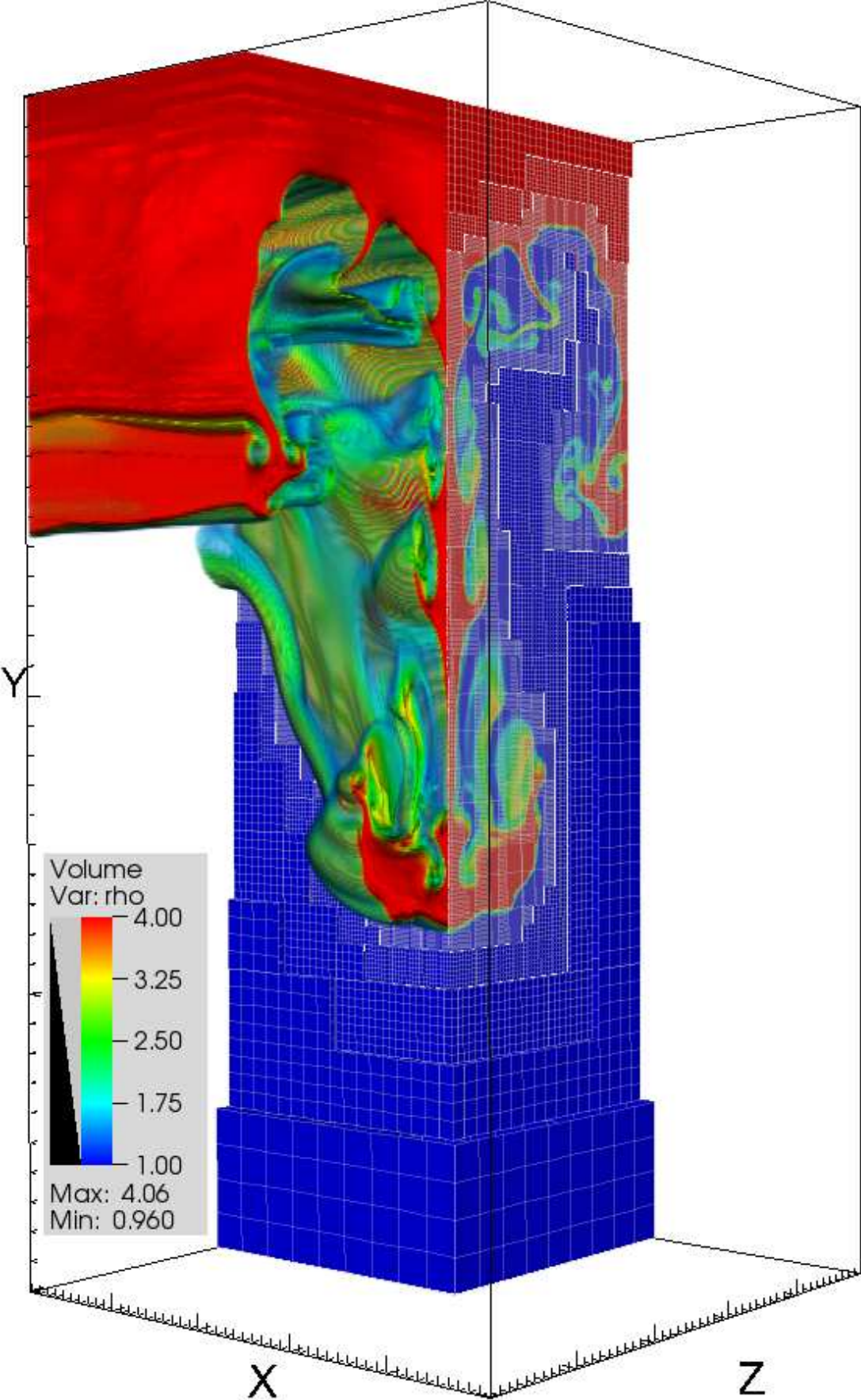}
\caption{\footnotesize Three-dimensional view of the Rayleigh-Taylor 
        instability problem showing density at $t=3$ for the un-magnetized 
        (left panel), weakly magnetized (middle panel) and strongly magnetized
        case (right panel) using $4$ levels of refinement.
        In each panel we also show a sliced boxed emphasizing the grid 
        refinement ratios.}
\label{fig:mhd_rt}
\end{figure*}
\begin{figure}
\centering
\includegraphics*[width=\columnwidth]{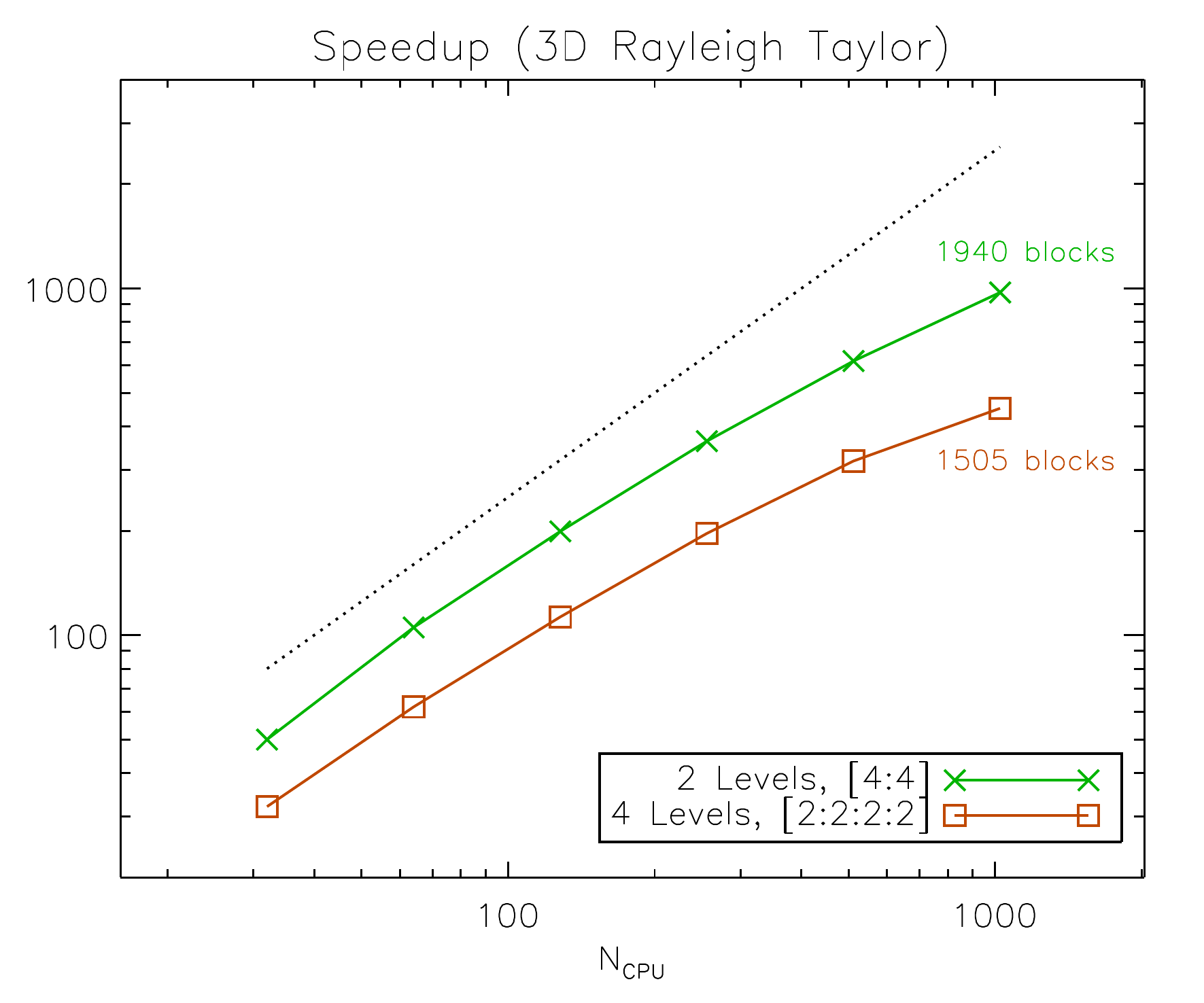}
\caption{\footnotesize Parallel speedup versus number of processors 
        $N_{\rm CPU}$ for the 3D Rayleigh-Taylor problem
        computed with 2 (green crosses) and 4 (red squares) refinement 
        levels between $t=2$ and $t=2.25$. The dotted line gives the ideal 
        scaling. The number of blocks on the finest level at $t=2$ is printed 
        above and below the corresponding curves.}
\label{fig:mhd_RTScaling}
\end{figure}
\fi
In this example we consider the dynamical evolution of two fluids of 
different densities initially in hydrostatic equilibrium.
If the lighter fluid is supporting the heavier fluid against gravity, the configuration is 
known to be subject to the Rayleigh-Taylor instability (RTI).
Our computational domain is the box spanned by $x,z\in[-1/2,1/2]$, 
$y\in[-3/2,1/2]$ with gravity pointing in the negative $y$ direction, i.e. 
$\vec{g}=(0,-1,0)$. 
The two fluids are separated by an interface initially lying in the $xz$ 
plane at $y=0$, with the heavy fluid $\rho=\rho_h$ at $y>0$ and 
$\rho=\rho_l$ at $y<0$.
In the current example we employ $\rho_h=4$, $\rho_l=1$ and specify the pressure from the 
hydrostatic equilibrium condition:
\begin{equation}
 p(y) = \frac{100}{\gamma} - \rho y \,,
\end{equation}
so that the sound crossing time in the light fluid is $0.1$.
The seed for instability is set by perturbing the velocity field 
at the center of the interface:
\begin{equation}
  v_y = - \frac{\exp\left[-(5r)^2\right]}{10\cosh(10y)^2}        
\end{equation}
where $r=\sqrt{x^2+z^2}$.
We assume a constant magnetic field $\vec{B}=(B_x,0,0)$ parallel to the 
interface and oriented in the $x$ direction with different field strengths,
$B_x=0$ (hydro limit), $B_x=0.2B_c$ (moderate field)
and $B_x = 0.6B_c$ (strong field). Here $B_c = \sqrt{(\rho_h - \rho_l)gL}$ 
is the critical field value above which instabilities are suppressed
\citep{SG07}.
We use the PPM method with the Roe Riemann solver and a Courant number
$C_a=0.45$. The base grid has $16\times32\times16$ cells and we perform
two sets of computations using i) $4$ refinement levels with a grid 
spacing ratio of $2$ and ii) $2$ refinement levels with a grid jump of $4$, in both 
cases achieving the same effective resolution ($256\times512\times256$).
Refinement is triggered using the second derivative error norm of 
density and a threshold value $\chi_{\rm ref} = 0.5$. 
Periodic boundary conditions are imposed at the $x$ and $z$ boundaries while 
fixed boundaries are set at $y=-3/2$ and $y=1/2$.

Results for the un-magnetized, weakly and strongly magnetized cases are shown 
at $t=3$ in Fig. \ref{fig:mhd_rt}.
In all cases, we observe the development of a central mushroom-shaped finger.
Secondary instabilities due to Kelvin-Helmholtz modes develop on the side and
small-scale structures are gradually suppressed in the direction of the field as
the strength is increased.
Indeed, as pointed out by \cite{SG07}, the presence of a uniform magnetic field
has the effects of reducing the growth rate of modes parallel to it although the
interface still remains Rayleigh-Taylor unstable in the perpendicular direction
due to interchange modes.
As a net effect, the evolution becomes increasingly anisotropic as the 
field strengthens.

Parallel performance is plotted, for the moderate field case, in Fig. 
\ref{fig:mhd_RTScaling} where we measure the speedup factors of the 4- and 
2- level computations versus the number of CPUs.
Here the speedup is defined by $S=T_1/T_{N_{\rm CPU}}$ where $T_1$ is the 
inferred running time relative to the 4-level computation on a single-processor.
The two-level case shows improved performance over the four-level 
calculation both in terms of CPU cost as well as parallel efficiency
$S/N_{\rm CPU}$.
Indeed the relative gain between the two cases approaches a factor of $2$ 
for an increasing number of CPUs. 
Likewise, the efficiency of the fewer level case remains above
$\gtrsim 0.8$ up to $512$ processors and drops to $\sim 0.61$ 
(versus $\sim 0.44$ for the 4-level case) at the largest number of 
employed cores ($1024$) which is less than the number of blocks on
the finest grid.

\subsection{Two-dimensional Shock-Cloud Interaction}
%
%
\ifx\IncludeEps\yes
\begin{figure*}\centering
\includegraphics*[width=0.9\textwidth]{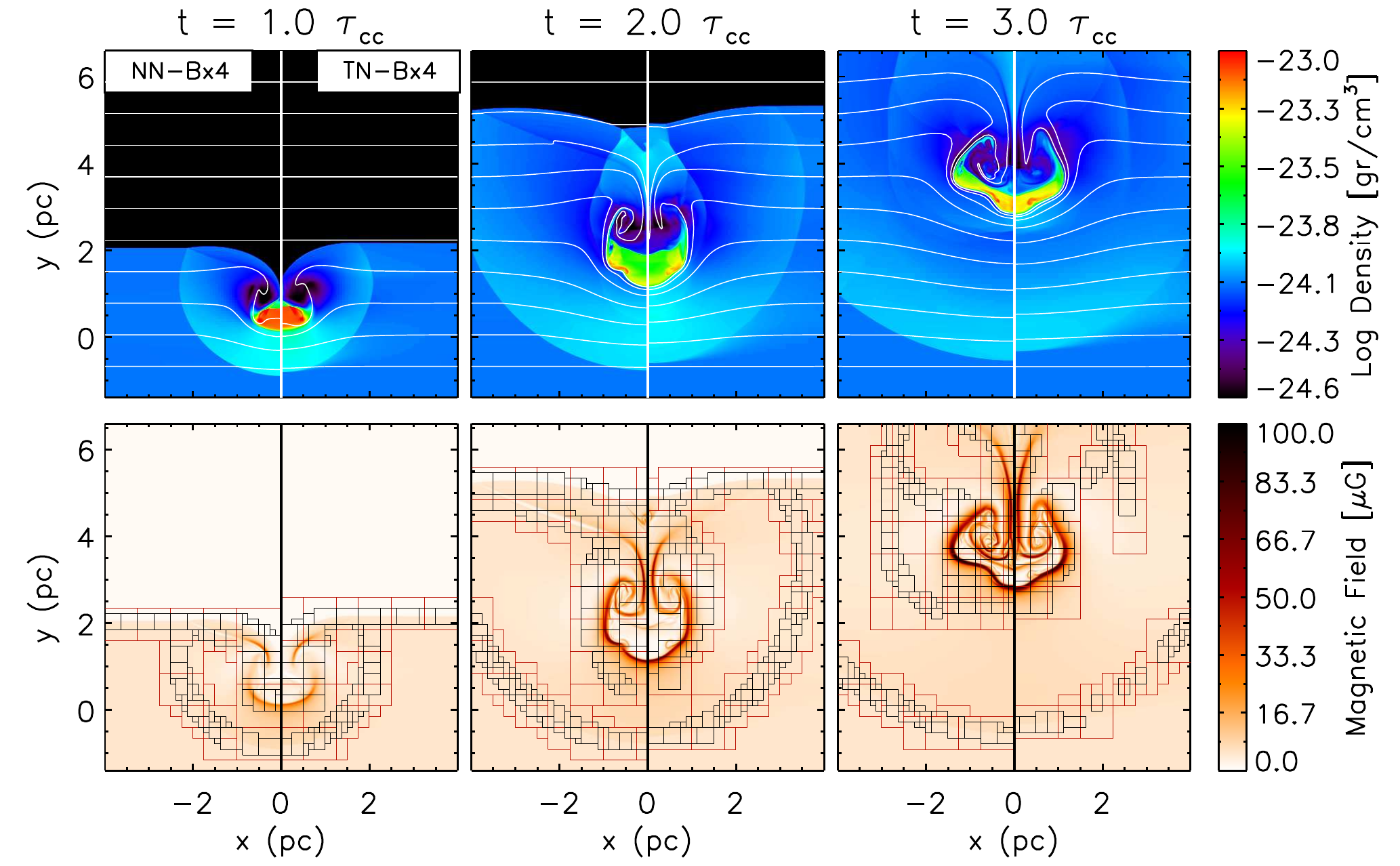}
\caption{\footnotesize Top panel: density maps (${\rm gr}/{\rm cm}^3$ in Log scale) 
         for the shock-cloud interaction at $t=1,2,3$ for a magnetic field 
         initially parallel to the shock interface (Bx4). Here the time unit
         is defined as the cloud crushing time and equal to $5,400\, {\rm yrs}$. 
         Bottom panel: magnetic field strength (in units of $10^{-6}$ G) with 
         the distribution of patches at levels $3$ and $4$ superimposed.
         The left and right half of each panel shows the results computed with the PPM
         method without (NN) and with (TN) thermal conduction, respectively.
         The resolution of the base grid is $32^2$ and $5$ levels of refinement are used.
}         
\label{fig:mhd_shockcloudBx4}
\end{figure*}
\begin{figure*}\centering
\includegraphics*[width=0.9\textwidth]{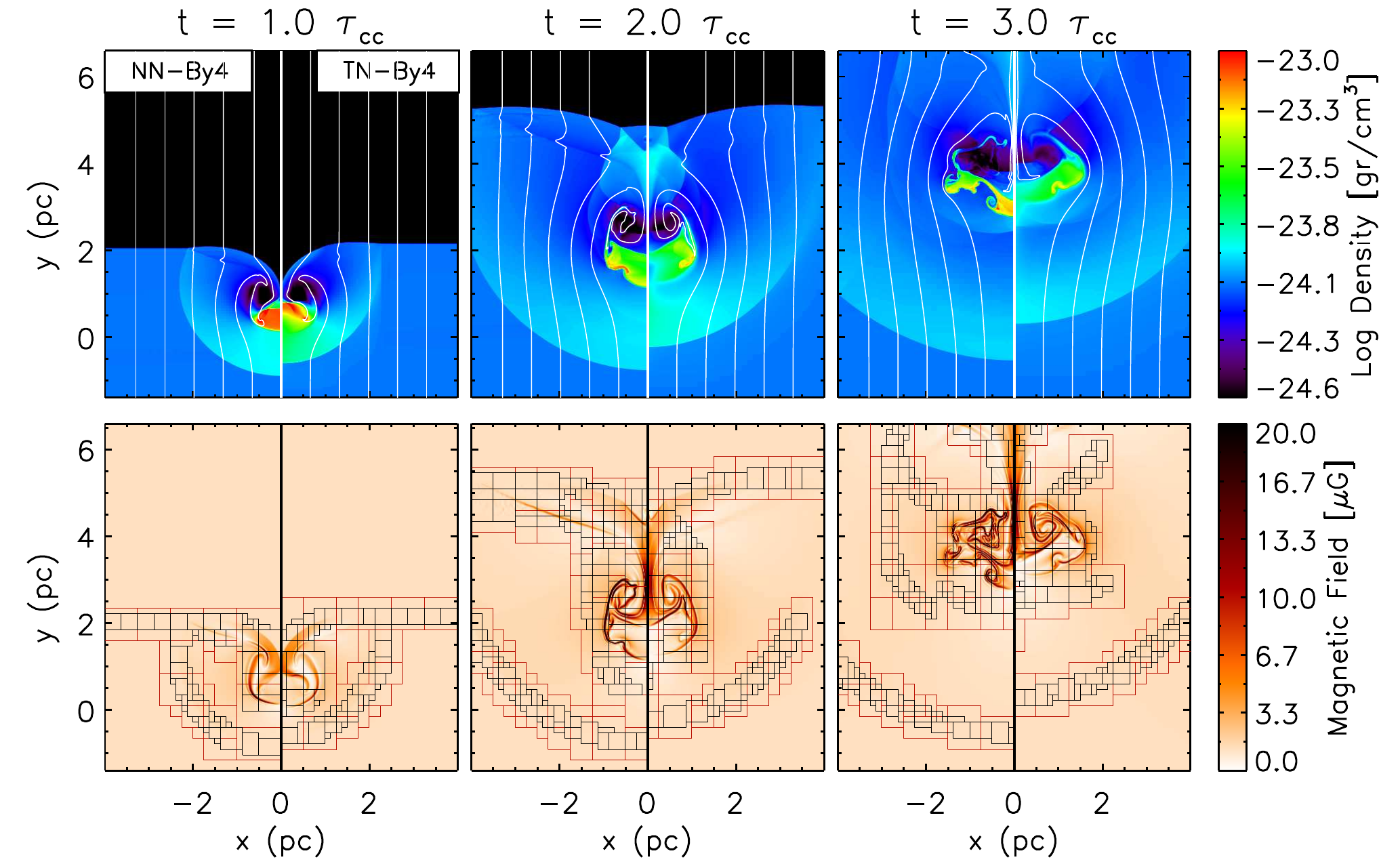}%
\caption{\footnotesize Same as Fig. \ref{fig:mhd_shockcloudBx4} but for a 
         field initially pointing in the $y$ direction (that is, perpendicular to 
         the shock front).}         
\label{fig:mhd_shockcloudBy4}
\end{figure*}
\fi
The shock-cloud interaction problem has been extensively studied and used as a standard 
benchmark for the validation of MHD schemes and inter-code comparisons (see 
\citealp{DW94, Balsara01, LD09} and reference therein).
In an astrophysical context, it adresses the fundamental issue of the complex morphology of 
supernova remnants as well as their interaction with the interstellar medium. 
Energy, mass and momentum exchange leading to cloud-crushing strongly depends on the orientation 
of the magnetic field and the resulting anisotropy of thermal conduction.

Following \cite{Orlando08}, we consider the two-dimensional Cartesian domain $x\in[-4,4]$, 
$y\in[-1.4\times6.6]$ with the shock front propagating in the positive $y$ direction and 
initially located at $y=-1$. Ahead of the shock, for $y>-1$, the hydrogen number density 
$n_{1}$ has a radial distribution of the form:
\begin{equation}
  n_1 = n_a - \frac{n_a - n_c}{\cosh\left[\sigma(r/r_{\rm cl})^\sigma\right]}
 \,,\quad
  r = \sqrt{x^2+y^2}
\end{equation}
where $n_c = 1\,{\rm cm}^{-3}$ is the hydrogen number density at the cloud's center, $n_a=0.1n_c$ is 
the ambient number density, $r_{\rm cl}=1\,{\rm pc}$ is the cloud's radius ($\sigma =10$) and $r$ 
is the radial distance from the center of the cloud.
The ambient medium has a uniform temperature $T_a=10^4\,{\rm K}$ in pressure equilibrium with the 
cloud, with a thermal pressure of $p_1 = 2k_Bn_a T_a$ (we assume a fully ionized gas).
Downstream of the shock, density and transverse components of the magnetic field are compressed by 
a factor of $n_2/n_1=(\Gamma+1)/(\Gamma - 1)$ while pressure and normal velocity are given by
\begin{equation}
  p_2     = 2n_2k_BT_s \,,\quad
  v_{y,2} = \sqrt{\frac{2}{\Gamma-1}\frac{2k_BT_s}{\mu m_H}} 
\end{equation}
where $\mu=1.26$ is the inverse of the hydrogen mass fraction and 
$T_s = 4.7\cdot 10^6\,{\rm K}$ is the post-shock temperature.
The normal component of the magnetic field ($B_y$) remains continuous 
through the shock front. 

We perform two sets of simulations with a magnetic field strength of
$|\vec{B}|=1.31 \;\mu{\rm G}$, initially parallel (case Bx4) or 
perpendicular (case By4) to the shock front.
Adopting the same notations as \cite{Orlando08}, we solve in each case
the MHD equations with (TN) and without (NN) thermal 
conduction effects for a total of 4 cases.
Thermal conductivity coefficients along and across the magnetic field lines 
(see Eq. \ref{eq:tc_flux}) are given, in c.g.s units, by
\begin{equation}
  k_\parallel = \DS 9.2\cdot10^{-7}T^{5/2} 
\,, \quad
  k_\perp     = \DS 5.4\cdot10^{-16} \frac{n_H^2}{\sqrt{T} \vec{B}^2} \,.
\end{equation}
The resolution of the base grid is set to $32^3$ points and $5$ levels of 
refinement are employed, yielding an equivalent resolution of $1024\times1024$. 
Upwind characteristic tracing (Section \ref{sec:np_char}) together with PPM 
reconstruction (Eq. \ref{eq:ppm1}) and the HLLD Riemann solver are chosen 
to advance the equations in time. Open boundary conditions are applied on
all sides, except at the lower $y$ boundary where we keep constant inflow 
values.
The CFL number is $C_a = 0.8$. Refinement is triggered upon density  
with a threshold $\chi_{\rm ref} = 0.15$.

In Fig. \ref{fig:mhd_shockcloudBx4} and \ref{fig:mhd_shockcloudBy4} we show 
the cloud evolution for the four different cases at three different times 
$t=1,2,3$ (in units of the cloud crushing time 
$\tau_{cc} = 5.4\cdot 10^3\,{\rm yrs}$).
Our results are in excellent agreement with those of \cite{Orlando08} 
confirming the efficiency of thermal conduction in suppressing the development
of hydrodynamic instabilities at the cloud borders. 
The topology of the magnetic field can be quite effective in reducing the 
efficiency of thermal conduction.
For a field parallel to the shock front (TN-Bx4), the cloud's expansion and 
evaporation are strongly limited by the confining effect of the enveloping
magnetic field.
In the case of a perpendicular field (TN-By4), on the contrary, thermal 
exchange between the cloud and its surroundings becomes more efficient in 
the upwind direction of the cloud and promotes the gradual heating of the core 
and the consequent evaporation in a few dynamical timescales.
In this case, heat conduction is strongly suppressed laterally by the 
presence of a predominantly vertical magnetic field.

\subsection{Radiative Shocks in Stellar Jets}
%
%
\ifx\IncludeEps\yes
\begin{figure*}[!ht]
 \begin{center}
  \begin{minipage}{0.38\textwidth}
   \includegraphics*[width=\textwidth]{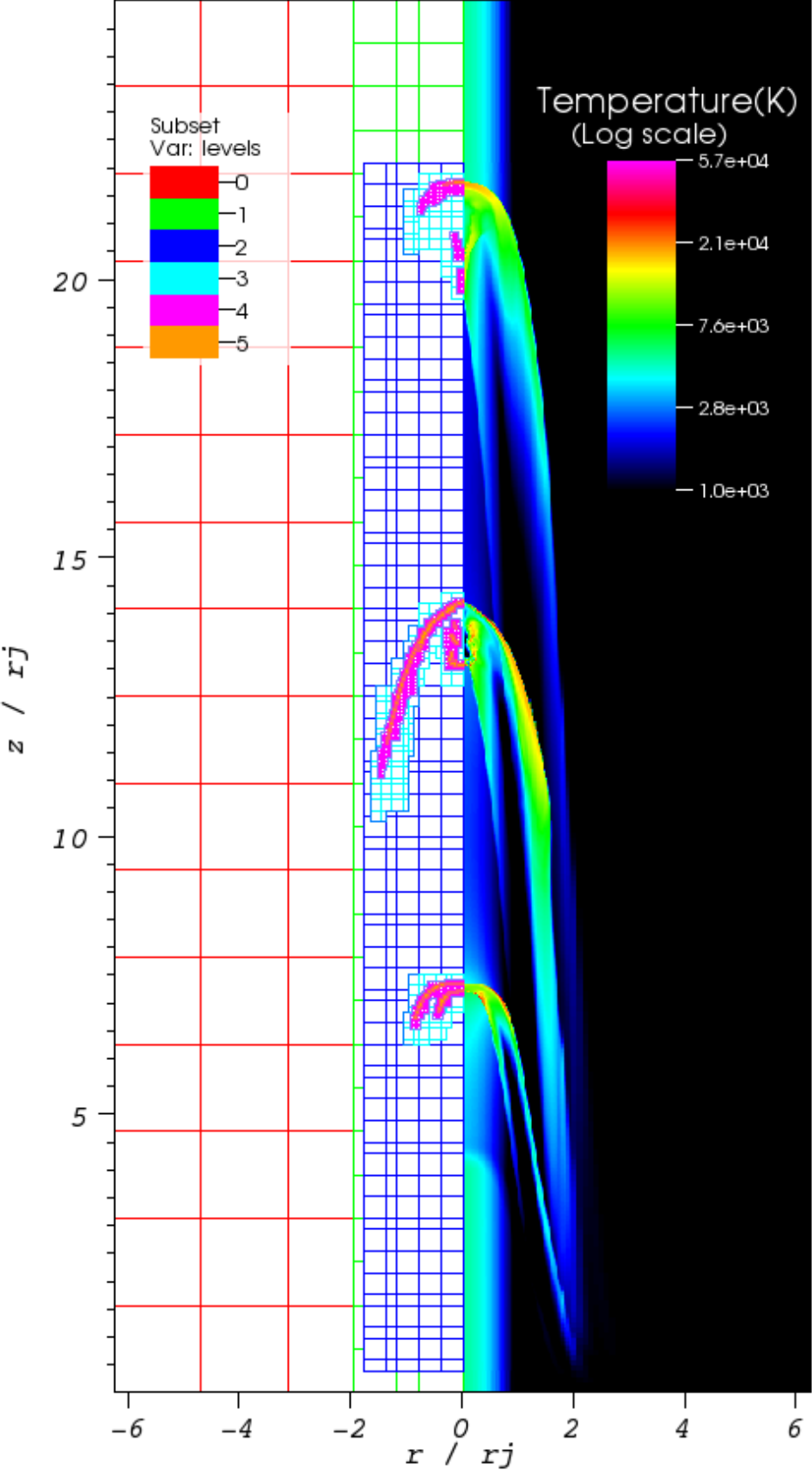} 
  \end{minipage}
  \begin{minipage}{0.58\textwidth}
   \begin{tabular}{ll}
    \includegraphics*[width=0.48\textwidth]{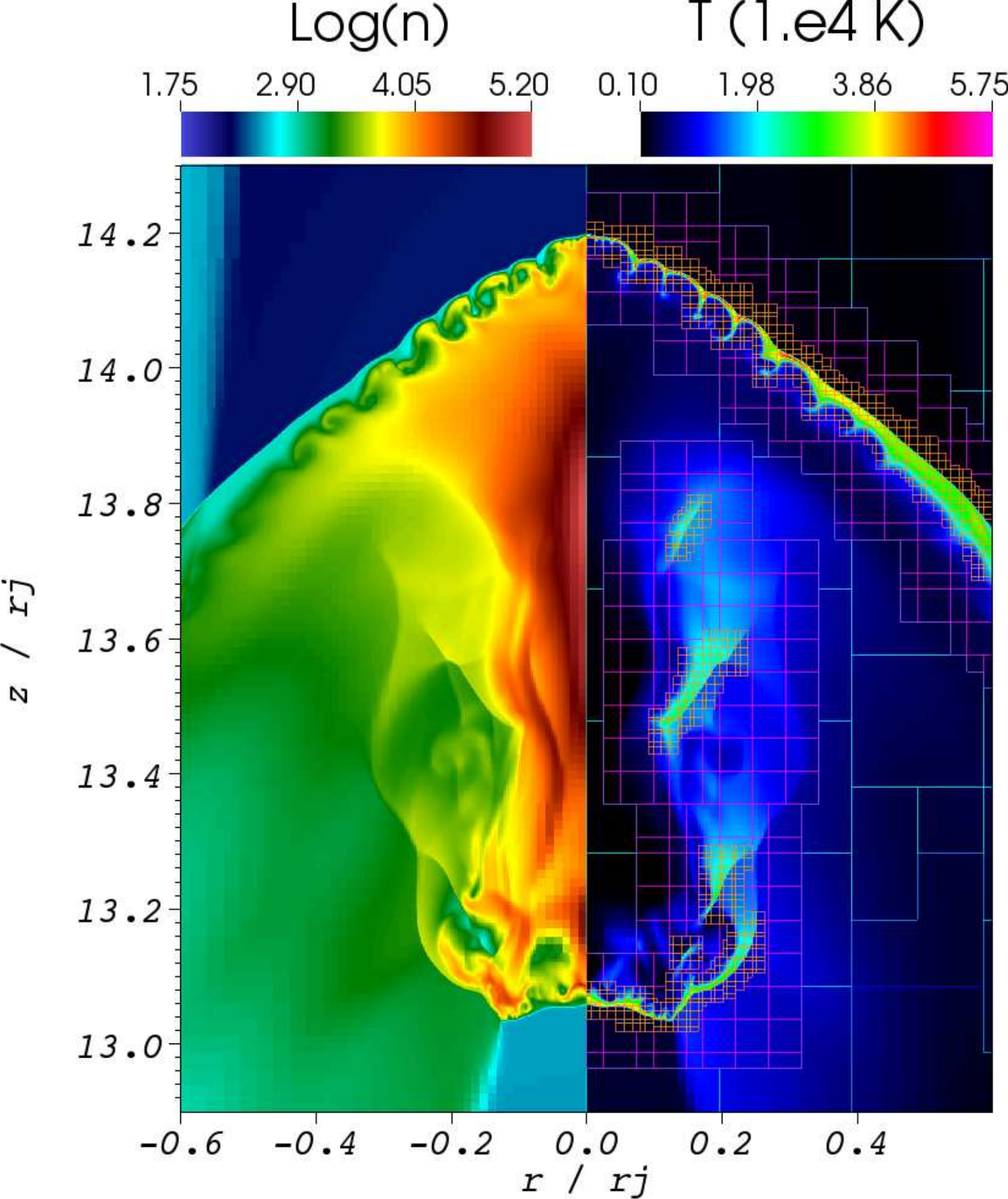} &
    \includegraphics*[width=0.48\textwidth]{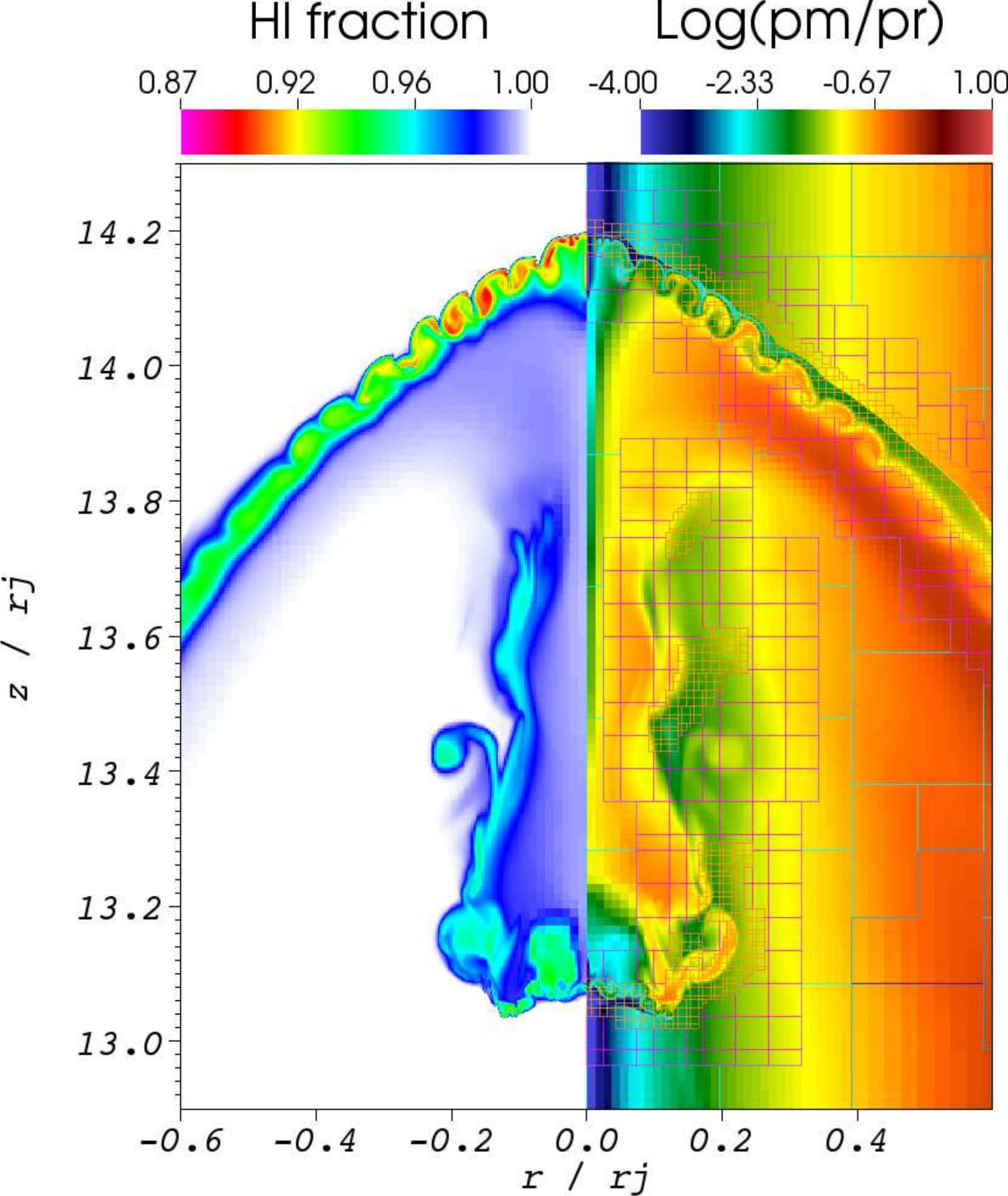}
\\ 
    \includegraphics*[width=0.48\textwidth]{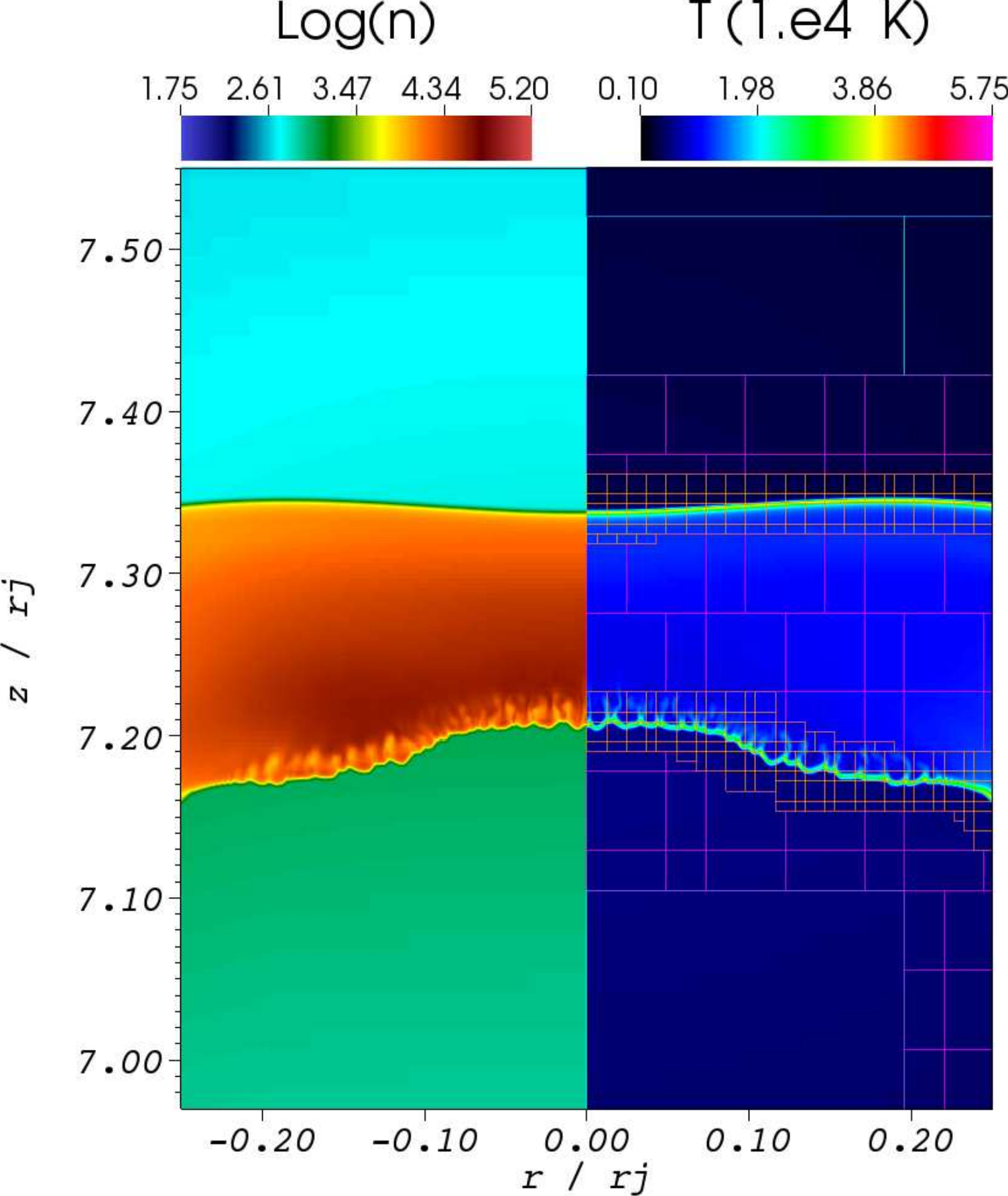} &
    \includegraphics*[width=0.48\textwidth]{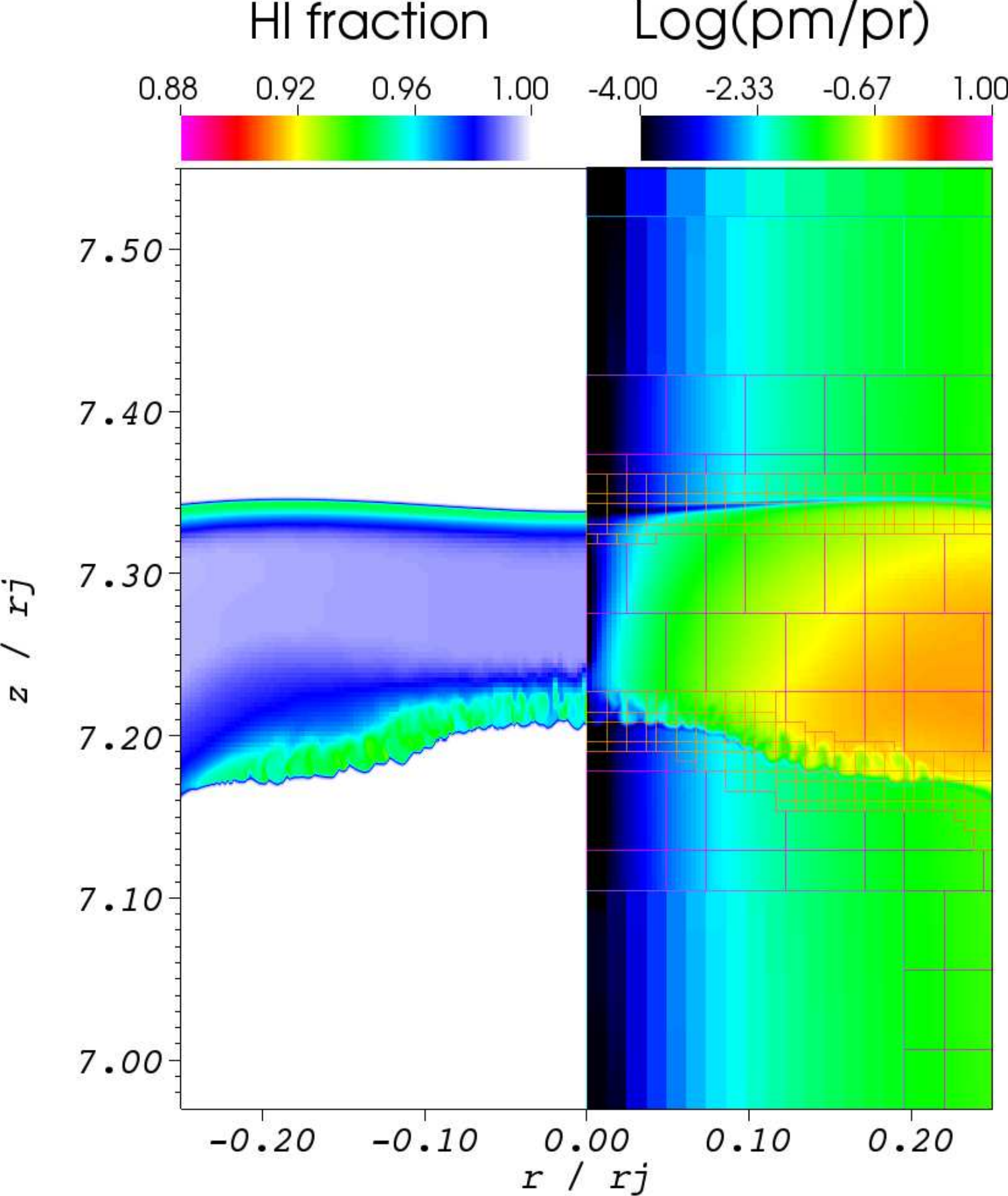}
   \end{tabular}
  \end{minipage} 
 \end{center}
\caption{\footnotesize Radiative jet structure at $t\approx 64.3\,{\rm yrs}$.
         On the left we display a full view of the temperature distribution 
         (right) and AMR block structure (left).
         The smaller panels to the right show closeup views of number density
	 (left half, in cm$^{-3}$), temperature (right half, in $10^4\,$ K), 
	 neutral hydrogen (left half) and magnetization $B_\phi^2/(2p)$ (right 
	 half) for the second (top panels) and third (bottom panels) shock, respectively.
         The AMR block structure is overplotted in the right half of each 
         panel.}         
\label{fig:mhd_radjet}
\end{figure*}
\begin{figure}\centering
\includegraphics*[width=\columnwidth]{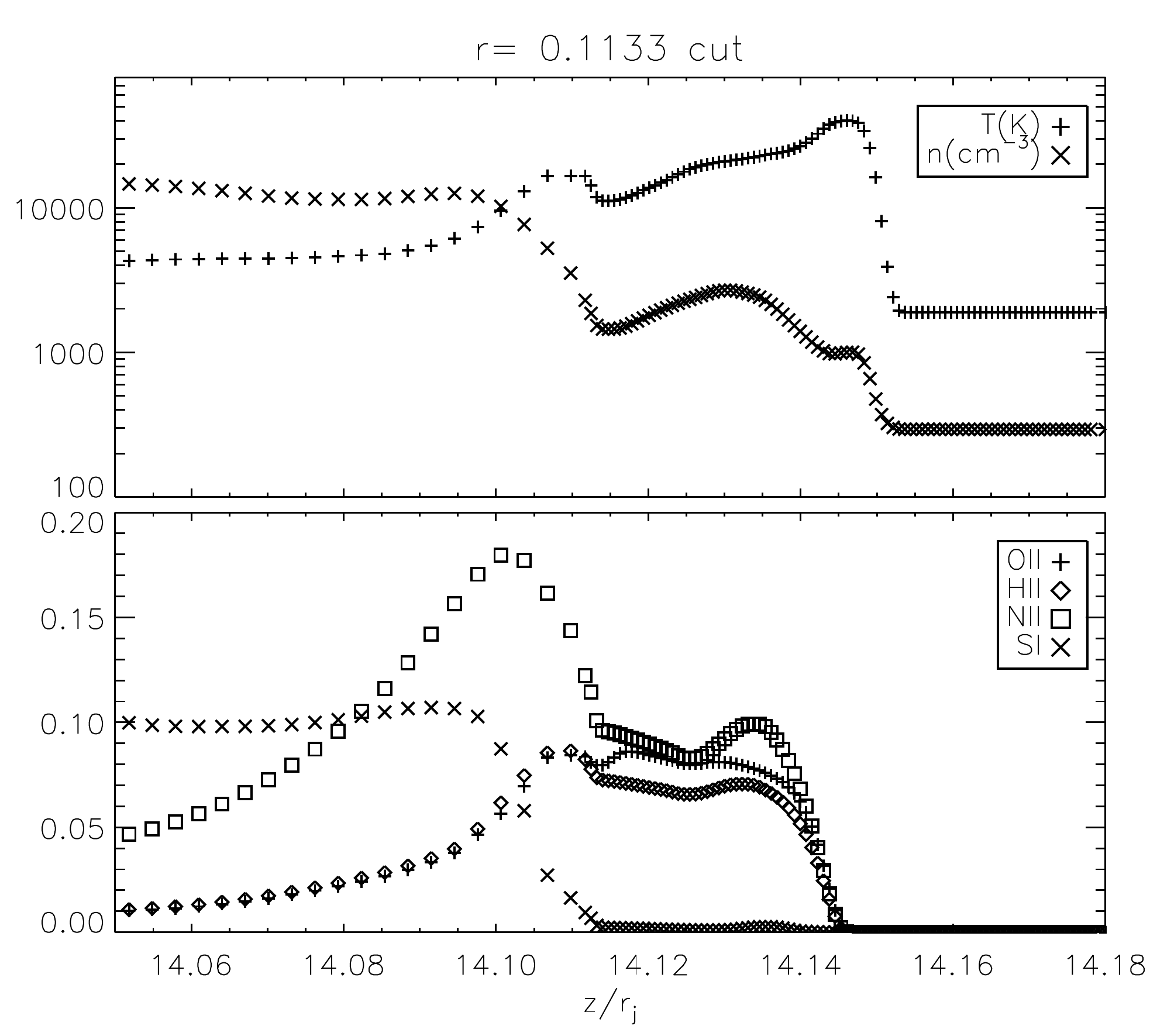}%
\caption{\footnotesize One dimensional enlarged cuts at $r\approx 0.1133r_j$ across 
         the middle shock in the radiative jet model at $t=64.3\,{\rm yrs}$.
         Temperature density distributions are plotted in the top 
         panel while the first ionization stage of O, H, N and neutral 
         Sulphur are plotted in the bottom panel.}         
\label{fig:mhd_radjetCuts}
\end{figure}
\fi
Radiative jet models with a variable ejection velocity are commonly adopted 
to reproduce the knotty structure observed in collimated outflows from 
young stellar objects.
The time variability leads to the formation of a chain of perturbations that 
eventually steepen into pairs of forward/reverse shocks 
\citep[the "internal working surfaces" see, for instance][]{dCR06,Raga07,Tes09}
which are held responsible for the typical spectral signature of these objects.
While these internal working surfaces travel down the jet, the emission 
from post-shocked regions occurs on a much smaller spatial and temporal 
scale thus posing a considerable challenge even for AMR codes.


As an example application, we solve the MHD equations coupled to the chemical
network described in \cite{Tes08}, evolving the time-dependent ionization and
collisionally-excited radiative losses of the most important atomic/ionic
species: H, He, C, N, O, Ne and S.
While hydrogen and helium can be at most singly-ionized, we only consider the
first three (instead of five) ionization stages of the other elements which
suffices for the range of temperatures and densities considered here. 
This amounts to a total of 19 additional non-homogeneous continuity/rate 
equations such as Eq. (\ref{eq:species}).
%
%

The initial condition consists of an axisymmetric supersonic collimated beam in
cylindrical coordinates $(r,z)$ in equilibrium with a stationary ambient medium.
Density and axial velocity can be freely prescribed as
\begin{equation}
 \rho(r) = \rho_a + \frac{\rho_j-\rho_a}{\cosh(r^4/r_j^4)}\,,\quad
 v_z(r)  = \frac{v_j}{\cosh(r^4/r_j^4)}\,,\quad
\end{equation}
where $\rho_j$ and $\rho_a=\rho_j/5$ are the jet and ambient mass 
density, $v_j = 200\,{\rm km/s}$ is the jet velocity and 
$r_j=2\cdot 10^{15}\,{\rm cm}$ is the radius.
The jet density is given by $\rho_j=n_j\mu_j m_a$ where $n_j=10^3$ is the total
particle number density, $\mu_j$ is the mean molecular weight and $m_a$ is the
atomic mass unit.
Both the initial jet cross section and the environment are neutral with the
exception of C and S which are singly ionized.
The steady state results from a radial balance between the Lorentz and 
pressure forces and has to satisfy the time-independent momentum 
equation:
\begin{equation}\label{eq:rad_equil}
  \frac{dp}{dr} = - \frac{1}{2}\frac{1}{r^2}\frac{d(rB_\phi)^2}{dr}\,
\end{equation}
where, for simplicity, we neglect rotations and assume a purely 
azimuthal magnetic field of the form 
\begin{equation}
B_\phi(r) = -\frac{B_m}{r}\sqrt{1 - \exp\left[-(r/a)^4\right]} \,.
\end{equation}
Here $a=0.9r_j$ is the magnetization radius and $B_m$ is a constant that 
depends on the jet and ambient temperatures.
Direct integration of Eq. (\ref{eq:rad_equil}) yields the equilibrium profile
\begin{equation}\label{eq:pequil}
  p(r) = p_j - \frac{1}{2}\frac{B_m^2\sqrt{\pi}
         \,\textrm{erf}(r^2/a^2)}{a^2}\,,
\end{equation}
where $p_j=\rho_jk_BT_j/(\mu_jm_a)$ is the jet pressure on the axis, $\mu_j$ is
the mean molecular weight, $m_a$ is the atomic mass unit, $k_B$ is the Boltzmann
constant and $T_j$ is the jet temperature. 
The value of $B_m$ is recovered by solving equation (\ref{eq:pequil}) 
in the $r\to\infty$ limit given the jet and ambient temperatures
$T_j=5\cdot 10^3\,{\rm K}$ and $T_a=10^3\,{\rm K}$, respectively.
Our choice of parameters is similar to the one adopted by 
\cite{Raga07}.

The computational domain is defined by $r/r_j\in[0,12.5]$, $z/r_j\in[0,25]$
with axisymmetric boundary conditions holding at $r=0$.
At $z=0$ we keep the flow variables constant and equal to the equilibrium
solution and add a sinusoidal time-variability of the jet velocity with a period
of $20\,{\rm yrs}$ and an amplitude of $40\,{\rm km/s}$ around the mean value. 
Free-outflow is assumed on the remaining sides.
We integrate the equations using linear reconstruction with the MC limiter, 
Eq. (\ref{eq:MClim}), and the HLLC Riemann solver.
The time step is computed from Eq. (\ref{eq:Ca}, \ref{eq:dt_cool}) 
using a Courant number $C_a=0.6$, a relative tolerance $\epsilon_c=0.02$
and following the considerations given in Section \ref{sec:cool_source}.
Shock-adaptive hybrid integration is used by locally switching to the more
diffusive MinMod limiter (Eq. \ref{eq:minmod}) and the HLL Riemann solver,
according to the mechanism outlined in Appendix \ref{app:MULTID_Flattening}.

The base grid consists of $128\times256$ zones and $5$ additional levels
with grid refinement jumps of $2:2:2:4:4$ are used.
At the effective resolution of $16384\times 32768$ zones the minimum length 
scale that can be resolved corresponds to $1.526\cdot10^{12} \,{\rm cm}$
($\approx 0.1$ AU), the same as model $M3$ of \cite{Raga07}.
Zones are tagged for refinement when the second derivative error norm of 
density exceeds the threshold value $\chi_{\rm ref} = 0.15$.
In order to enhance the resolution of strongly emitting shocked regions, we 
prevent levels higher than $2$ to be created if the temperature does not 
exceed $5\cdot10^3 \times \ell$ where $\ell$ is the level number.

The results are shown, after $\approx 63.4$ years, in Figure
\ref{fig:mhd_radjet} where the steepening of the perturbations leads to the 
formation of forward/reverse shock pairs.
Radiative losses become strongly enhanced behind the shock fronts where 
temperature attains larger values and the compression is large.
This is best illustrated in the closeup views of Fig
\ref{fig:mhd_radjet} where we display density, temperature, hydrogen
fraction and magnetization for the second (upper panel) and third shocks
(lower panel), respectively located at $z \approx 14.2$ and 
$z\approx 7.34$ (in units of the jet radius).
Owing to a much shorter cooling length, $t_c \sim p/\Lambda$, a thin 
radiative layer forms the size of which, depending on local temperature
and density values, becomes much smaller than the typical advection scale.
An extremely narrow one-dimensional cut shows, in Figure 
\ref{fig:mhd_radjetCuts}, the profiles of temperature, density 
and ionization fractions across the central radiative shock, resolved 
at the largest refinement level.
Immediately behind the shock wave, a flat transition region with
constant density and temperature is captured on $\sim 6$ grid points
and has a very small thickness $\sim 0.005 r_j$.
A radiatively cooled layer follows behind where temperature drops and 
the gas reaches the largest ionization degree.
Once the gas cools below $10^4$ K, radiative losses become negligible and
the gas is accumulated in a cold dense adiabatic layer.
A proper resolution of these thin layers is thus crucial for correct 
and accurate predictions of emission lines and intensity ratios in 
stellar jet models \citep{TMMB11}. 
Such a challenging computational problem can be tackled only by means of 
adaptive grid techniques.


\section{Relativistic MHD Tests}
\label{sec:RMHD_Test}
%
%
%
%

In this section we apply \PlCh to test problems involving relativistic
magnetized flows in one, two and three dimensions.
Both standard numerical benchmarks and applications will be considered.
By default, the conservative MUSCL-Hancock predictor scheme 
(Eq. \ref{eq:MUSCL-Hancock}) together with linear reconstruction on primitive 
variables are used during the computation of the normal predictors.

\subsection{One dimensional Shock-Tube}
%
%

\ifx\IncludeEps\yes
\begin{figure}[!t]\centering
\includegraphics*[width=\columnwidth]{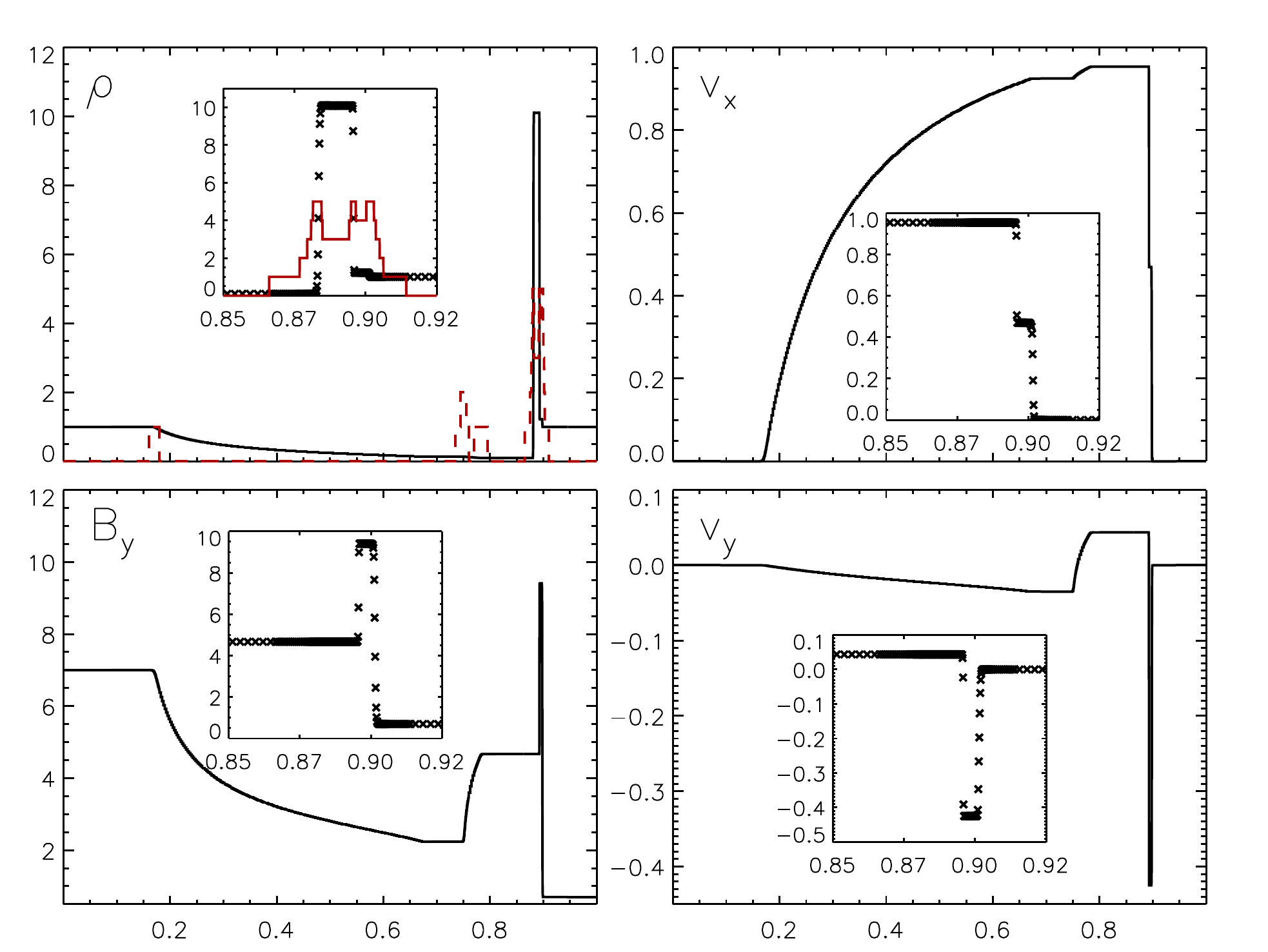}
\caption{\footnotesize 
         Density (top left), longitudinal velocity (top right),
         $y$ components of magnetic field and velocity (bottom left and
         right) for the one-dimensional relativistic magnetized 
         shock tube problem at t=0.4. We employ a base grid of $400$ 
         cells with $5$ levels of refinement with consecutive jump ratios
         of $4:2:2:2:2$ yielding an equivalent resolution of $25600$ zones.
         The grid hierarchy is shown in the top left panel (dashed red line)
         while magnifications of the thin shell are plotted, for each quantity,  
         using symbols in the interior plot windows.}         
\label{fig:rmhd_shocktube}
\end{figure}
\fi
Our first test consists of an initial discontinuity located at $x=0.5$ 
separating two regions of fluids characterized by 
\begin{equation}
  \left(B_y,\,  B_z, \, p\right)=\left\{\begin{array}{ll}
  (  7,\,   7,\, 10^3) & \quad {\rm for}\quad x < 0.5 \\ \noalign{\medskip}
  (0.7,\, 0.7,\, 0.1)  & \quad {\rm for}\quad x > 0.5  \,
\end{array}\right. 
\end{equation}
(see \citealp{Balsara01, MB06,HKM08} and references therein).
The fluid is initially at rest with uniform density $\rho=1$ and the 
longitudinal magnetic field is $B_x= 10$.

We solve the equations of relativistic MHD with the ideal gas law ($\Gamma=5/3$) 
using the 5-wave HLLD Riemann solver of \cite{MUB09} and $C_a=0.6$.
The computational domain $[0,1]$ is discretized using $5$ levels of refinement 
with consecutive grid jump ratios of $4:2:2:2:2$  starting from a base grid 
(level $0$) of $400$ grid zones (yielding an effective resolution of $25600$ 
zones). 
Refinement is triggered upon the variable $\sigma = (B_y^2+B_z^2)/D$ using
Eq. (\ref{eq:RefCrit}) with a threshold $\chi_r = 0.1$.
At $t=0.4$ the resulting wave pattern (Fig. \ref{fig:rmhd_shocktube}) is 
comprised of two left-going rarefaction fans (fast and slow) and two 
right-going slow and fast shocks. 
The presence of magnetic fields makes the problem particularly challenging 
since the contact wave, slow and fast shocks propagate very close to each other, 
resulting in a thin relativistic blast shell between $x\approx 0.88$ and 
$x\approx 0.9$.
\begin{deluxetable}{cccccc}
\tabletypesize{\footnotesize}
\tablecaption{CPU performance for the one-dimensional RMHD shock-tube.}
\tablewidth{0pt}
\tablehead{\multicolumn{2}{c}{Static Run} &  \multicolumn{3}{c}{AMR Run} &
           \colhead{Speedup} \\ 
           \cline{1-2} \cline{3-5} \\
           \colhead{$N_x$} & \colhead{Time (s)} & \colhead{Level} & 
           \colhead{Ref ratio} & \colhead{Time (s)} & \colhead{}}
\startdata
1600      & 6.1    &  1  & 4 & 1.8  &  3.4\\
3200      & 29.9   &  2  & 2 & 3.8  &  7.9\\
6400      & 124.5  &  3  & 2 & 8.2  &  15.2\\
12800     & 502.3  &  4  & 2 & 18.1 &  27.8\\ 
25600     & 2076.1 &  5  & 2 & 41.2 &  50.4
\enddata
\tablecomments{The first and second columns give the number of points 
              $N_x$ and corresponding CPU for the static grid run (no AMR).
              The third, fourth and fifth columns give, respectively, the number  
              of levels, the refinement ratio and CPU time for the AMR run at 
              the equivalent resolution.
              The last column shows the corresponding speedup factor.}
\label{tab:rmhd_shocktube_timing}
\end{deluxetable}

In Table \ref{tab:rmhd_shocktube_timing} we compare, for different resolutions,
the CPU timing obtained with the static grid version of the code versus 
the AMR implementation at the same effective resolution, 
starting from a base grid of $400$ zones.
In the unigrid computations, halving the mesh size implies approximately a 
factor of four in the total running time whereas only a factor of two in the AMR
approach. 
At the large resolution employed here, the overall gain is approximately $50$.
This example confirms that the resolution of complex wave patterns in RMHD can 
largely benefit from the use of adaptively refined grids.

\subsection{Inclined Generic Alfv\'en Test}
%
%
%

%

\ifx\IncludeEps\yes
\begin{figure*}[!ht]\centering
\includegraphics[width=0.8\textwidth]{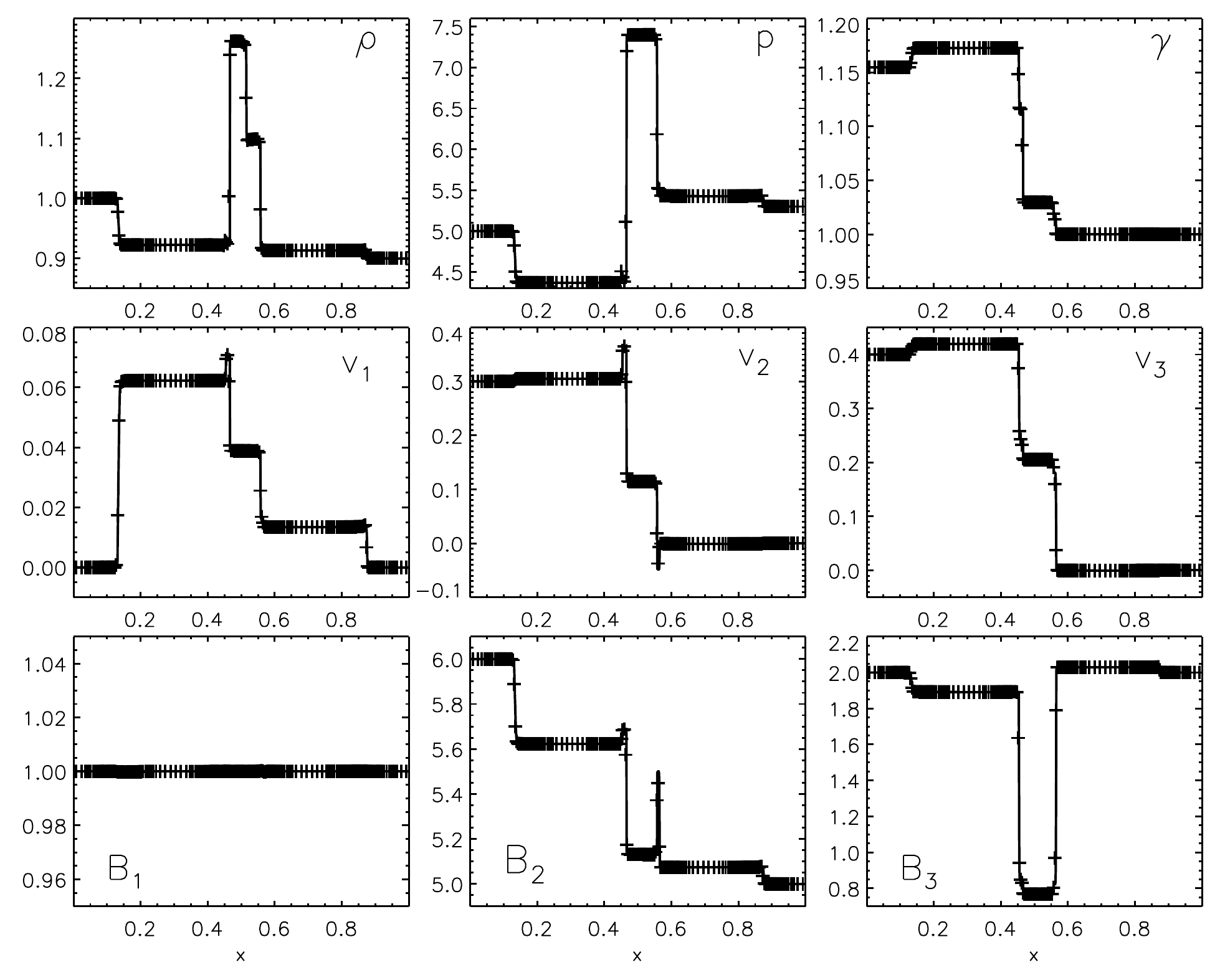}
\caption{\footnotesize Horizontal cuts for the rotated inclined Alfv\'en test
         at $t=0.4/\sqrt{2}$ (symbols) and for the 1-D reference solution at 
         $t=0.4$ (solid line). The top row shows, from left to right, 
         proper density, gas pressure and Lorentz factor. 
         In the middle and bottom rows we plot the components of velocity and 
         magnetic field normal ($v_1$ and $B_1$) and transverse ($v_2,v_3$ 
         and $B_2, B_3$) to the surface of discontinuity.}         
\label{fig:rmhd_inclinedTube}
\end{figure*}
\begin{figure}[!ht]\centering
\includegraphics*[width=\columnwidth]{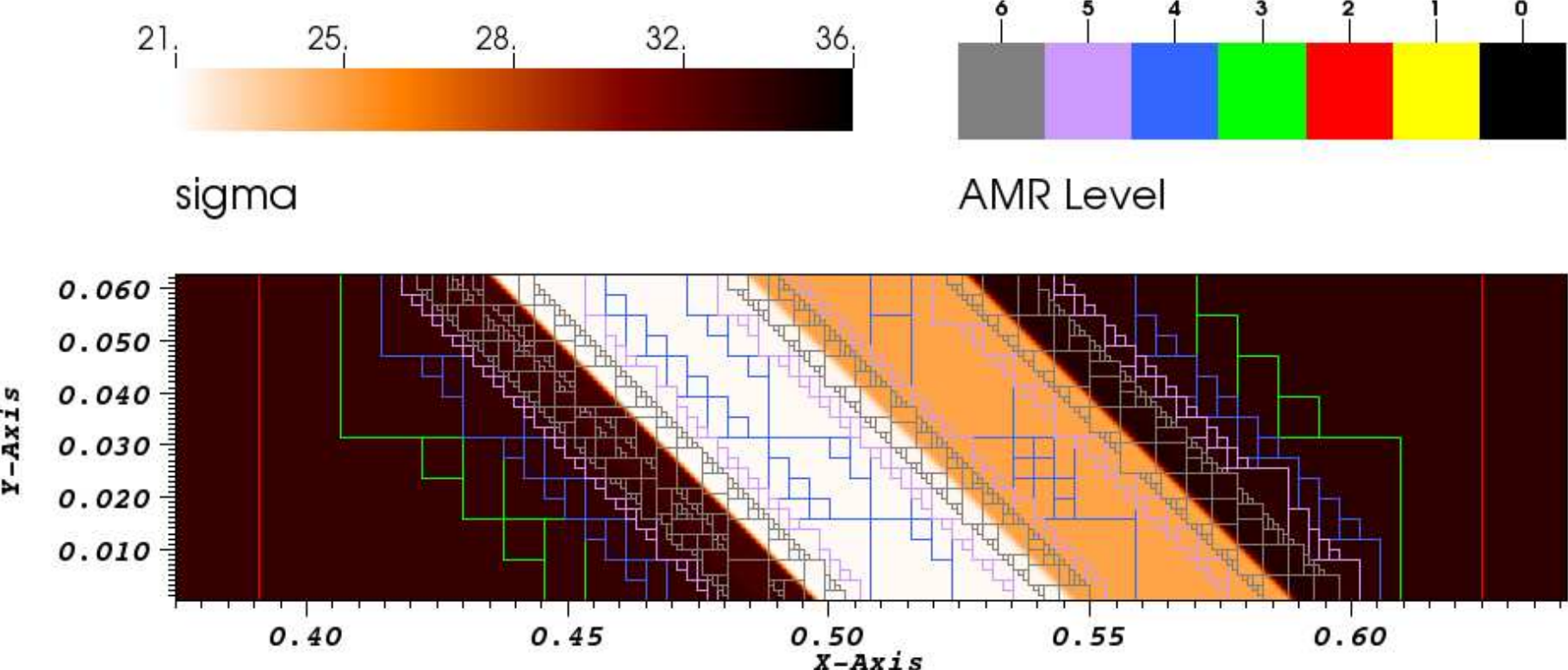}         
\caption{\footnotesize A closeup of the central region in the 
         inclined Alfv\'en test at $t=0.4/\sqrt{2}$ showing 
         $\sigma=\vec{B}^2/D$ with the overplotted block distribution.
         The base grid has $64\times 4$ zones and $5$ levels of refinement
         are used.}         
\label{fig:rmhd_inclinedTubeLevels}
\end{figure}
\fi
The generic Alfv\'en test \citep{GR06,MUB09} consists of the following 
non planar initial discontinuity
\begin{equation}\label{eq:genericAlfven}
\left\{
\begin{array}{lclr}
\vec{V}_L & = & \DS \left(1, 0,0.3,0.4, 1, 6, 2, 5\right)^T 
              & \mathrm{for}\quad  x_1 < 1/2 \,, \\ \noalign{\medskip}
\vec{V}_R & = & \DS \left(0.9, 0,0,0, 1, 5, 2, 5.3\right)^T 
              & \mathrm{for}\quad  x_1 > 1/2  \,,
\end{array}\right.
\end{equation}
where, as in Section \ref{sec:Compound}, 
$\vec{V}=(\rho, v_1, v_2, v_3, B_1, B_2, B_3, p)$ is given in 
the frame of reference aligned with the direction of motion $x_1$.
The ideal equation of state (\ref{eq:gammaEos}) with $\Gamma = 5/3$ is adopted.

Here we consider a two-dimensional version by rotating the discontinuity 
front by $\pi/4$ with respect to the mesh and following the evolution 
until $t=0.4\cos\pi/4$.
The test is run on a coarse grid of $64\times 4$ zones covering the domain 
$x\in[0,1]$, $y\in[-1/32, 1/32]$ with 6 additional levels of refinement
resulting in an effective resolution of $4096\times 256$ zones
(a factor of 2 in resolution is used between levels).
In order to trigger refinement across jumps of different nature, we 
define the quantity $\sigma=\vec{B}^2/D$ together with Eq. (\ref{eq:RefCrit}) 
and a threshold value $\chi_r = 0.03$.
The Courant number is $C_a = 0.6$ and the HLLD Riemann solver is used
throughout the computation.
Boundary conditions assume zero-gradients at the $x$ boundaries
while translational invariance is imposed at the top and bottom 
boundaries where, for any flow quantity $q$, we set $q(i,j) = q(i\pm1,j\mp1)$. 

The breaking of the discontinuity, shown in Fig. \ref{fig:rmhd_inclinedTube},
leads to the formation of seven waves including a left-going fast rarefaction,
a left-going Alfv\'en discontinuity, a left-going slow shock, a tangential
discontinuity, a right-going slow shock, a right-going Alfv\'en discontinuity 
and a right-going fast shock.
Our results indicate that refined regions are created across discontinuous 
fronts which are correctly followed and resolved, in excellent agreement 
with the reference solution (obtained on a fine one-dimensional grid).
Note also that the longitudinal component of the magnetic field, $B_1$, does not
present spurious jumps (Fig. \ref{fig:rmhd_inclinedTube}) and shows very small 
departures from the expected constant value.
In the central region, for $0.4\lesssim x\lesssim 0.6$, rotational 
discontinuities and slow shocks are moving slowly and very adjacent to each 
other thus demanding relatively high resolution to be correctly captured.
With the given prescription, grid adaptation efficiently provides adequate 
resolution where needed. This is best shown in the closeup of the central 
region, Fig. \ref{fig:rmhd_inclinedTubeLevels}, showing $\sigma$ together with 
the AMR block structure.

A comparison of the performance between static and adaptive grid computations
is reported in Table \ref{tab:rmhd_inclinedtube_timing} using different 
resolutions and mesh refinements. At the resolution employed here, the 
gain is approximately a factor of $9$.

\begin{deluxetable}{cccccccccc}
\tabletypesize{\footnotesize}
\tablecaption{CPU performance for the two-dimensional inclined generic Alfv\'en test.}
\tablewidth{0pt}
\tablehead{\multicolumn{2}{c}{Static Run} &  & \multicolumn{4}{c}{AMR Run} &
           \colhead{Speedup} \\ 
           \cline{1-2} \cline{4-7} \\
           \colhead{$N_x$} & \colhead{Time (s)} &    &
           \colhead{Level} & \colhead{Ref ratio} & \colhead{\#Blocks} &
           \colhead{Time (s)} & \colhead{}}
\startdata
128      &   1.8   &  & 1  & 2 &   4  & 1.1    & $1.6$\\
256      &   10.8  &  & 2  & 2 &  14  & 6.1    & $1.8$\\
512      &   75.2  &  & 3  & 2 &  44  & 33.0   & $2.3$\\
1024     &  557.8  &  & 4  & 2 & 126  & 185.9  & $3.0$\\ 
2048     & 4390.5  &  & 5  & 2 & 306  & 900.0  & $4.9$\\
4096     & 35250.0 &  & 6  & 2 & 761  & 4346.9 & $8.1$
\enddata
\tablecomments{The base grid for the AMR computation is $64\times 4$ zones.}
\label{tab:rmhd_inclinedtube_timing}
\end{deluxetable}
 
\subsection{Relativistic Rotor Problem}
%
%
%
%
\ifx\IncludeEps\yes
\begin{figure*}\centering
\includegraphics[width=0.9\textwidth]{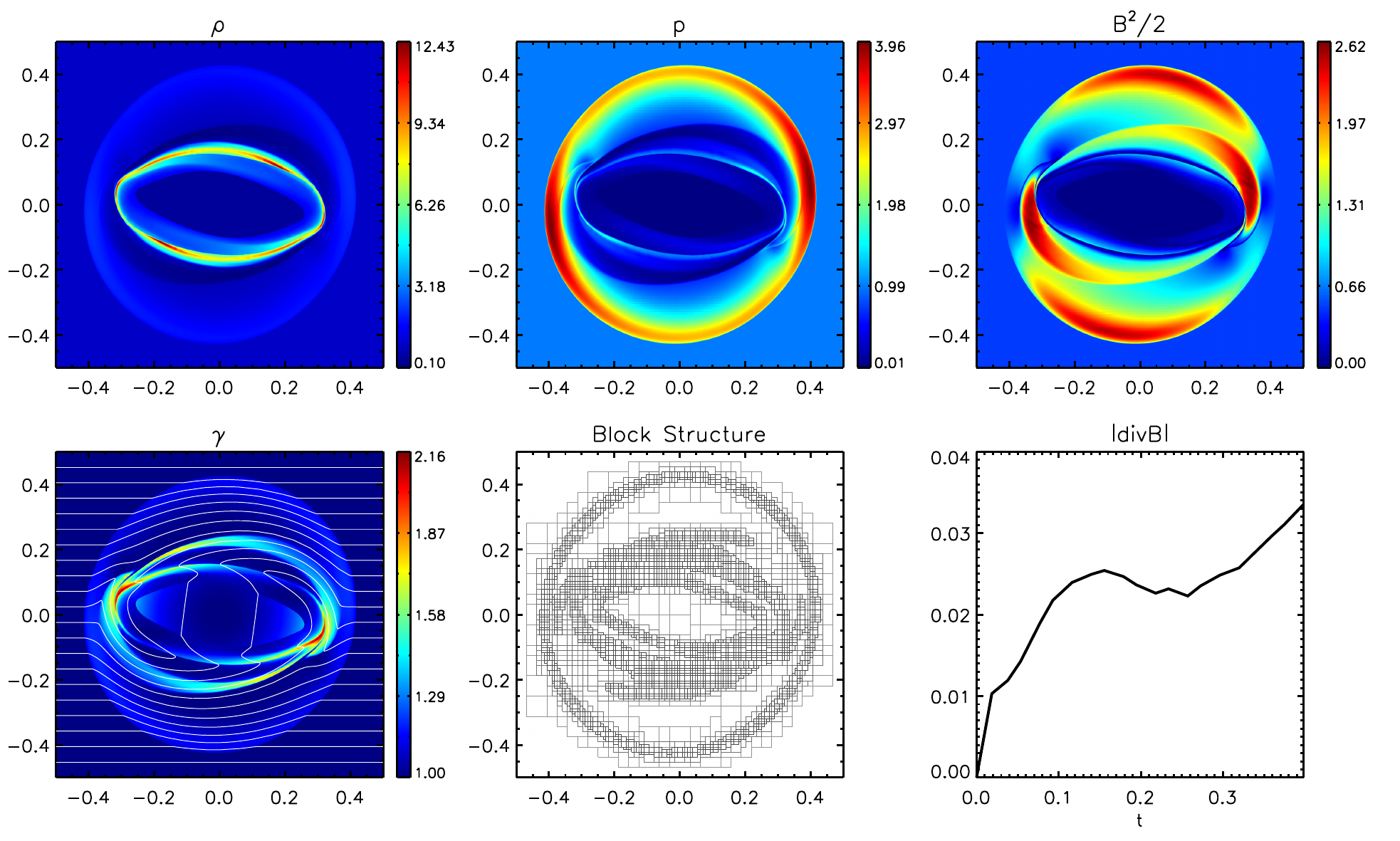}
\caption{\footnotesize The relativistic rotor problem at $t=0.4$ using 
         $6$ levels of refinement on a base grid with $64^2$ zones. 
         From left to right, top to bottom: density, gas pressure, 
         magnetic field, density, Lorentz factor (magnetic field lines 
         are overplotted), block distribution and domain-average of 
         $|\nabla\cdot\vec{B}|$.}         
\label{fig:rmhd_rotor}
\end{figure*}
\begin{figure}[!t]\centering
\includegraphics*[width=\columnwidth]{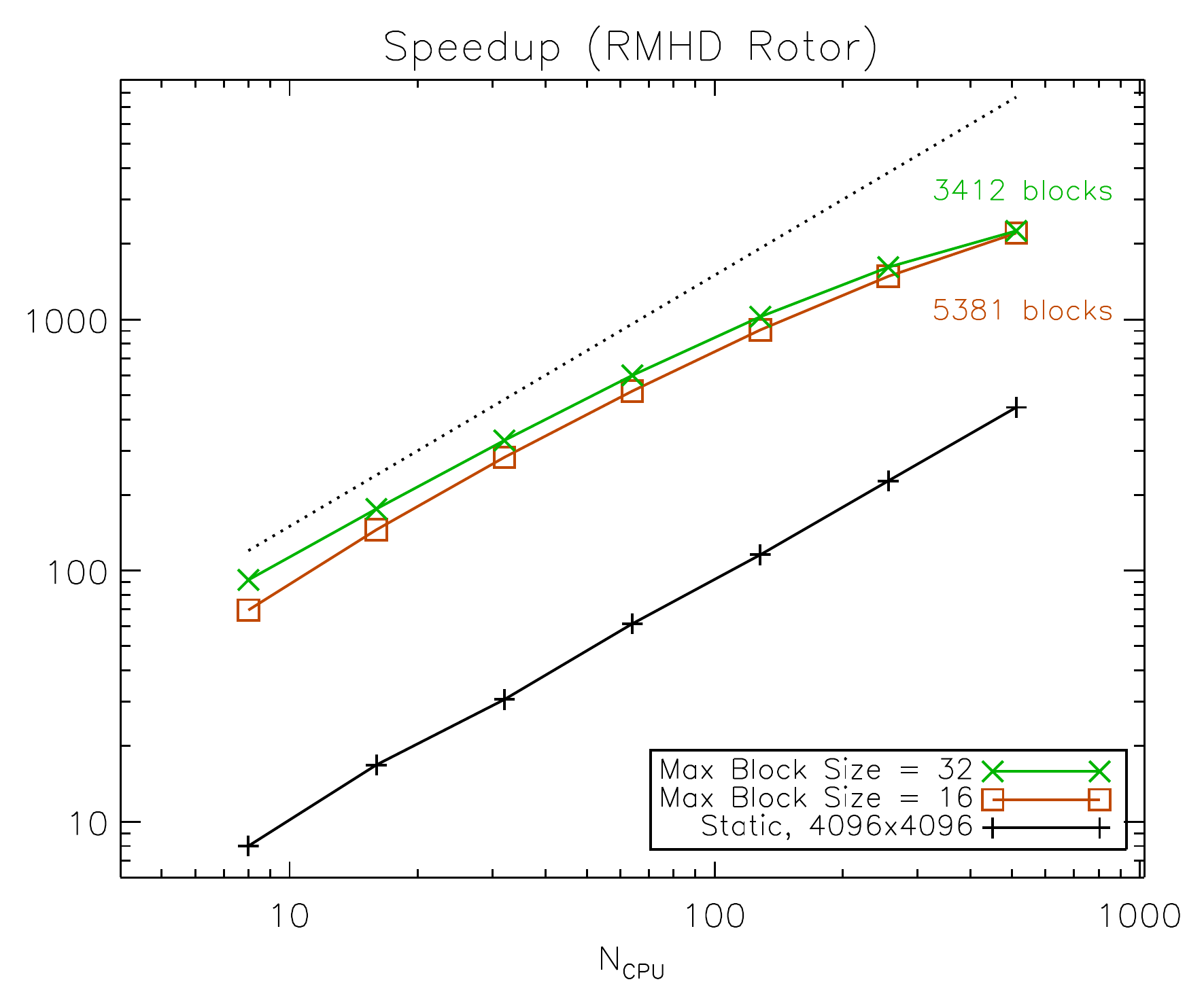}
\caption{\footnotesize Parallel scalings for the 2D relativistic
          rotor from 8 to 512 CPUs.
          The red and green lines (squares and crosses,
          respectively) give the speedup factors corresponding to computations
          with maximum block sizes of $16$ and $32$ zones, respectively.
          The number of blocks on the finest level ($\ell = 6$) at the end of 
          integration is reported above and below the corresponding curve.
          The dotted black line gives the perfect scaling.}         
\label{fig:rmhd_rotorScaling}
\end{figure}
\fi
The two-dimensional rotor problem \citep{dZBL03, HKM08, DBTM08} consists of 
a rapidly spinning disk embedded in a uniform background medium ($\rho=p=1$) 
threaded by a constant magnetic field $\vec{B}=(1,0,0)$.
Inside the disk, centered at the origin with radius $R = 0.1$, the density 
is $\rho=10$ and the velocity is prescribed as $\vec{v}=\omega(-y,x,0)$ 
where $\omega=9.95$ is the angular frequency of rotation.
The computational domain is the square $x,y\in[-\frac{1}{2},\frac{1}{2}]$ 
with outflow (i.e. zero-gradient) boundary conditions.
Computations are performed on a base grid with $64^2$ grid points and $6$ 
levels of refinement using the HLL Riemann solver and a CFL number $C_a = 0.5$. 
Zones are tagged for refinement whenever the second derivative error norm of 
${\cal E}/(\rho\gamma)$, computed with (\ref{eq:RefCrit}), exceeds 
$\chi_r = 0.15$.
The effective resolution amounts therefore to $4096^2$ zones, the largest 
employed so far (to the extent of our knowledge) for this particular problem.

Results are shown at $t=0.4$ in Fig. \ref{fig:rmhd_rotor} where one can see an
emerging flow structure enclosed by a circular fast forward shock
($r\approx 0.425$) traveling into the surrounding medium.
An inward fast shock bounds the innermost oval region where density has been 
depleted to lower values. 
The presence of the magnetic field slows down the rotor and the maximum Lorentz
factor decreases from $10$ to $\approx 2.2$. 
The numerical solution preserves the initial point symmetry and density 
corrugations, present in lower resolution runs, seem drastically
reduced at this resolution, in accordance with the findings of \cite{HKM08}.
Divergence errors are mostly generated in proximity of the outer fast shock and
are damped while being transported out of the domain at the speed of light in
the GLM formulation.
We monitor the growth of such errors by plotting the volume-average 
of $|\nabla\cdot\vec{B}|$ as a function of time showing (bottom panel 
in Fig. \ref{fig:rmhd_rotor}) that the growth of monopole errors is limited 
to the same values of \cite{HKM08} (the factor of $4$ comes from the fact that 
our computational domain is twice as small).

Parallel performance for this particular test is shown in Fig. 
\ref{fig:rmhd_rotorScaling} plotting the speedup, defined as
$S=8T^{\rm u}_{8}/T_{N_{\rm CPU}}$, where $T^{\rm u}_8$ is the execution 
time of the uniform grid run performed at the same effective resolution 
($4096^2$) on $8$ processors while $T_{N_{\rm CPU}}$ is the running time
measured with $N_{\rm CPU}$ processors.
AMR scaling has been quantified for computations using maximum 
patch sizes of $16$ and $32$ grid points resulting, respectively, 
in $11921$ and $6735$ total number of blocks at the end of integration.
Approximately half of it belongs to the finest level, as reported in 
Fig \ref{fig:rmhd_rotorScaling}.
In general, the AMR approach offers a $5$ to $10$ speedup gain in terms
of execution time over the uniform, static grid approach.
For $N_{\rm CPU} < 256$ the parallel efficiency, measured as $S/N_{\rm CPU}$,
is larger than $0.7$ while it tends to be reduced for more processors. 
Computations carried with fewer blocks (i.e. larger block sizes) tend to be
more efficient when fewer CPUs are in use since inter-processor
communications are reduced.
However, this cost becomes sensibly larger at $512$ (or more) 
processors resulting in a loss of efficiency.

\subsection{Cylindrical Blast Wave}
%
%

\ifx\IncludeEps\yes
\begin{figure*}\centering
\includegraphics[width=0.9\textwidth]{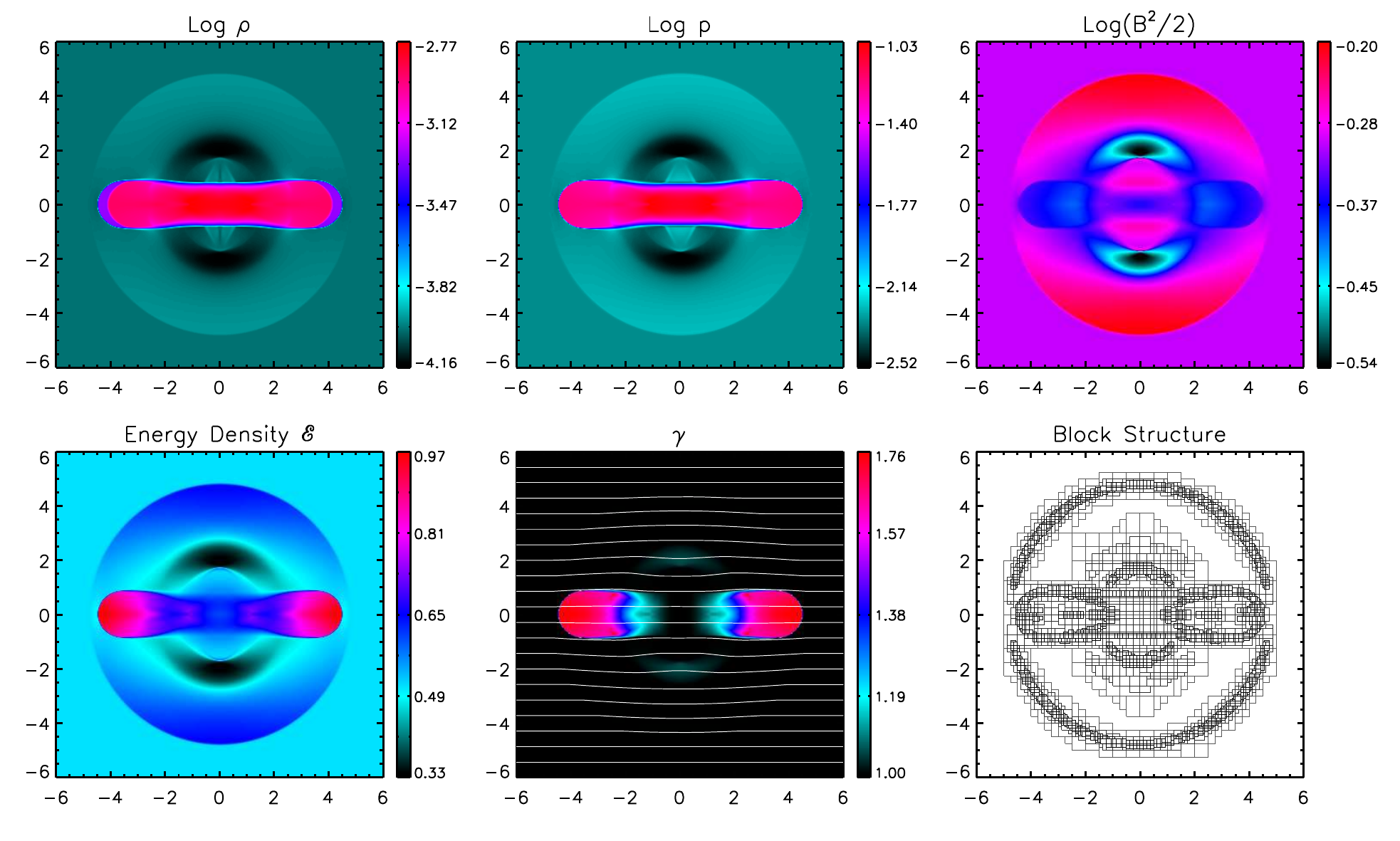}
\caption{\footnotesize Cylindrical relativistic blast wave at $t=4$. 
         The base level grid has $48^2$ zones and $5$ levels of refinement 
         are used. 
         The different maps show proper density (top left, log scale), thermal
         pressure (top right)
         magnetic field strength (top right, log scale)
         density (bottom left, log scale) and Lorentz factor 
         (bottom right). Refinement levels $3$ and $4$ are overplotted on
         the first plot while magnetic field lines overlap in the latter.}         
\label{fig:rmhd_blast2D}
\end{figure*}
\fi
Strong symmetric explosions in highly magnetized environments can become 
rather arduous benchmarks probing the code's ability to evolve strongly
magnetized shocks \citep[see, for instance][]{K99, Leis05, MB06, dZBL03, BS11}. 
Failures may lead to unphysical densities or pressures whenever the 
scheme does not introduce adequate dissipation across oblique discontinuities 
and/or if the divergence-free condition is not properly controlled.

The two dimensional setup considered here consists of the Cartesian 
square $[-6,6]\times[-6,6]$ initially filled with constant density and 
pressure values $\rho=10^{-4}$ and $p=5\cdot 10^{-3}$ and threaded by a 
constant horizontal magnetic field with strength $B_0=1$.
A cylinder with radius $r=0.8$ and centered at the origin 
delimits a higher density and pressure region where 
$\rho=10^{-2}$, $p=1$. The ideal EoS with with $\Gamma = 4/3$ is used. 
Our choice of parameters results in a low $\beta$ plasma ($2p/B^2=10^{-2}$) 
and corresponds to the strongly magnetized cylindrical explosion discussed by 
\cite{BS11}.
We point out that this configuration differs from the one discussed in 
\cite{K99} and \cite{MB06} by having the ambient pressure ten times larger.
This allows to run the computation without having to introduce any specific 
change or modification in the algorithm, such as shock flattening or 
redefinition of the total energy.
We adopt a base grid of $48$ zones in each direction with outflow boundary 
conditions holding on all sides.
Integration proceeds until $t=4$ using the HLLC Riemann solver 
\citep{MB06} using $5$ levels of refinement, reaching an effective resolution
of $1536^2$ zones.
Total energy triggers refinement with a threshold of $\chi_{\rm ref} = 0.025$.

The over-pressurized region, Fig. \ref{fig:rmhd_blast2D}, sets an anisotropic 
blast wave delimited by a weak fast forward shock propagating almost radially.
The explosion is strongly confined along the $x$ direction by the magnetic 
field where plasma is accelerated to $\gamma_{\max} \approx 1.76$.
The inner structure is delimited by an elongated structure enclosed by a 
slow shock adjacent to a contact discontinuity. 
The two fronts blend together as the propagation becomes perpendicular to the 
field line. Since the problem is purely two dimensional, rotational 
discontinuties are absent.
The block distribution, shown in the lower panel of Fig. \ref{fig:rmhd_blast2D},
shows that refinement clusters around the outer discontinuous 
wave as well as the multiple fronts delimiting the extended inner region.
The total number of blocks is $2258$ with the finest level giving a relative
contribution of $\sim 0.55$ and a corresponding volume filling factor of 
$\sim 0.12$.
The numerical solution retains the expected degree of symmetry and the
agreement with earlier results demonstrate the code can robustly   
handle strongly magnetized relativistic shocks within the 
GLM-MHD formalism. 
Results obtained with $4$ processors show that calculation done with 
AMR is faster than the static grid runs by a factor of $\sim 4.5$.

\subsection{Spherical Blast Wave}
%
%
\ifx\IncludeEps\yes
\begin{figure*}\centering
\includegraphics*[width=0.46\textwidth]{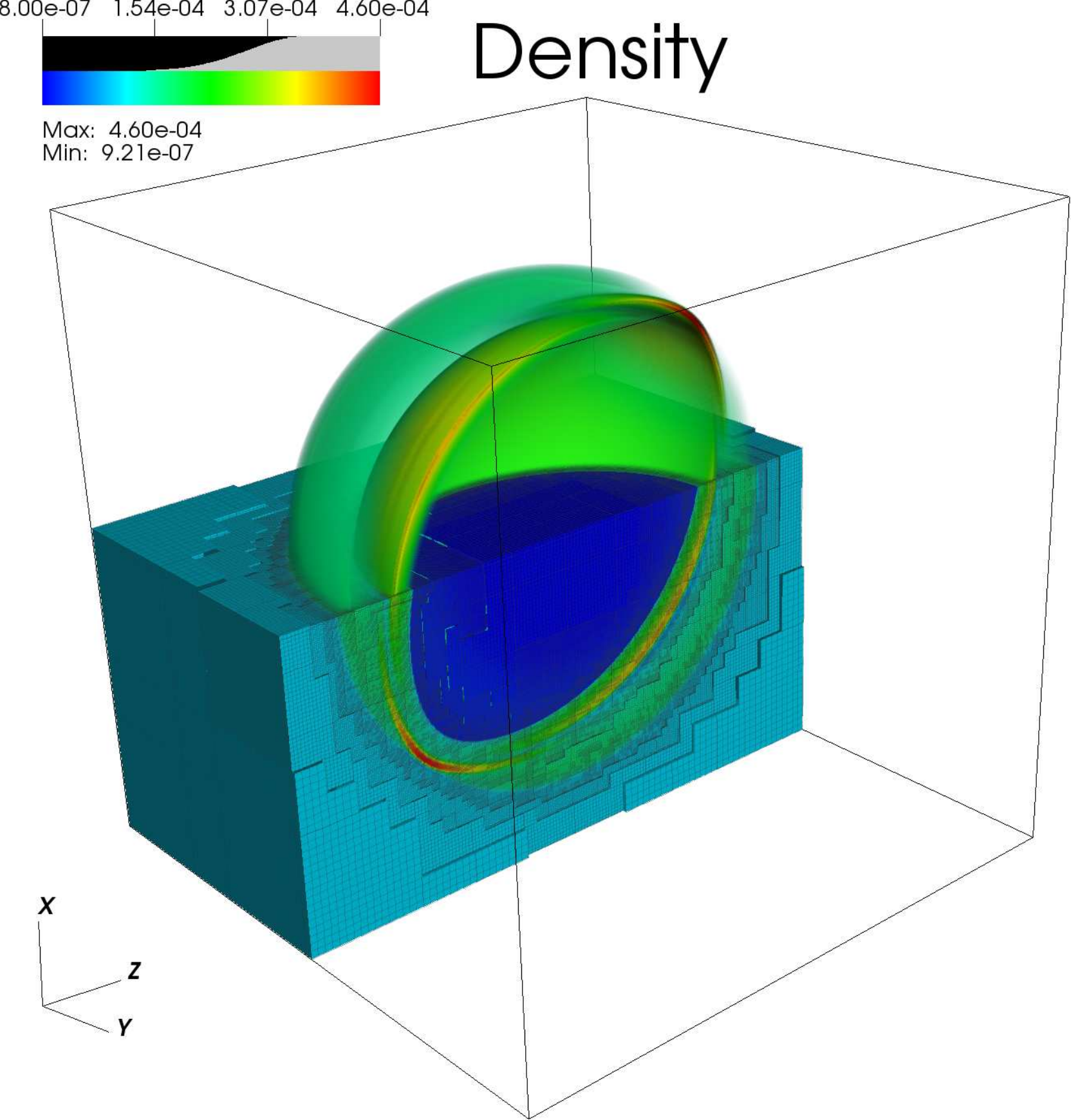}
\includegraphics*[width=0.46\textwidth]{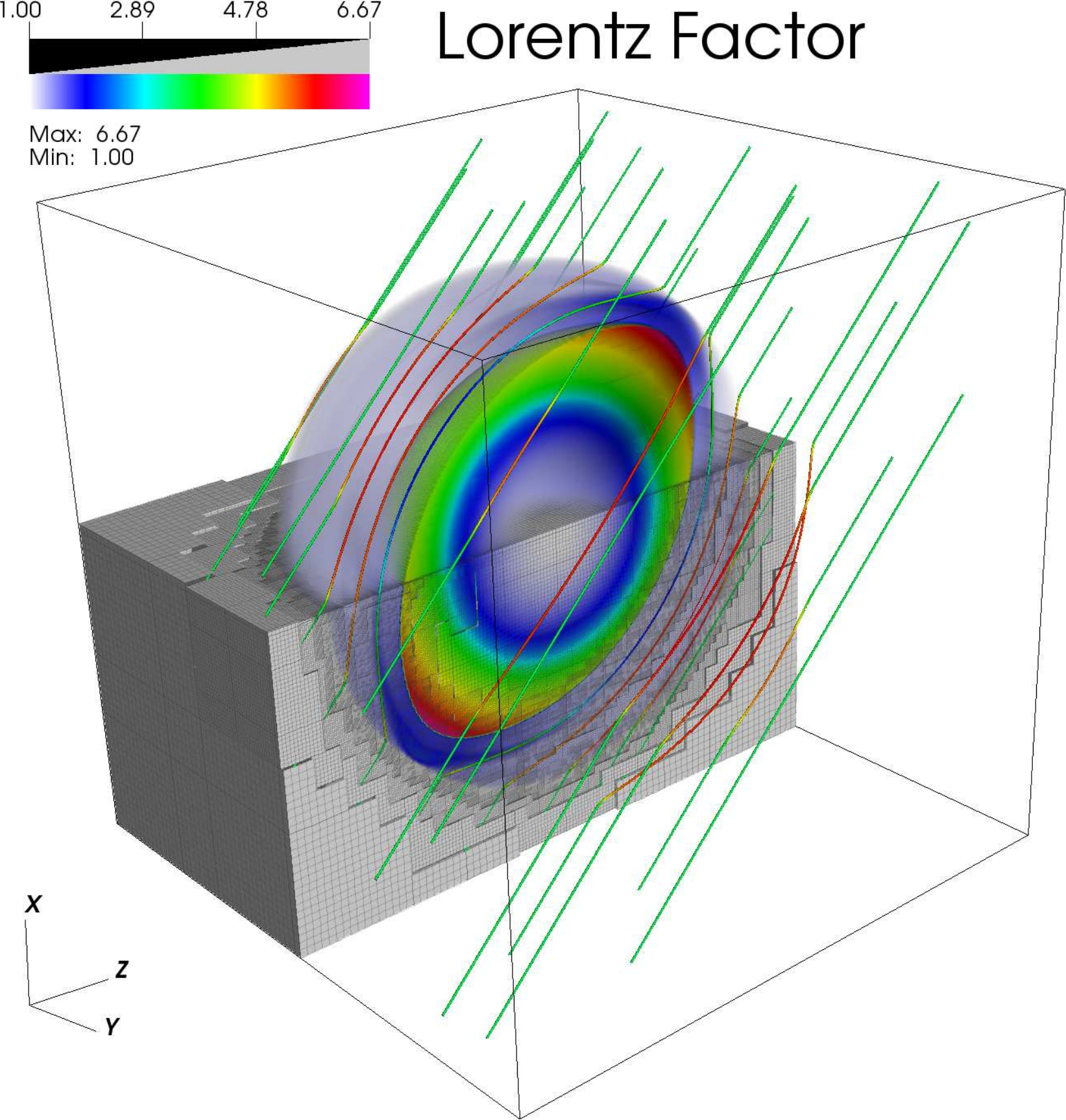}
\includegraphics*[width=0.46\textwidth]{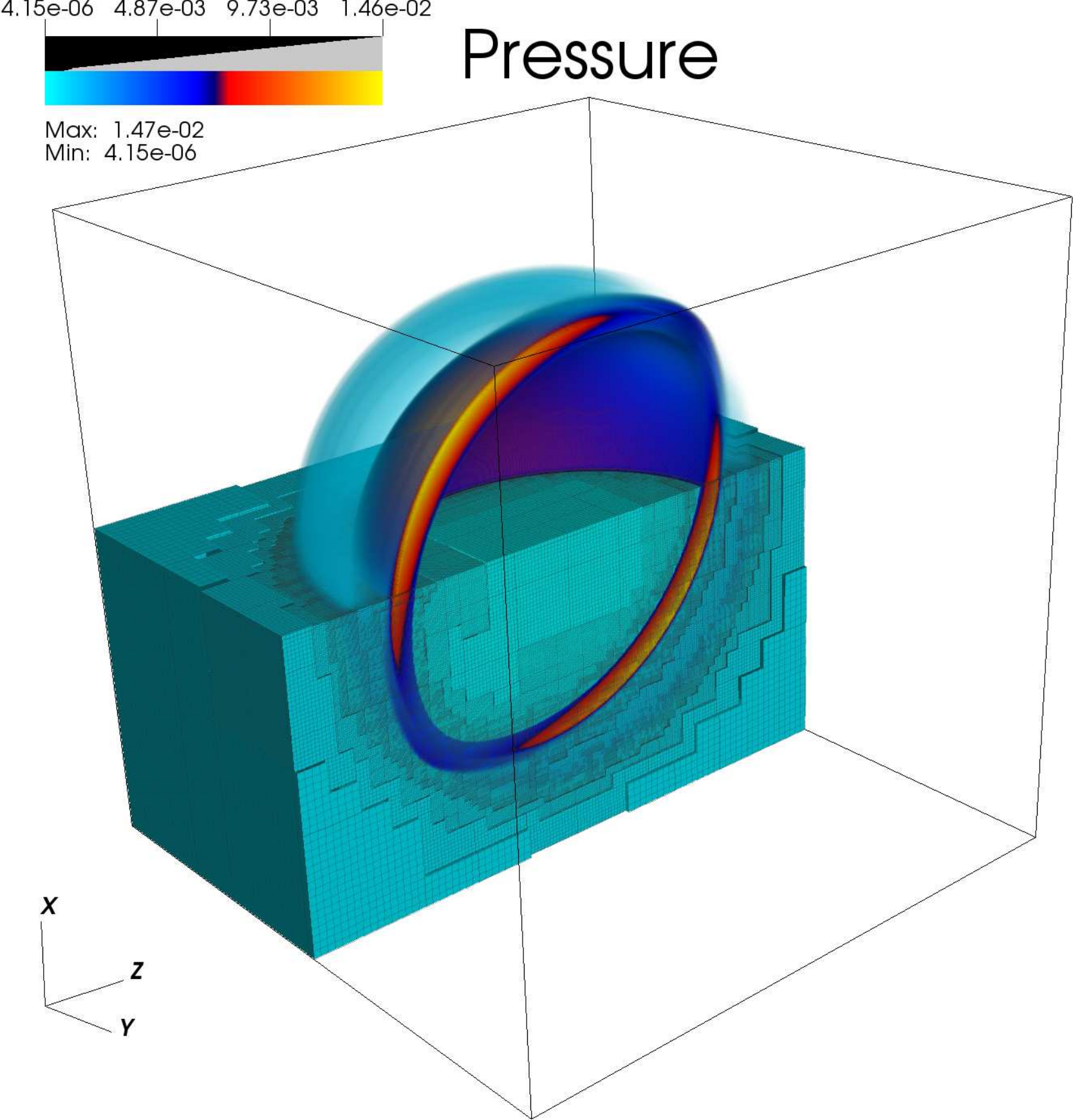}
\includegraphics*[width=0.46\textwidth]{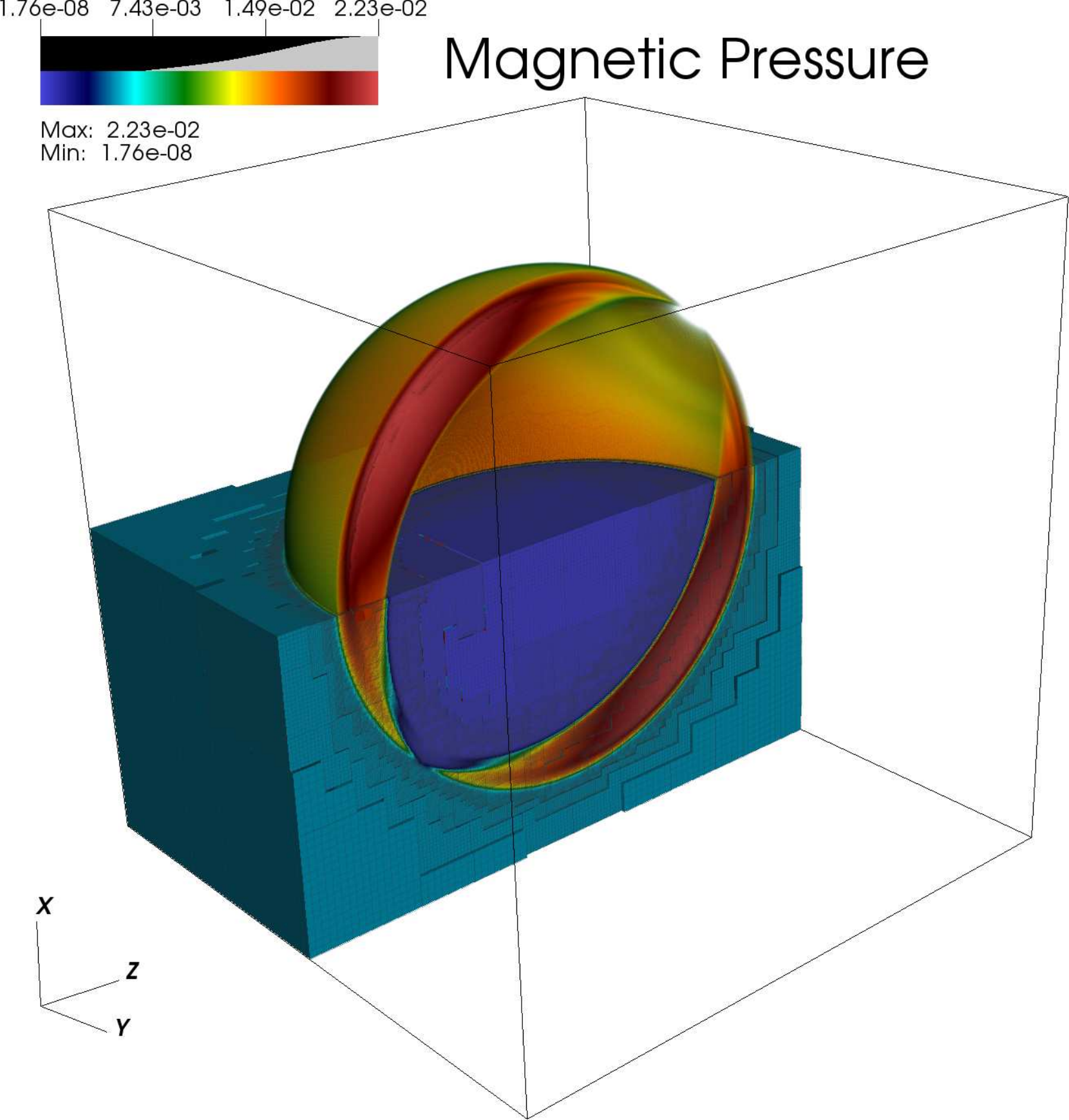}
\caption{\footnotesize Density slice-cut for the relativistic magnetized blast wave in
         three dimensions at $t=4$. Four levels of refinement are used to achieve
         an effective resolution of $640^3$. The box in the lower half-semispace 
         emphasize jump ratios across levels. Oblique magnetic fieldlines are 
         overplotted.}         
\label{fig:rmhd_blast3D}
\end{figure*}
\begin{figure}\centering
\includegraphics*[width=\columnwidth]{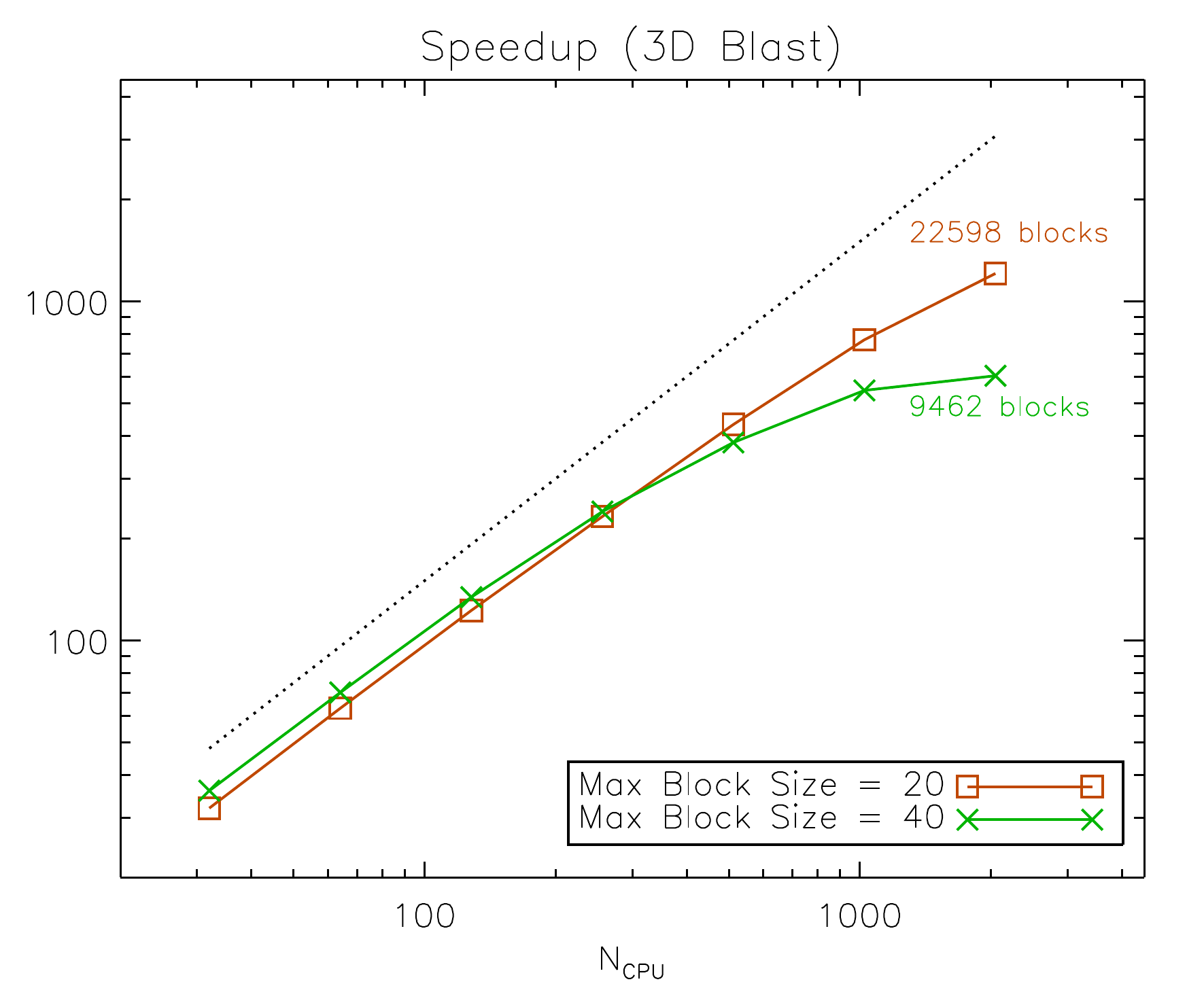}
\caption{\footnotesize Parallel scalings for the three-dimensional relativistic
          blast wave problem from 32 to 2048 processors. 
          The red and green lines (squares and crosses,
          respectively) give the speedup factors corresponding to "frozen-flux"
          computations with maximum block sizes of $20$ and $40$ zones,
          respectively.
          The number of blocks on the finest level, reported above and below 
          each line, stays constant in time.
          Ideal scaling is given by the dotted black line.}         
\label{fig:rmhd_Blast3DScaling}
\end{figure}
\fi
In the next example we consider a three-dimensional extension of the 
cylindrical blast wave problem presented in \cite{ECHO07}.
The initial condition consists of a sphere with radius $r=0.8$ centered at 
the origin filled with hot gas having $\rho=10^{-2}$, $p=1$ 
(the ideal EoS with $\Gamma = 4/3$ is used).
Outside this region, density and pressure decrease linearly reaching the
constant ambient values $\rho=10^{-4}$, $p=5\cdot 10^{-3}$ for $r\ge 1$.
The plasma is at rest everywhere and a constant magnetic field is set oblique 
to the mesh, i.e., $\vec{B}=B_0(\hvec{e}_x + \hvec{e}_y)/\sqrt{2}$ with 
$B_0=0.1$.
The computational domain is the Cartesian box $[-6,6]^3$ covered by a base grid 
of $40$ zones in each direction with outflow boundary conditions holding on all 
sides.
Using the HLLD Riemann solver, we follow integration until $t=4$ using 
$4$ levels of refinement (effective resolution $640^3$) and a CFL number
$C_a=0.4$. Zones are tagged for refinement upon density fluctuations 
with a threshold value of $\chi_{\rm ref} = 0.15$.

Results shown in Figure \ref{fig:rmhd_blast3D} reveal an oval-shaped 
explosion enclosed by an outer fast shock propagating almost radially. 
A strong reverse shock confines a prolate spheroidal region where 
magnetic field has been drained and inside which expansion takes place 
radially.
The largest Lorentz factor $\sim 6.3$ is attained in the direction of 
magnetic field lines close to the ellipsoid poles.
 
In order to measure parallel performance on a fixed number of blocks,
we have integrated the solution array starting from $t=3.5$ for a fixed 
number of steps by enforcing zero interface flux.
Results are shown in Fig. \ref{fig:rmhd_Blast3DScaling} where we compare two 
sets of "frozen-flux" computations with maximum block sizes of 20 and 40 mesh
points corresponding, respectively, to 27661 (22598 on the finest level) and 
11229 (9462 on the finest level) total number of blocks.
In the former case, the larger number of patches allows to reach a better 
efficiency (more than $0.75$) whereas the latter case performs worse for 
increasing number of processors.
Even if the finest level has a number of blocks still larger than the maximum number of CPUs
employed, many boxes are much smaller than the maximum block size allowed. The ideal workload
per processor (i.e. the number of cells of the finest level divided by the number of processors)
is comparable to the maximum block size in the 1024 CPUs run and becomes smaller in the 2048
CPUs run. In this latter case, the workload of the processors integrating the larger blocks 
is therefore larger than the ideal one and most of the CPUs must wait for these processors to complete
the integration. This could explain the poor parallel performance of this case.


\subsection{Kelvin-Helmholtz flow}
%
%

\ifx\IncludeEps\yes
\begin{figure*}\centering
\includegraphics[width=0.8\textwidth]{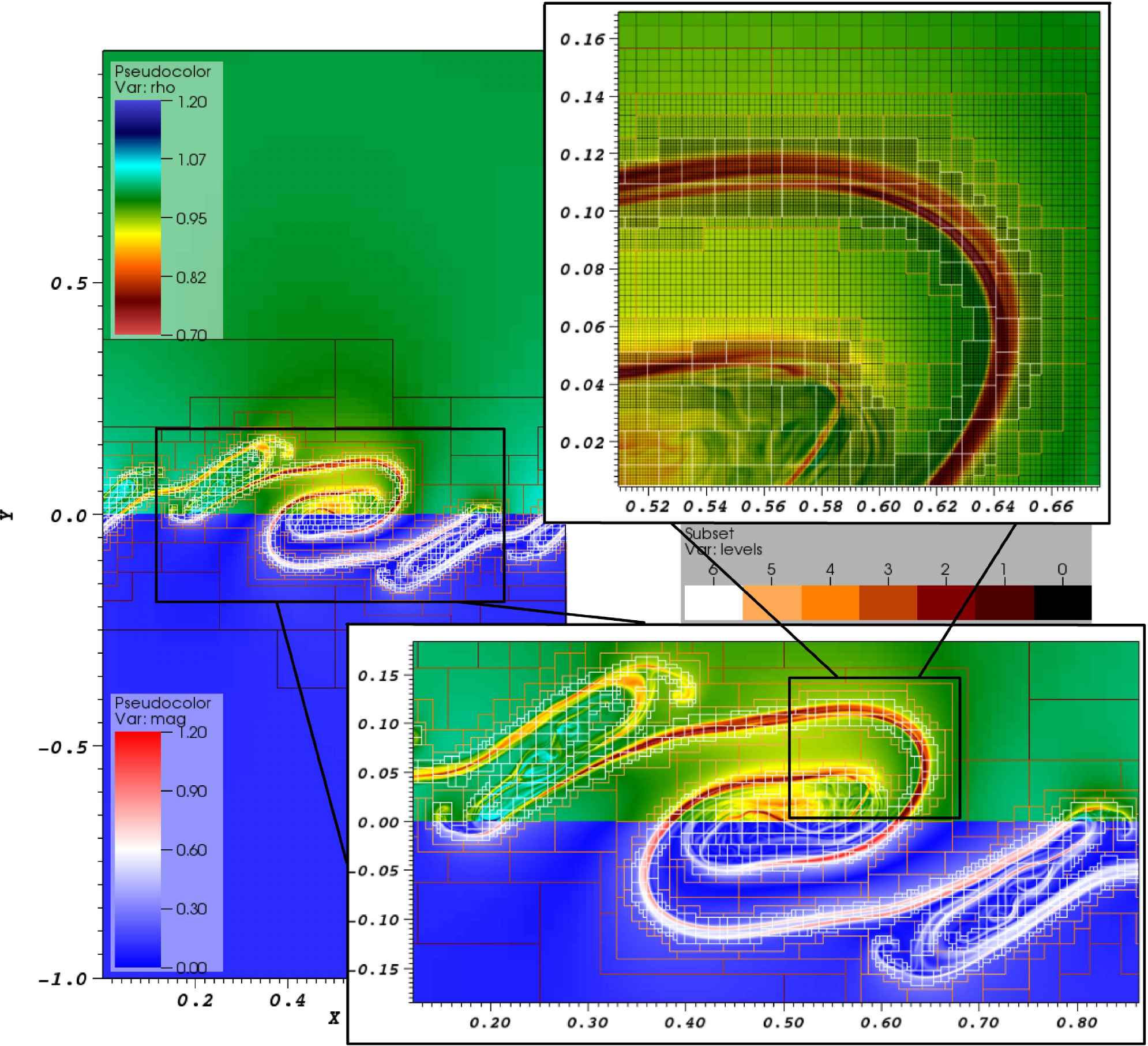}
\caption{\footnotesize Relativistic Kelvin-Helmholtz instability at $t=5$
         using $6$ additional levels of refinement starting from a base
         grid of $32\times 64$.
         The panel on the left shows density (upper half) and the quantity
         $(B_x^2+B_y^2)^{\HALF}/B_z$ (lower half) on the whole computational
         domain. Selected regions are enlarged in the two panels on the right.}         
\label{fig:rmhd_kh}
\end{figure*}
\begin{figure}\centering
\includegraphics*[width=\columnwidth]{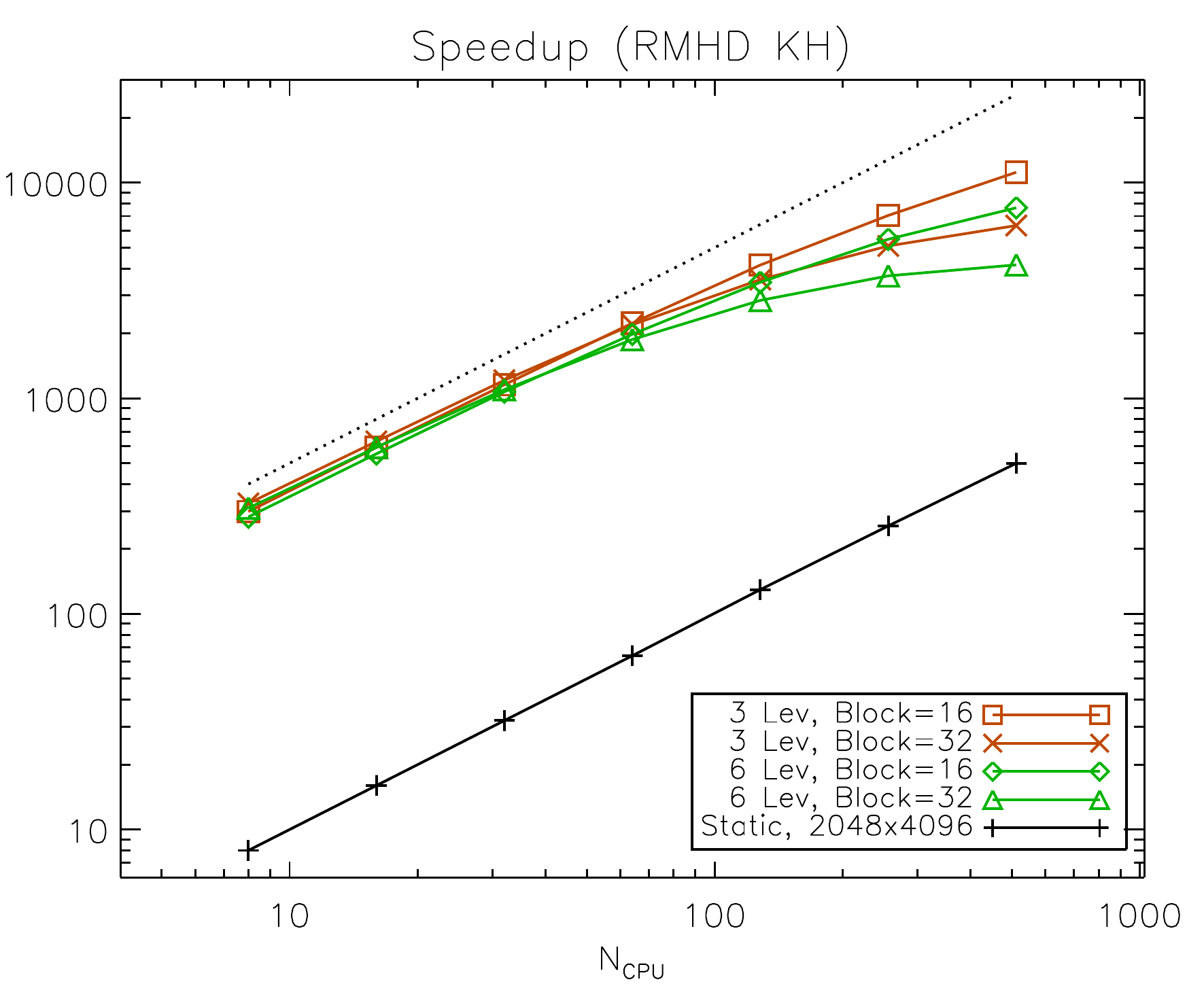}
\caption{\footnotesize Parallel speedup up to 512 processors for the
        two-dimensional relativistic Kelvin-Helmholtz instability problem.
        Speedup is computed as the ratio $8T_8^u/T_{N_{\rm CPU}}$ between
        the execution time of the static uniform grid run at the same effective
        resolution ($2048\times4096$) on $8$ processors and the running time
        measured with $N_{\rm CPU}$ processors. Red lines corresponds
        to the 3-level run using max grid sizes of 16 (squares) and 32 (crosses).
        Green lines corresponds the the 6-level compuations.}
\label{fig:rmhd_KHScaling}
\end{figure}
\fi
In the next example \citep{BdZ06, MUB09} we consider a two-dimensional planar
domain with $x\in[0,1]$, $y\in[-1,1]$ initially filled with a (relativistically)
hot gas with constant density $\rho=1$ and pressure $p=20$.
An initially perturbed shear velocity profile of the form
\begin{equation}
 (v_x, v_y, v_z) = \left[V_0\tanh\frac{y}{\alpha},
                         \epsilon \sin(2\pi x)\exp\left(-\frac{y^2}{\beta^2}\right),
                         0\right]
\end{equation}
is set across the domain, where $V_0=1/4$ is the nominal flow velocity,
$\epsilon=V_0/100$ is the amplitude of the perturbation while 
$\alpha = 0.01$, $\beta = 0.1$.
The TM equation of state, Eq. (\ref{eq:TMEos}), is used during the evolution
to recover the gas enthalpy from density and pressure.
The magnetic field is initially uniform and prescribed in terms of the 
components parallel and perpendicular to the plane of integration:
\begin{equation}
  (B_x, B_y, B_z) = \left(\sqrt{2\sigma_{\rm pol}p},0,
                          \sqrt{2\sigma_{\rm tor}p}\right)
\end{equation}
where $\sigma_{\rm pol}=0.01$, $\sigma_{\rm tor} = 1$.
We perform the integration on a base grid of $32\times 64$ cells with $6$ 
additional levels of refinement, reaching an effective resolution 
of $2048\times 4096$ zones.
The Courant number is set to $C_a=0.8$ and refinement is triggered whenever
Eq. (\ref{eq:RefCrit}) computed with total energy density exceeds $\chi_r = 0.14$.
Open boundary conditions hold at the lower and upper $y$ boundaries while
periodicity is imposed in the $x$ direction. 

Fig. \ref{fig:rmhd_kh} shows density and magnetic field distributions at 
$t=5$ which approximately marks the transition from the linear phase to 
the nonlinear evolution \citep[see the discussion in][]{MUB09}.
Both density and magnetic field are organized into filaments that wrap 
around forming a number of vortices arranged symmetrically with respect to 
the central point.
Refined levels concentrate mainly around the location of the fluid 
interface and accurately follow the formation of the central vortex and the 
more elongated adjacent ones. 
The computation preserves the initial point-symmetry even if the grid
generation algorithm does not necessarily do so.
The lower resolution outside this region contributes to damp acoustic waves
as they travel towards the outer boundaries and eventually reduces the amount 
of spurious reflections.

In Fig. \ref{fig:rmhd_KHScaling} we compare the parallel speedup between a 
number of adaptive grid calculations using either $6$ levels (grid jump of
2) or $3$ levels (grid jump of 4) and the equivalent uniform grid run with 
$2048\times4096$ zones.
Generally, the AMR approach is $\approx 40$ faster than the fixed grid run. 
In addition, the 3-level computation seems to perform somewhat 
better than the 6-level run although the number of blocks on the finest 
levels is essentially the same at the end of computation 
(1085 and 1021, respectively).
Parallel scaling sensibly deteriorates when the number of blocks per 
processor becomes less than $3$ or $4$, in accordance with the results of
previous tests.

\subsection{Three dimensional Shock-Cloud Interaction}
%
%

\ifx\IncludeEps\yes
\begin{figure*}\centering
\includegraphics*[width=0.45\textwidth]{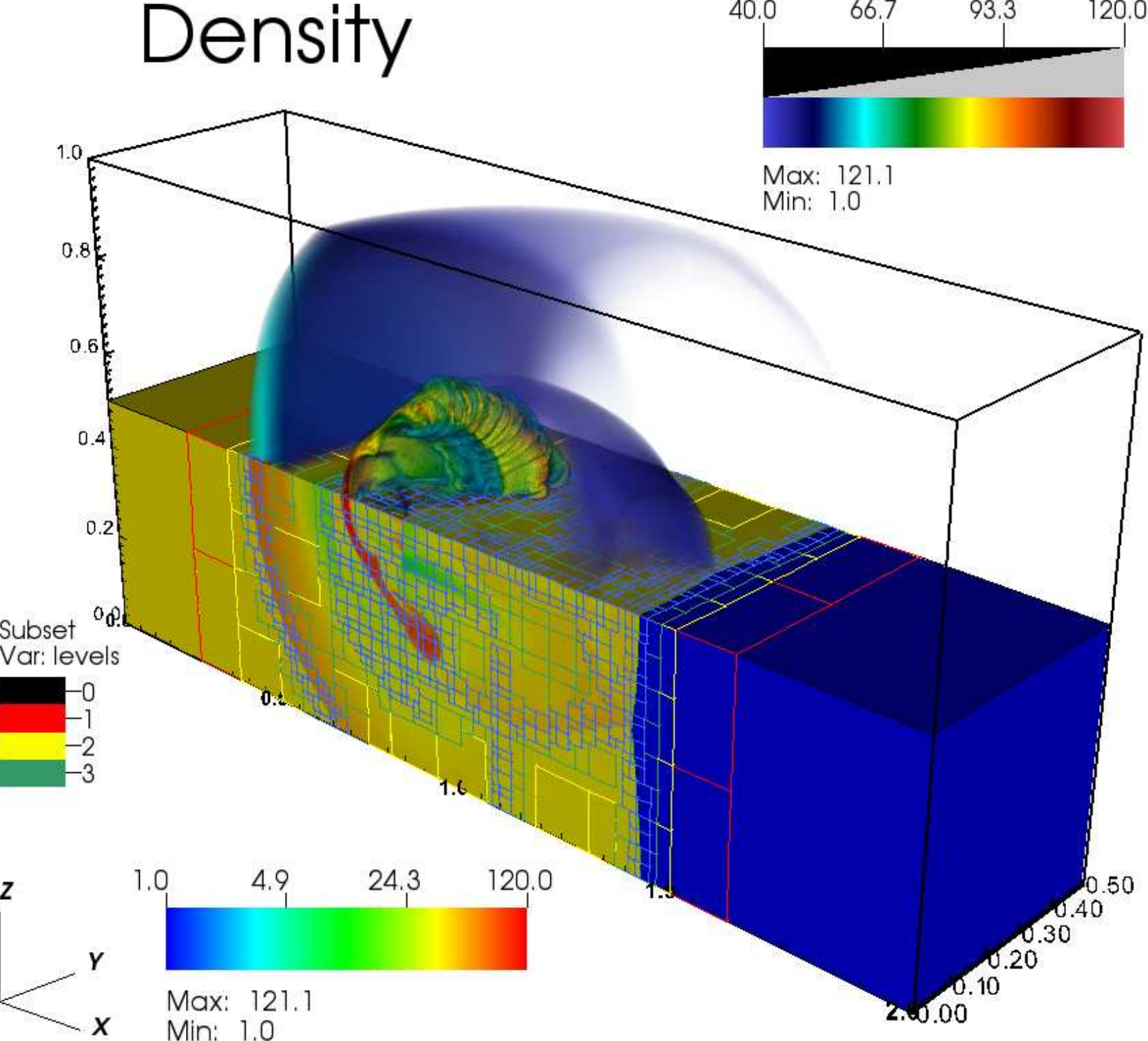}%
\includegraphics*[width=0.45\textwidth]{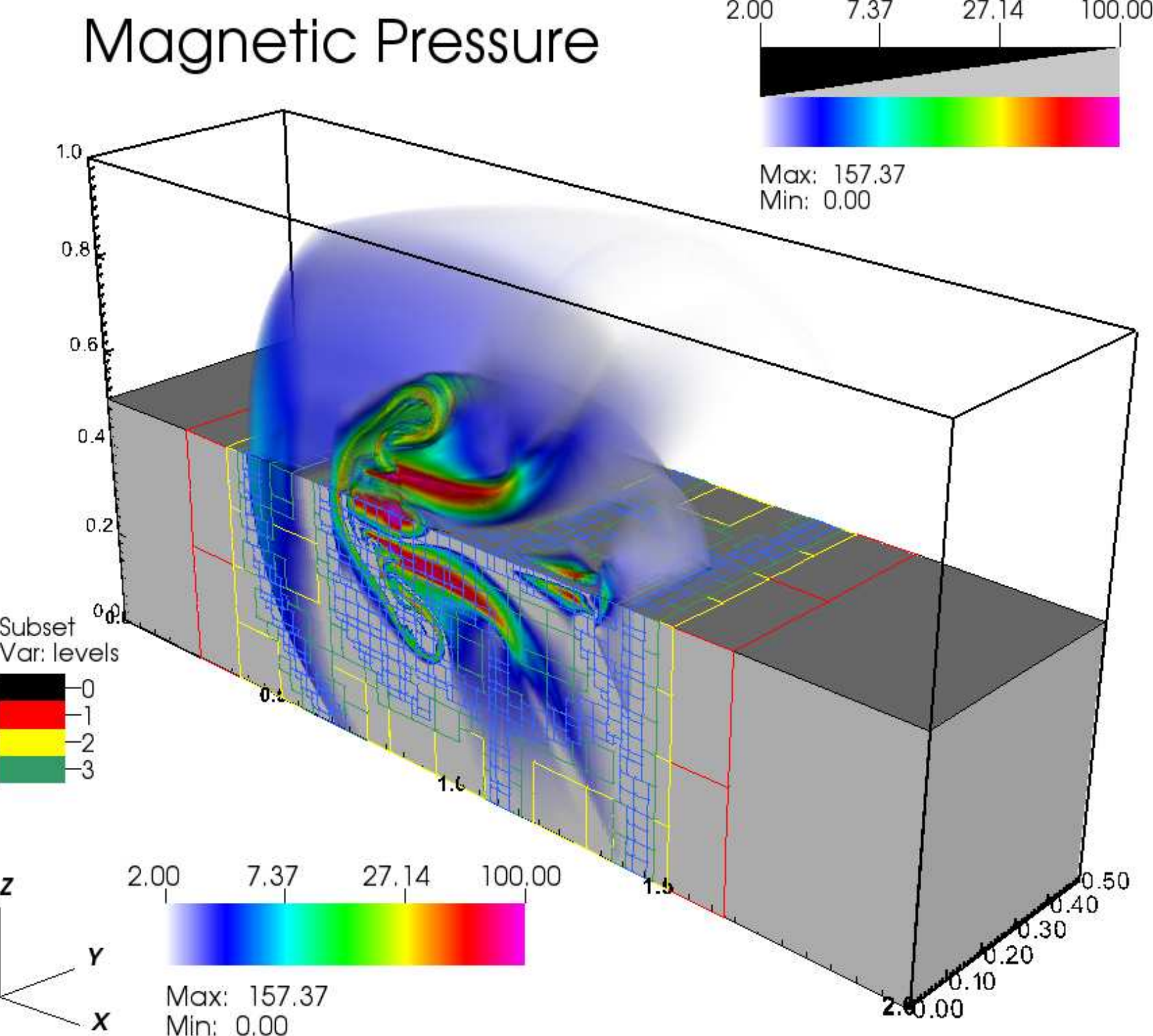}
 \caption{\footnotesize Vertical cut in the $xz$ plane showing the interaction 
          of a magnetized relativistic shock with a density cloud at $t=1$. 
          Density and magnetic pressure $\vec{B}^2/2$ are shown in the 
          left and right panels, respectively.
          For each quantity, a volumetric rendering is shown in the upper half
          whereas an intensity color map in the middle vertical and horizontal 
          planes displays below.
          The base grid corresponds to $64\times 32^2$ and $4$ levels
          of refinement are employed (effective resolution $1024\times 512^2$).}
 \label{fig:rmhd_shockcloud}
\end{figure*}
\fi
In the next example we consider a three-dimensional extension of
the planar relativistic shock-cloud interaction originally presented 
in \cite{MB06}. The initial condition consists of a high density ($\rho=10$) 
spherical clump (radius $0.15$) centered at $(0.8,0,0)$ adjacent to a shock 
wave located at $x=0$ with downstream and upstream values given by 
$(\rho,v_x,B_z,p)= (42.5942,\, 0,\, -2.12971,\, 127.9483)$ for $x<0$ and
$(\rho,v_x,B_z,p)=(1,\, -\sqrt{0.99},\, 0.5,\, 10^{-3})$ for $x\ge 0$.
The remaining quantities are set to zero.
The computational domain is the box $x\in[0,2]$, $y,z\in[-\HALF, \HALF]$.
At the base level we fix the resolution to $64\times32\times32$ zones and
employ $4$ levels of refinement equivalent to an effective resolution 
of $1024\times512\times512$.
The refinement criterion (\ref{eq:RefCrit}) uses the conserved density for 
zone tagging with a threshold $\chi_r = 0.2$. 
The equations of relativistic MHD are solved using the TM EoS 
\citep{MMcK07}, the HLLD Riemann solver and linear reconstruction 
with the harmonic limiter (\ref{eq:harmonic_lim}).
Shock adaptive hybrid integration provides additional numerical dissipation
in proximity of strong shock waves by switching the integration scheme to the 
HLL Riemann solver and the MinMod limiter (Eq. \ref{eq:minmod}), according to 
the strategy described in Appendix \ref{app:MULTID_Flattening}.
For efficiency purposes, we take advantage of the symmetry across the 
$xy$ and $xz$ planes to reduce the computation to one quadrant only.

Fig. \ref{fig:rmhd_shockcloud} shows a three-dimensional rendering of 
density and magnetic pressure through a cross-sectional view of the $xy$ and 
$xz$ planes at $t=1$.
The collision generates a fast, forward bow shock propagating ahead and a 
backward reverse shock transmitted back into the cloud.
The cloud becomes gradually wrapped by the incident shock into a
mushroom-shaped shell reaching a large compression ($\rho_{\max}\approx 121$,
$\vec{B}^2/2\approx 157$).
Grid refinement concentrates mainly around the incident and forward shocks
and the tangential discontinuities bordering the edge of the cloud. 

\subsection{Propagation of Three-dimensional Relativistic Magnetized Jet}
%
%

\ifx\IncludeEps\yes
\begin{figure*}
\centering
\includegraphics*[width=0.4\textwidth]{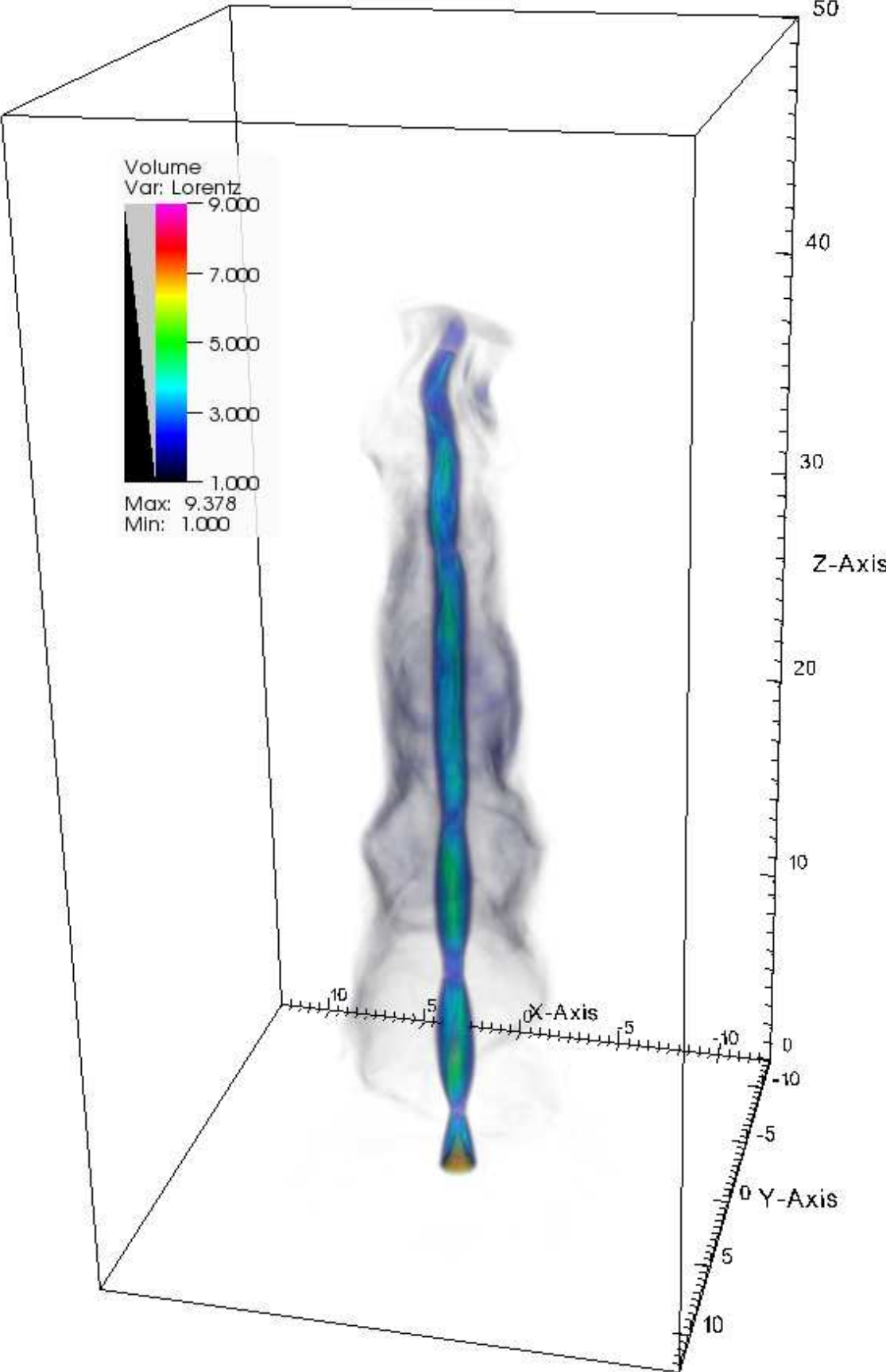}%
\includegraphics*[width=0.4\textwidth]{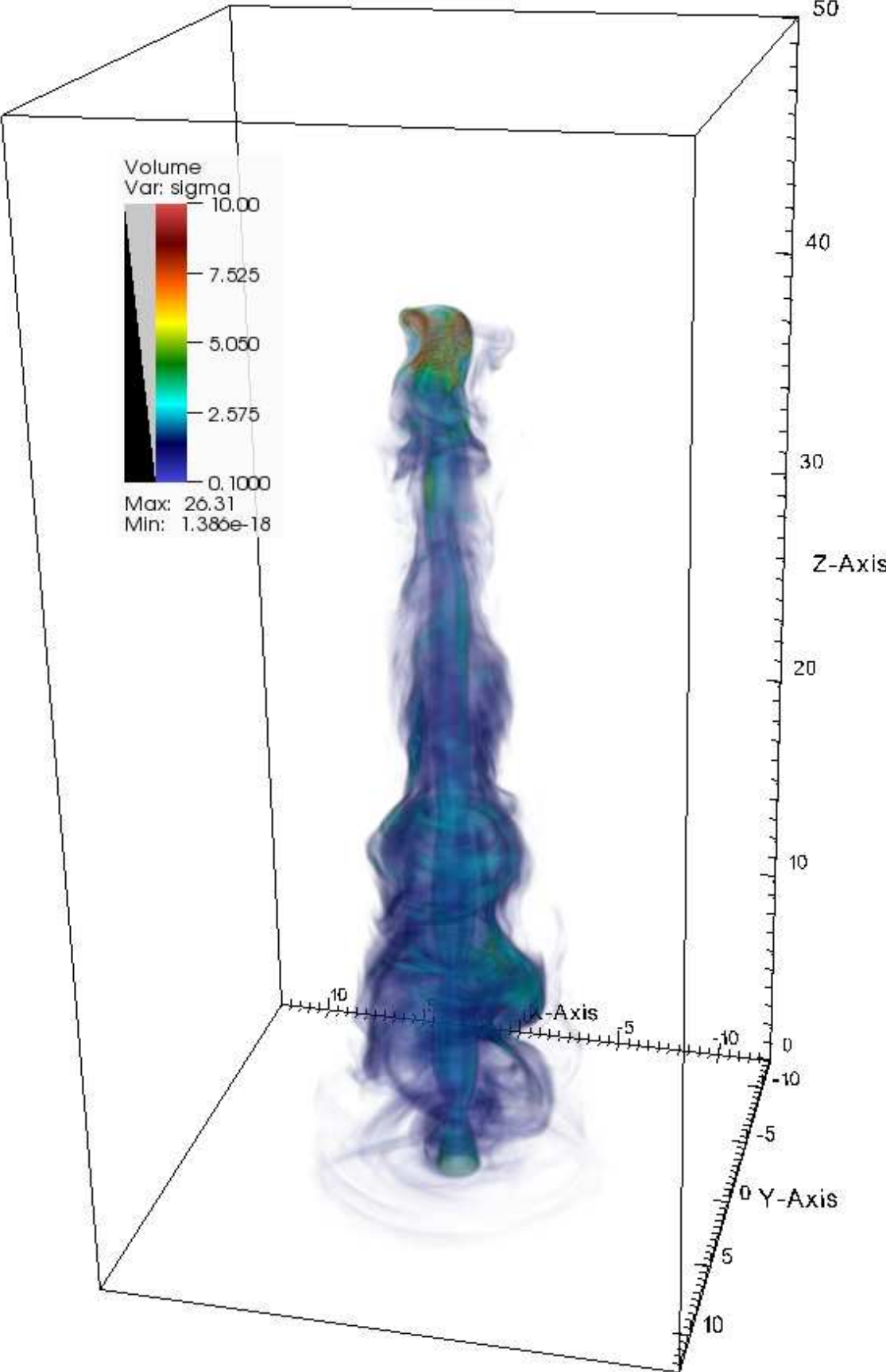}
\includegraphics*[width=0.4\textwidth]{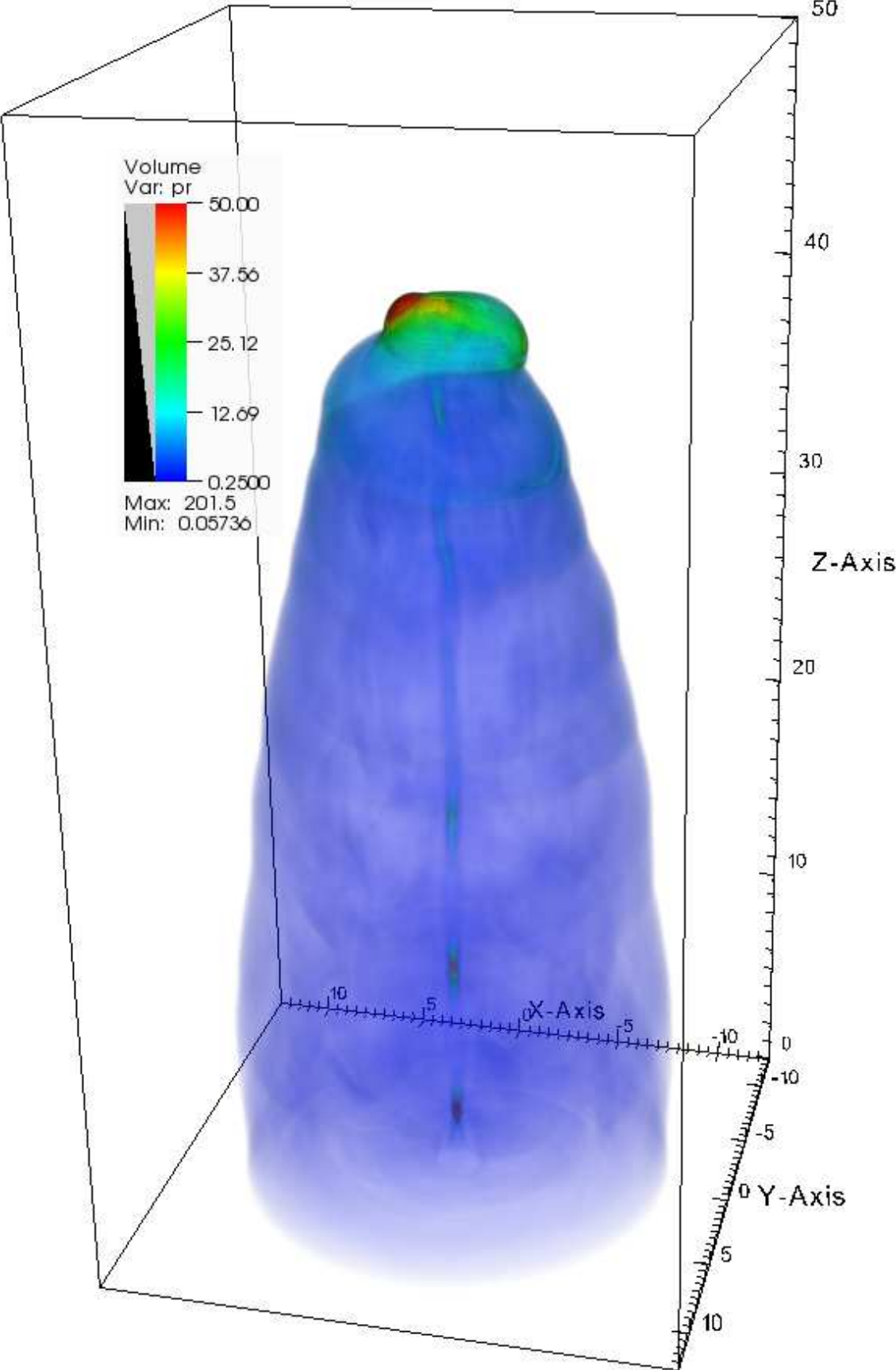}%
\includegraphics*[width=0.4\textwidth]{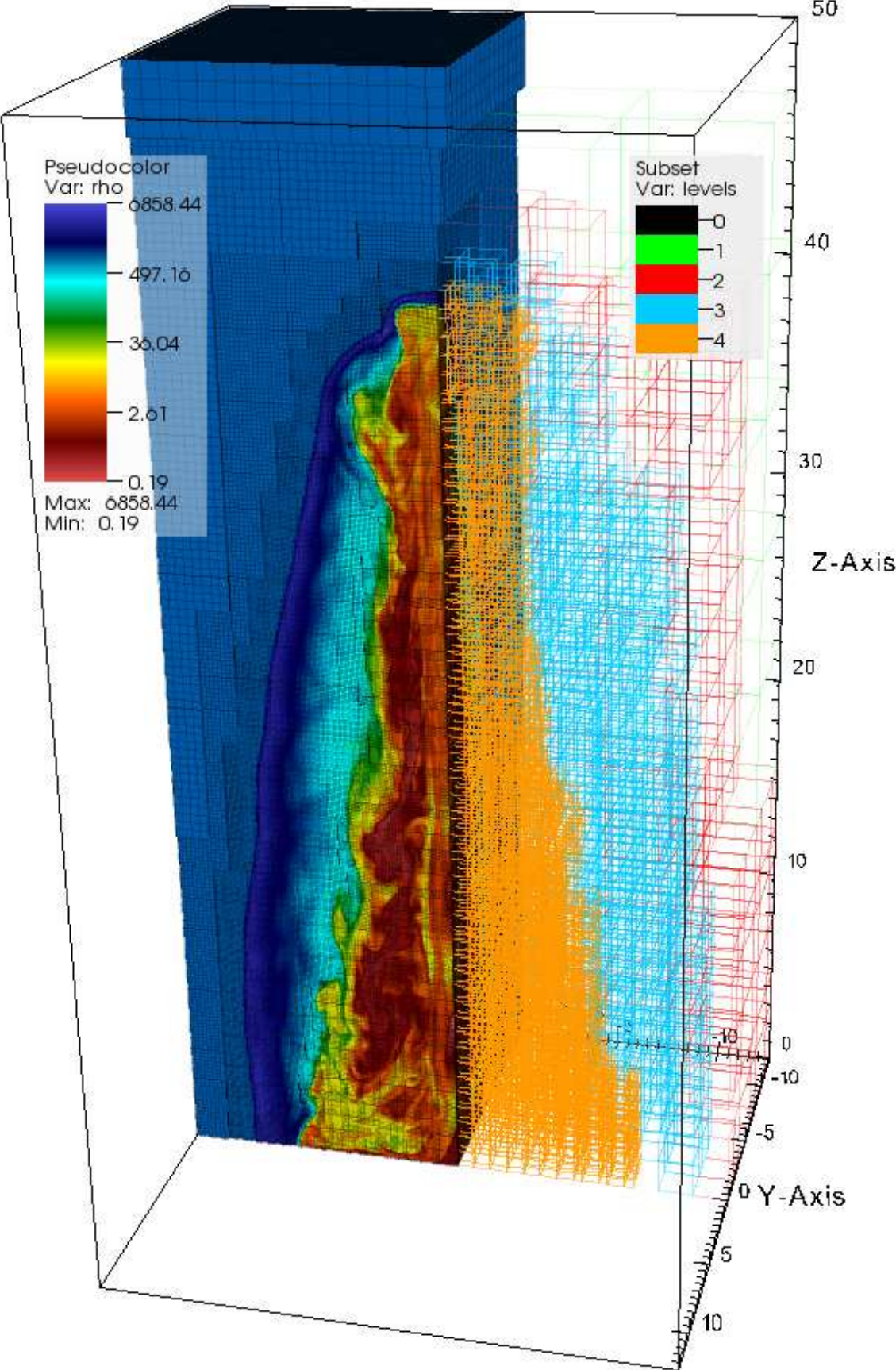}
 \caption{\footnotesize Three dimensional visualization of the 
          relativistic magnetically dominated jet at $t=130$
          using 4 levels of refinement.
          In the top panels we show the distribution of the Lorentz
          factor (left) and magnetic to kinetic energy ratio 
          $|\vec{B}|^2/(\gamma^2\rho)$ (right).
          In the bottom panels we show the thermal pressure distribution 
          (left) and a slice cut of density together with the grid structure 
          (right).}
 \label{fig:rmhd_jet}
\end{figure*}
\fi

As a final application, we discuss the three-dimensional propagation of a 
relativistic magnetized jet carrying an initially purely azimuthal magnetic field.
Indeed, the presence of a substantial toroidal component of the field is commonly 
invoked and held responsible for the acceleration and collimation of jets
from active galactic nuclei.
If on one hand cylindrical MHD configuration are expected to be unstable to 
reflection, Kelvin-Helmholtz and current driven modes, on the other hand 
astrophysical jets appear to be quite stable thus posing an unsolved issue,
see \cite{MRBFM10} and reference therein.

Following \cite{MUB09} we set the box $x,y\in[-12.5,12.5]$, $z\in[0,50]$ as
our computational domain, initially filled with constant uniform density and 
pressure $\rho_a$ and $p_a$.
The beam is injected at the lower boundary for $z\le0$, $r=\sqrt{x^2+y^2} \le 1$
with a density $\rho_j$, a longitudinal velocity corresponding
to a Lorentz factor $\gamma_j$ and an azimuthal magnetic field given by 
\begin{equation}
  B_\phi = \left\{\begin{array}{ll} 
   \gamma_jb_m r/a &\quad{\rm for}\quad r < a     \,,\\ \noalign{\medskip}
   \gamma_jb_m a/r &\quad{\rm for}\quad a < r < 1 \,,\\ \noalign{\medskip}
   0               &\quad{\rm otherwise}\,,
\end{array}\right.
\end{equation}
where $b_m$ is the magnetic field strength in the fluid's frame and $a=0.5$ is
the magnetization radius. 
The pressure profile is found by imposing radial momentum balance across
the beam giving 
\begin{equation}
  p(r) = p_a + b_m^2\left[1 - \min\left(\frac{r^2}{a^2},1\right)\right]\,,
\end{equation}
where $p_a$ is the ambient pressure defined in terms of the sonic Mach number
$M$:
\begin{equation}\label{eq:rel_jet_press}
 p_a = \frac{\rho_jv_j^2(\Gamma -1)}{\Gamma(\Gamma -1)M^2 - \Gamma v_j^2}\,,
\end{equation}
with $\Gamma = 5/3$ although the TM equation of state (\ref{eq:TMEos}) is 
used during the evolution.
The actual value of $b_m$ may be found by prescribing the magnetization 
parameter $\sigma_\phi$ defined as the ratio between beam-averaged magnetic 
energy density and thermal pressure. 
The final result yields \citep{MUB09}:
\begin{equation}
  b_m = \sqrt{\frac{-4p_a\sigma_\phi}{a^2(2\sigma_\phi-1+4\log a)}}\,.
\end{equation}
For the present simulation we adopt $\gamma_j = 7$, $M=3$, 
$\sigma_\phi = 1.3$, $\rho_j=1$ and $\rho_a = 10^3$ thus corresponding to 
a magnetically dominated, underdense  relativistic jet.

We follow the simulation up to $t=130$ (in units of the speed of light crossing
of the jet radius), starting with a base grid of $32\times 32\times 64$ 
with $4$ levels of refinement.
The refinement ratio between consecutive levels is 2 so that, at the finest
resolution, we have $\approx 20$ points per beam radius.
Zones belonging to levels 0,1 and 2 are tagged for refinement using the 
normalized second-derivative Eq. (\ref{eq:RefCrit}) of the total energy 
density with a threshold $\chi_r = 0.25$.
Zones at the finest level grid are instead created by using 
$\vec{B}^2/(\gamma\rho)$ in place of the total energy density.
The rationale for choosing this selective rule is to provide the 
largest resolution on the jet material only, still being able to track 
the sideway expansion of the cocoon at lower resolution.
We use the HLLD Riemann solver with the harmonic limiter 
(\ref{eq:harmonic_lim}) except at strong shocks where we employ the 
multidimensional shock dissipation switch outlined in Appendix 
\ref{app:MULTID_Flattening}.
The Courant number is $C_a=0.3$.

The jet structure, Fig. (\ref{fig:rmhd_jet}), shows that the flow 
maintains a highly relativistic central spine running through 
multiple recollimation shocks where jet pinching occurs. 
At the jet head the beam decelerates to sub-relativistic velocities 
favoring the formation of a strongly magnetized termination shock
where $\vec{B}^2\approx 26\gamma^2\rho$.
Here, the presence of a predominant toroidal magnetic field induces current
driven (CD) kink instability which are responsible for the observed jet 
wiggling and symmetry breaking (see \citealp{MRBFM10}).
This test shows that \PlCh can be effectively used in simulations involving
relativistic velocities, highly supersonic flows and strongly magnetized
environments.

At $t=130$ the finest level grid has a volume filling factor of $\approx 11$
per cent with 10761 blocks while level $3$ occupies $\approx 41$ per cent
of the total volume with 5311 blocks.
For this particular application, the benefits offered by grid adaptivity are
more evident at the beginning of the computation when most of the computational
zones in the domain are unrefined.
In this respect, we have found that the CPU time increases (for $t \le 130$)
with the number of steps $n$ approximately as $a+bn+cn^2+dn^3$ where
$a\approx 1.46$, $b\approx 0.079$, $c\approx 5.03\cdot10^{-4}$ and
$d\approx 2.38\cdot 10^{-6}$.
This suggests that, as long as the filling factor remains well below $1$, the
AMR calculation with the proposed refinement criterion should be at least
$\sim 2.5$ times faster than an equivalent computation using static mesh
refinement and considerably larger if compared to a uniform grid run at the
effective resolution.

\section{Summary}
\label{sec:summary}
%
%
%
%

In this paper, we have presented a cell-centered implementation of the 
PLUTO code for multi-dimensional adaptive mesh refinement (AMR) computations
targeting Newtonian and relativistic magnetized flows.
A block structured approach has been pursued by taking full
advantage of the high-level parallel distributed infrastructure 
available in the CHOMBO library.
This choice provides the necessary level of abstraction in delivering 
AMR functionality to the code allowing, at the same time, full 
compatibility with most of the modular implementations already available 
with the static grid version.
This eases up the process of adding or replacing physics modules as 
long as they comply with the interface requirements.

A novel extension to incorporate diffusion terms such as viscosity, resistivity
and heat conduction without the need for operator splitting, has been illustrated
in the context of the dimensionally unsplit corner transport upwind (CTU)
time stepping scheme.
In addition, the scheme has also been extended to the realm of relativistic 
MHD (RMHD) using a characteristic projection-free, MUSCL-Hancock normal 
predictor step. 
The integration scheme retains second-order spatial and temporal accuracy 
requiring 6 Riemann solvers per cell per step. 
Interface states are calculated using the piecewise parabolic method 
(PPM) integration although alternatives based on weighted essentially 
non-oscillatory (WENO) or linear slope-limited schemes are also available.
The proposed cell-centered version of PLUTO-CHOMBO enforces the divergence-free
condition by augmenting the system of equations with a generalized Lagrange
multiplier \citep[GLM, see][]{Dedner02}, in the implementation outlined by
\cite{MT10}. 
The choice of the GLM-MHD formalism has proven to be a convenient starting 
point in porting a significant fraction of the static grid code implementations 
to the AMR framework.
Although future extensions will also consider other strategies to 
enforce the $\nabla\cdot\vec{B}=0$ condition, the proposed formulation 
offers substantial ease of implementation and a viable robust 
alternative to a constrained transport approach which typically requires 
additional care for prolongation and restriction operations at fine-coarse 
interfaces \citep[see, for instance,][]{Balsara01, Balsara04, AstroBEAR09}.

A suite of several test problems including standard numerical benchmarks and 
astrophysical applications for MHD and RMHD has been selected to assess the 
efficiency of PLUTO-CHOMBO in resolving complex flow patterns viable through an 
adaptive grid approach in 1, 2 and 3 dimensions.
Examples include simple shock-tube problems, resistive sheets, radiative 
and thermally conducting flows, fluid instabilities and blast wave 
explosions in highly magnetized environments.
The computational saving offered by an adaptive grid computation has been shown, 
for many of the selected test, to be several times or even order of magnitudes
faster than a static grid integration.
Parallel performance, evaluated for some of the proposed tests, suggests
good scalability properties for two- and three-dimensional problems as long
as the number of grids per processor is larger than a factor between 
$2$ and $3$.

Both the static and AMR version of \PLUTO are distributed as  
a single software package publicly available at 
\texttt{http://plutocode.ph.unito.it/} while the CHOMBO library can be freely 
downloaded from \texttt{https://seesar.lbl.gov/anag/chombo/}.                        
The AMR version of \PLUTO has been designed to retain salient features 
characterizing the static grid version \citep{PLUTO} such as
modularity - the possibility to easily combine different numerical schemes 
to treat different physics-, portability and user-friendliness.
While stable code releases along with extensive documentation and benchmarks 
are made available on the WEB, the code is being actively developed and future 
development will address additional new physical aspects as well as improved 
numerical algorithms.



\acknowledgements
This work has been supported by the PRIN-INAF 2009 grant.
We acknowledge the CINECA Awards N. HP10CJ1J54 and N. HP10BHHHEJ, 2010 under
ISCRA initiative for the availability of high performance computing 
resources and support.
Intensive parallel computations were performed on the 4.7 GHz IBM Power 6 
p575 cluster running AIX 6 or the 0.85 GHz IBM BlueGene/P cluster running CNL.

\appendix
\section{Discretization of the Thermal Conduction Flux}
\label{app:satTC}
%
%
%
%
%

The discretization of the heat conduction flux, Eq. (\ref{eq:tc_flux}),
reflects the mixed parabolic/hyperbolic mathematical nature of the underlying 
differential operators, as anticipated in Section \ref{sec:nonideal}.
From the diffusion limit, $|\vec{F}_{\rm class}|/q \to 0$, we compute 
$\vec{F}_{\rm class}$ at cell interfaces using second-order accurate central 
difference expressions for $\nabla T$, e.g., at constant $x$-faces,
\begin{equation}
 \left.\pd{T}{x}\right|_{i+\HALF} \approx \frac{T_{i+1,j}-T_{i,j}}{\Delta x} 
 \,,\qquad
 \left.\pd{T}{y}\right|_{i+\HALF} \approx 
  \frac{(T_{i+1,j+1}+T_{i,j+1}) - (T_{i+1,j-1}+T_{i,j-1})}{4\Delta y}
\end{equation}
and similarly at constant $y$- and $z$- faces.
In the purely hyperbolic limit, $|\vec{F}_{\rm class}|/q\to\infty$, we see that 
$\vec{F}_c \to q \hat{\vec{t}}$ where 
$\hat{\vec{t}} = \vec{F}_{\rm class}/|\vec{F}_{\rm class}|$ is a unit vector in 
the direction of the heat flux.
To gain more insights, we re-write the one-dimensional energy equation in this 
limit by keeping only pressure-related terms, that is,
\begin{equation}\label{eq:sat_eq}
  \pd{}{t}\left(\frac{p}{\gamma-1}\right) - \pd{}{x}\left(5\phi\rho c_{\rm iso}^3t_x\right) = 0 \,,
\end{equation}
where $c_{\rm iso} = \sqrt{p/\rho}$ is the isothermal speed of sound
and $t_x = \hat{\vec{e}}_x\cdot\hat{\vec{t}}$.
Equation (\ref{eq:sat_eq}) is a nonlinear advection equation of 
the form $\partial_tu + \partial_x f=0$ where $u = p/(\gamma-1)$ and 
$f = -qt_x$ is the 
hyperbolic saturated flux.
A stable discretization is therefore provided by adopting an upwind scheme
\citep{BTH08} such as
\begin{equation}\label{eq:sat_flux}
  q_{i+\HALF} = \left\{\begin{array}{ll}
\DS   \frac{5\phi}{\sqrt{\rho_{i+\HALF}}}p_L^{3/2} & \qquad{\rm if}\quad  t_x < 0 \,,\\ \noalign{\medskip}
\DS   \frac{5\phi}{\sqrt{\rho_{i+\HALF}}}p_R^{3/2} & \qquad{\rm otherwise}   \,.\\ \noalign{\medskip}
\end{array}\right.
\qquad
\end{equation}
In the previous equation $\rho_{i+\HALF}=(\rho_L+\rho_R)/2$, $p_L$ and $p_R$ are 
computed from the left/right input states for the Riemann problem.
Finally, we define the thermal conduction flux at constant $x$-faces 
(for instance) as the harmonic mean between the two regimes,
\begin{equation}
  \left(\hat{\vec{e}}_x\cdot\vec{F}_c\right)_{i+\HALF} = 
  \frac{q_{i+\HALF}}{|\vec{F}_{\rm class}|_{i+\HALF} + q_{i+\HALF}}
  \left(\hat{\vec{e}}_x\cdot\vec{F}_{\rm class}\right)_{i+\HALF}
\end{equation}
where $q_{i+\HALF}$ is computed from Eq. (\ref{eq:sat_flux}).

\section{Multidimensional Shock Detection and Adaptive Hybrid-Integration}
\label{app:MULTID_Flattening}
%
%
%
%
%

In regions of strong shocks or gradients spurious numerical oscillations 
may arise and quickly lead to the occurrence of unphysical states
characterized, for example, by negative pressures, energies, densities or, 
in the case of relativistic flows, superluminal speeds.
Such episodes are usually limited to very few grid zones and are originated 
by either insufficient numerical diffusion or lack of a physical solution of 
the Riemann problem.

To circumvent this potential hitch, \PLUTO provides a safeguard built-in 
mechanism that i) flag zones that may potentially reside inside a 
strong shock wave and ii) introduces additional numerical dissipation by 
locally replacing the integration scheme with a more diffusive one.
For these reasons, most of the interpolation schemes and Riemann solvers 
available with \PLUTO embody a hybrid selective mechanism that allows
to switch, if required, to a more dissipative choice.
This process is controlled by a 3D array of integers where each element 
represents a set of flags that can be individually turned on or off by 
simple bitwise operations.
Zones are flagged according to a flexible shock-detection criterion,
\begin{equation}
\frac{\left|\Delta_x q\right|}{\min\left(q_{i+1},q_{i-1}\right)} +
\frac{\left|\Delta_y q\right|}{\min\left(q_{j+1},q_{j-1}\right)} +
\frac{\left|\Delta_z q\right|}{\min\left(q_{k+1},q_{k-1}\right)}
 > \epsilon_q
 \qquad{\rm\emph{and}}\qquad
  \nabla\cdot\vec{v} < 0\,,
\end{equation}
where $\Delta_d$ is a standard central difference operator, $q$ is thermal
(default) or magnetic pressure, $\vec{v}$ is the 
velocity and $\epsilon_q$ is a free adjustable parameter (default is 5).
The first condition detects zones within a strong gradient while the 
second one is switched on where compressive motion takes place.
When both conditions are met, we flag the zone $\{i,j,k\}$ to be updated
with the HLL Riemann solver. 
At the same time we also flag all neighboring zones whose interpolation 
stencil includes the shocked zone to be interpolated using 
slope-limited reconstruction with the MinMod limiter.
We note that both flags may be selected and combined independently 
and, in any case, preserve the second-order accuracy of the scheme.


\end{document}